\documentclass[twocolumn,twoside,aps,superscriptaddress,prb,floatfix]{revtex4-2}
\usepackage[caption=false]{subfig}
\usepackage{graphicx}
\usepackage{times}
\usepackage{amssymb}
\usepackage{amsfonts}
\usepackage{amsmath}
\usepackage{amsbsy}
\usepackage{upgreek}
\usepackage{bm}
\usepackage{natbib}
\usepackage{comment}
\usepackage{xcolor}

\usepackage[hidelinks]{hyperref}
\hypersetup{
    colorlinks,
    citecolor=blue,
    filecolor=blue,
    linkcolor=blue,
    urlcolor=blue
}

\newcommand{\bra}[1]{\left\langle    {#1} \right|}
\newcommand{\ket}[1]{\left|          {#1} \right\rangle}
\newcommand{\braket}[3]{\left\langle {#1} \left| {#2}\right|#3 \right\rangle}
\newcommand{\brkt}[1]{\left\langle   {#1} \right\rangle}

\newcommand{\brktprod}[2]{\left\langle {#1}\!\left| \right.\! {#2} \right\rangle}
\newcommand{\brakettext}[3]{\langle {#1} | {#2} |#3 \rangle}
\newcommand{\brkttext}[1]{\langle   {#1} \rangle}

\newcommand{\PHDG}{^{\vphantom{\dagger}}}

\newcommand{\ZZ}{{\hspace{-0.75em}\phantom{\big)}}_z}
\newcommand{\vvv}[1]{\mbox{\boldmath$\mathrm{#1}$}}
\renewcommand{\vec}[1]{\mathrm{\bf{#1}}}

\newcommand{\sig}[2]{\hat{\sigma}^{#1}_{#2}}
\newcommand{\sigvec}[1]{\hat{\bm{\sigma}}^{\vphantom{y}}_{#1}}

\newcommand{\sigcrosvec}[2]{\sigvec{#1}\times\sigvec{#2}}
\newcommand{\sigcros}[3]{\left(\sigcrosvec{#1}{#2}\right)_{#3}}
\newcommand{\sigcroszz}[2]{\left(\sigcrosvec{#1}{#2}\right)\ZZ}

\newcommand{\sigcroscomp}[5][-]{\left(\sigvec{#2}\times\left(\sigvec{#3}{#1}\sigvec{#4}\right)\right)_{#5}}

\newcommand{\sigmiss}[4][-]{\sigvec{#2}\cdot\sigvec{#3}{#1}\sig{#4}{#2}\sig{#4}{#3}}
\newcommand{\sigmisscomp}[5][-]{\sigvec{#2}\cdot\left(\sigvec{#3}{#1}\sigvec{#4}\right)-\sig{#5}{#2}\left(\sig{#5}{#3}{#1}\sig{#5}{#4}\right)}

\newcommand{\sigdotcomp}[4][-]{\sigvec{#2}\cdot\left(\sigvec{#3}{#1}\sigvec{#4}\right)}

\newcommand{\spin}[2]{\hat{S}^{#1}_{#2} }

\newcommand{\del}[2]{\delta^{\vphantom{x}}_{#1,#2}}
\newcommand{\codel}[2]{\delta^{#1,#2}_{\vphantom{x}}}
\newcommand{\lev}[3]{\varepsilon^{#1,#2,#3}_{\vphantom{x}}}

\newcommand{\ud}{\,\mathrm{d}}

\newcommand{\im}{\mathrm{i}}
\newcommand{\e}{\textrm{e}}

\newcommand{\VB}{v_{\mathrm{B}}}

\DeclareMathOperator{\Tr}{Tr}

\begin{document}

\title{Directional scrambling of quantum information in helical multiferroics}

\author{M.~Sekania}
\email[]{Mikheil.Sekania@physik.uni-halle.de}
\affiliation{Institut f\"ur Physik, Martin-Luther Universit\"at Halle-Wittenberg, 06099 Halle/Saale, Germany}
\author{M.~Melz}
\affiliation{Institut f\"ur Physik, Martin-Luther Universit\"at Halle-Wittenberg, 06099 Halle/Saale, Germany}
\author{N.~Sedlmayr}
\affiliation{Institute of Physics, Maria Curie-Sk\l{}odowska University, Plac Marii Sk\l{}odowskiej-Curie 1, PL-20031 Lublin, Poland}
\author{Sunil~K.~Mishra}
\affiliation{Department of Physics, Indian Institute of Technology (Banaras Hindu University), Varanasi - 221005, India}
\author{J.~Berakdar}
\affiliation{Institut f\"ur Physik, Martin-Luther Universit\"at Halle-Wittenberg, 06099 Halle/Saale, Germany}

\begin{abstract}
Local excitations as carriers of quantum information spread out in the system in ways governed by the underlying interaction and symmetry. Understanding this phenomenon, also called quantum scrambling, is a prerequisite for employing interacting systems for quantum information processing.
The character and direction dependence of quantum scrambling can be inferred from the out-of-time-ordered commutators (OTOCs) containing information on correlation buildup and entanglement spreading. 
Employing OTOC, we study and quantify the directionality of quantum information propagation in oxide-based helical spin systems hosting a spin-driven ferroelectric order.
In these systems, magnetoelectricity permits the spin dynamics and associated information content to be controlled by an electric field coupled to the emergent ferroelectric order.
We show that topologically nontrivial quantum phases, such as chiral or helical spin ordering, allows for electric-field controlled anisotropic scrambling and a direction-dependent buildup of quantum correlations.
Based on general symmetry considerations, we find that starting from a pure state (e.g., the ground state) or a finite temperature state is essential for observing directional asymmetry in scrambling.
In the systematic numerical studies of OTOC on finite-size helical multiferroic chains, we quantify the directional asymmetry of the scrambling and verify the conjectured form of the OTOC around the ballistic wavefront.
The obtained direction-dependent butterfly velocity $\VB(\vec{n})$ provides information on the speed of the ballistic wavefront.
In general, our calculations show an early-time power-law behavior of OTOC, as expected from an analytic expansion for small times.
The long-time behavior of OTOC reveals the importance of \text{(non-)integrability} of the underlying Hamiltonian as well as the implications of conserved quantities such as the $z$-projection of the total spin.
The results point to the potential of spin-driven ferroelectric materials for the use in solid-state-based quantum information processing.
\end{abstract}
\date{\today}

\maketitle


\section{Introduction and general considerations}

A sudden quench in a parameter entering a many-body Hamiltonian results in a reshuffling of quantum information during the subsequent time evolution \cite{Heyl2018,Heyl2013,Vosk2014,Eisert2015,Ponte2015,Azimi2014,Azimi2016,Peschel2009}. 
Although unitary dynamics is reversible, meaning a closed system remembers its initial state, local information can disperse into many-body quantum entanglements and correlations that are distributed over the entire system and become inaccessible to local measurements, i.e. the initial local information is scrambled \cite{Sekino2008,Lashkari2013}.
This concept goes along the dynamics of thermalization in closed quantum systems \cite{Deutsch1991,Srednicki1994,Tasaki1998,Rigol2008,Sekino2008,Lashkari2013,Sirker2014,Sedlmayr2018} and has recently been discussed as a tool for characterizing chaos in black holes, for example \cite{Hayden2007,Shenker2014,Kitaev2014,Kitaev2015,Maldacena2016}.
While a precise definition of quantum scrambling is somehow elusive, the
out-of-time-order correlation functions are mathematically well-defined and offer a compelling witness of scrambling.

Considering two operators $\hat{W}$ and $\hat{V}$ which act as local perturbations on the system with a Hamiltonian $\hat{H}$, the {\em out-of-time-ordered commutator} (OTOC) is defined as
\begin{equation}
\label{eq:OTOC1}
 C(t)
 =
 \brkt{
   [
     \hat{W}(t)
     ,
     \hat{V}
   ]^\dag
   [
     \hat{W}(t)
     ,
     \hat{V}
   ]
 },
\end{equation}
where 
${\hat{W}\left(t\right)=U^\dagger(t)\hat{W}U(t)}$ and ${\hat{V}\left(t\right)}$ are the Heisenberg pictures of $\hat{W}$ and $\hat{V}$, respectively, and ${\hat{U}(t)=\exp(-\im\hat{H}t)}$.
The angle brackets $\brkt{\cdot}$ in Eq.~(\ref{eq:OTOC1}) denote either the expectation value on a pure state of interest (typically quantum-mechanical ground state), ${\brkttext{\cdot}\equiv\brakettext{\psi}{\cdot}{\psi}}$, or a finite-temperature thermal average, $\brkttext{\cdot}\equiv\Tr(\hat{\rho}\,\cdot)$, with a density matrix
$\hat{\rho}=\e^{-\beta \hat{H}}/\Tr(\e^{-\beta \hat{H}})$, the inverse temperature ${\beta=1/T}$, and the Boltzmann constant scaled to ${k_B=1}$.
Expansion of $C(t)$ in $t$ contains both time-ordered and out-of-time-order correlators, hence the name OTOC.
In the case of unitary operators $\hat{V}$ and $\hat{W}$, one can rewrite OTOC in the alternative but equivalent form ${C(t)=2(1-\mathrm{Re}[F(t)])}$ with
\begin{equation}
 \label{eq:OTOC1_FT}
  F(t)=\brkt{ \hat{W}^{\dag}(t)\hat{V}^{\dag}\hat{W}(t)\hat{V}},
\end{equation}
which is also referred to as OTOC (where ``C'' stands for correlator) in the literature but in what follows, we use Eq.~\eqref{eq:OTOC1} for this abbreviation.
For a pure state, $\ket{\psi}$, $F(t)$ relates to the fidelity of the process when the order of applied operators is reversed, ${\ket{\phi_1}=\hat{W}(t)\hat{V}\ket{\psi}}$, ${\ket{\phi_2}=\hat{V}\hat{W}(t)\ket{\psi}}$, and OTOC is $F(t)=\brktprod{\phi_2}{\phi_1}$.


The concept of OTOC was first introduced in the late '60s by Larkin and Ovchinnikov \cite{Larkin1969} in the context of quasi-classical approaches to quantum systems. It received renewed interest recently \cite{Kitaev2014,Shenker2015,Kitaev2015} as it offers a quantifiable perspective on the emergence of quantum chaos and information propagation in quantum many-body systems \cite{Maldacena2016,Aleiner2016,Roberts2015,Roberts2016}. Although the OTOC was originally proposed for diagnosing quantum chaos, recently, OTOC has found use in studying the dynamics of quantum many-body systems. 
Due to fundamental relevance but also to the importance for quantum information processing, research on entanglement and quantum-information delocalization by OTOC is steadily increasing \cite{Yao2016,Hosur2016,Yoshida2019,Roberts2017,Zhou2019,Fan2017,Swingle2017,Chen2017,Slagle2017,Smith2019,Huang2017,HeLu2017,Wei2018,Bohrdt2017,Chapman2018,Swingle2016,Yao2016,Zhu2016,Campisi2017,Danshita2017,Garttner2017,Dag2019,Patel2017,Shen2017,Heyl2018a,DagSun2019,Grozdanov2018,Iyoda2018,Abeling2018,Dora2017,LinMotrunich2018_2,LinMotrunich2018,Riddell2019,Khemani2018VDLE,Khemani2018,Keyserlingk2018,Nahum2018,XuSwingle2019,Klug2018,Rakovszky2018,Syzranov2018,Iyoda2018,McGinley2019,Chen2020,Campo2017,Halpern2018}, including many-body-localized (MBL) systems \cite{Swingle2017,Fan2017,Chen2017,Huang2017,Bohrdt2017,HeLu2017,Slagle2017,Wei2018,Smith2019}, Luttinger-liquids \cite{Dora2017,LinMotrunich2018_2,Abeling2018}, and random unitary circuits \cite{Roberts2017,Khemani2018,Keyserlingk2018,Nahum2018,Zhou2019,XuSwingle2019}.

OTOCs carry multifaceted information. For example, OTOCs may serve as indicator for 
static \cite{Shen2017,Heyl2018a,DagSun2019} and dynamical phase transitions \cite{Heyl2018a,Chen2020}, and 
 are useful in distinguishing between many-body and Anderson localization \cite{Bohrdt2017,Chen2017,Fan2017,Huang2017,HeLu2017,Slagle2017,Swingle2017,Chapman2018,Wei2018,Smith2019}.
OTOC can also be related \cite{Hosur2016,Yao2016,Fan2017,Roberts2017,Yoshida2019,Zhou2019} to the second R\`enyi entropy of an appropriately defined subsystem. 
The quasiprobability behind the OTOC and its connection to the pseudorandomness has been studied in Refs.~\cite{Halpern2017,Halpern2018,Alonso2019,Roberts2017}.
While scrambling can be captured by entropic terms (cf. \cite{Shenker2014,Hosur2016}) or measured in terms of the tripartite information of a subsystem \cite{Hosur2016,Iyoda2018}, OTOCs are more accessible experimentally. 
Several experimental realizations have been discussed in the literature \cite{Roberts2016,Swingle2016,Yao2016,Zhu2016,Campisi2017,Danshita2017,Garttner2017,Halpern2018,Dag2019}.
 Early experiments are based on a variety of quantum-simulator platforms such as nuclear spins of molecules \cite{Li2017,Wei2018,Chen2020}, trapped ions \cite{Garttner2017,Landsman2019,Joshi2020}, and ultracold gases \cite{Meier2019}.
OTOC, Eq.~\eqref{eq:OTOC1}, is closely related to another probe of chaos, namely to the thermal average of Loschmidt echo signals \cite{Yan2020} providing a link to the familiar diagnostic that captures the dynamical aspect of chaotic behavior in the time domain and is accessible to experimental studies.

It is instructive to consider the semi-classical interpretation of scrambling.
Considering a chaotic system and taking $V$ and $W$ as the canonical momentum ${V\equiv p}$ and coordinate operators ${W(t)\equiv q(t)}$, for the short times ${C(t)=\hbar^{2}\exp\left(2\lambda_{L}t\right)}$ applies \cite{Maldacena2016}. 
The scrambling time is specified in terms of the classical Lyapunov exponent $\lambda_{L}$ and is equal to the Ehrenfest time $\tau\approx(1/\lambda_{L})\ln(1/\hbar)$. 
The Lyapunov exponent $\lambda_L$ is unbounded for classical systems.
For bounded operators and unitary evolution, however, OTOC is also bounded. Hence, it cannot diverge exponentially and saturates \cite{Kukuljan2017}.
Nevertheless, at short-times, before the saturation is reached, an exponential growth of OTOC may occur with a Lyapunov exponent bounded with the conjectured value $\lambda_L \leqslant 2\pi k_B T/\hbar$ \cite{Maldacena2016}.
This behavior is found in semiclassical and large-$N$ models \cite{Aleiner2016,Kitaev2015,Maldacena2016,XuSwingle2019} but not in physical systems with local Hamiltonians and finite on-site degrees of freedom.
Quantum systems that saturate this bound are known as fast scramblers \cite{Roberts2015,Fu2016,MaldacenaStanford2016}.
In contrast, a range of models with local Hamiltonians and finite on-site degrees of freedom exhibit a power-law early-time growth instead of exponential \cite{Swingle2017,Chen2017,Dora2017,Huang2017,Slagle2017,Fan2017,Deng2017,LinMotrunich2018,LinMotrunich2018,LinMotrunich2018_2,Khemani2018VDLE,Riddell2019,Smith2019}, and are therefore known as slow scramblers.

In local-Hamiltonian systems with spatial structure,
the maximum rate at which correlations build up is limited by the
Lieb-Robinson (LR) bound \cite{Lieb1972}, defined for local bounded operators $\hat{V}_{\vec{x}}$, $\hat{W}_{\vec{0}}$, with an initial support at ${\vec{x}}$, and ${\vec{0}}$, respectively, as
\begin{equation}
 \label{eq:LR}
  \lim_{\substack{t\rightarrow \infty \\ |{\vec{x}}|>vt}} \left\|\left[\hat{W}_{\vec{0}}(t),\hat{V}_{\vec{x}}\right]\right\|\e^{\mu(v)t}=0.
\end{equation}
It applies for all ${v>v_{\mathrm{LR}}}$, with ${\mu(v)>0}$ a positive increasing function.
The Lieb-Robinson velocity $v_{\mathrm{LR}}$ is the minimum speed for which Eq.~\eqref{eq:LR} holds, and it defines an emergent ``light-cone'' causality from local dynamics on a lattice \cite{Calabrese2009}. It is {\em a state-independent} microscopic velocity set by the magnitude of couplings in the Hamiltonian.
The function $\mu(v)$ bounds the exponential decay rate along the different constant velocity rays ${|\vec{x}|=vt> v_{\mathrm{LR}} t}$ outside the light-cone.

Based on LR bound, given by Eq.~\eqref{eq:LR}, the velocity-dependent Lyapunov exponent $\lambda(\vec{v})$ can be introduced \cite{Khemani2018VDLE} which quantifies the exponential growth or decay rate of the OTOC along a given velocity ($\vec{v}$) rays, ${\vec{x}=\vec{v}t}$:
\begin{equation}
 \label{eq:OTOC1_LV}
  C_{\vec{x}=\vec{v}t}(t)
  \sim
  \e^{\lambda(\vec{v})t}\,,
\end{equation}
with generally state dependent ${\lambda(\vec{v})}$.
From this perspective, the OTOC~\eqref{eq:OTOC1_LV} can be viewed as {\em a state-dependent} LR bound, 
which is helpful for studies of zero or finite temperature dynamics (cf. Ref.~\cite{Roberts2016,Han2019}).
For infinite temperature Eq.~\eqref{eq:OTOC1_LV} and LR become equivalent.
Here OTOC, as well as $\lambda(\vec{v})$, are explicit functions of the direction of the velocity.
For the one-dimensional case, this corresponds to the ``left'' or ``right'' direction.
The explicit directional dependence is relevant when we consider helical systems or helical-states and evaluate OTOC.

In spatially local systems that exhibit a ballistic spread of quantum information (linear light cone), a universal form of OTOC for the region close to the wavefront has been conjectured \cite{XuSwingle2019,Khemani2018VDLE}
\begin{equation}
 \notag
  C_{\vec{x}=\vec{v}t}(t) \sim \exp\left(-c_1\frac{(|\mathbf{x}|/\VB(\vec{\hat{n}}) - t)^{\alpha}}{t^{\alpha-1}}\right)\,,
\end{equation}
or equivalently \cite{Khemani2018VDLE}
\begin{equation}
 \label{eq:OTOC1_alpha_vb}
  C_{\vec{x}=\vec{v}t}(t) \sim \exp\left(-c_2(v-\VB(\vec{\hat{n}}))^\alpha t \right)\,,
\end{equation}
which includes the velocity-dependent Lyapunov exponent ${\lambda(\vec{v})} = {-c_2(v-\VB(\vec{\hat{n}}))^\alpha}$ and the direction-dependent butterfly velocity $\VB(\vec{\hat{n}})$ characterizing the speed of the ballistic spreading of OTOC.
The natural upper bound of this velocity is given by LR velocity, ${\VB(\vec{\hat{n}})\leqslant v_{\mathrm{LR}}}$.
The existence of a negative velocity-dependent exponent outside the wavefront also follows directly from the LR type bounds \cite{Roberts2016,Han2019} which also applies to nonchaotic integrable systems that display ballistic operator spreading \cite{Calabrese2009}.
Generally, disorder can impede correlation buildup and a different ansatz for localized systems is needed.
For example, in noninteracting disordered systems correlations (including OTOC) do not spread beyond the localization length \cite{Chen2017,Huang2017}, and these systems satisfy the so-called zero-velocity LR bound \cite{Hamza2012},
whereas, in MBL systems, they extend beyond the localization length, exhibiting the so-called logarithmic light-cone behavior \cite{Huang2017,Deng2017,Slagle2017,Smith2019}.
The shape of the wavefront Eq.~\eqref{eq:OTOC1_alpha_vb} is characterized by a single parameter $\alpha$ that depends on the studied system.
${\alpha=1}$, a simple exponential growth, is characteristic only for semiclassical or large-$N$ models, e.g., Sachdev-Ye-Kitaev (SYK) model \cite{Sachdev1993} and chains of coupled SYK dots with large-$N$ local dimension \cite{Kitaev2015,MaldacenaStanford2016,Schmitt2019,Gu2017,XuSwingle2019}, exhibiting a sharp wavefront.
These cases are reminiscent of the classical butterfly effect.
${\alpha>1}$, is generally attributed to the broadening of the wavefront during the propagation, and is typical for lattice systems with local interactions.
${\alpha = 2}$, implying a diffusive broadening of the wavefront, is found for random circuit models in one dimension \cite{Khemani2018VDLE,Nahum2018,LinMotrunich2018_2,XuSwingle2019,Keyserlingk2018},
whereas, ${\alpha=3/2}$ applies to general noninteracting systems with translational invariance \cite{Xu2020,LinMotrunich2018,Khemani2018VDLE,XuSwingle2019,Riddell2019}.

Most reported studies are done for systems with symmetric (direction independent) correlation buildup and entanglement spreading.
Anyonic statistics is found to induce asymmetric spreading of quantum information with asymmetric OTOC and light cones in nonequilibrium dynamics of Abelian anyons in a one-dimensional system \cite{LiuGarrison2018}.
Quite recently, similar findings have been made for parafermion (non-Abelian anyons) chains, even for inversion-invariant Hamiltonians~\cite{Zhang2021}.
Some early studies~\cite{Stahl2018} demonstrated the possibility of asymmetric scrambling for explicitly constructed Hamiltonian comprised out of solely asymmetric local interaction terms.
Reference~\cite{Zhang2020} presented a family of integrable Hamiltonians with asymmetric information spreading showing that anyonic particle statistics is not a necessary condition. Asymmetric transport prevails even when interaction terms are considered that render the system non-integrable.
For the latter case, the left/right butterfly velocities were also obtained by fitting the shape of OTOC near the ballistic wavefront to the universal form given by Eq.~\eqref{eq:OTOC1_alpha_vb}.

We are interested in a possible control of correlation buildup and entanglement spreading of quantum spin excitations in oxide-based chains~\cite{Masuda2004,*Masuda2005,Gippius2004,*Drechsler2005,Papagno2006,Mihaly2006,Park2007,Naito2007,Schrettle2008,Miyata2021,Sundaresan2021,Ruff2019,Enderle_2005,Guo_2018} such as $\text{LiCu}_2\text{O}_2$~\cite{Masuda2004,*Masuda2005,Gippius2004,*Drechsler2005,Papagno2006,Mihaly2006,Park2007} and $\text{LiCuVO}_4$~\cite{Naito2007,Schrettle2008}, which host helical spin ordering resulting in an emergent spin-driven ferroelectric phase~\cite{Katsura2005,Mostovoy2006,Sergienko2006,Tokura2014}. Such systems are not only interesting for use in solid-state-based quantum information processing but also provide a bridge to the broader class of magnetoelectrics and multiferroics that have a variety of (spin)electronic applications~\cite{Huang2020,Spaldin2019,Chen2019,Chu_2018,Fiebig2016,Wang2003,Ramesh2007,Bibes2008,Fiebig2005,Hemberger2007,Menzel2012,Cheong2007,Chotorlishvili2013,Khomeriki2016,Stagraczynski2017}. External electric and magnetic fields can affect, in a controlled way, the chiral order in spin-driven magnetoelectrics~\cite{PhysRevLett.95.087206,PhysRevLett.102.057604,doi:10.1126/science.1242862,doi:10.1126/science.1260561}.
Here, we envisage the use of these external fields  to create/control the directionality of information scrambling (equivalently spatial information spreading) and to study the form of the information spreading and its character.

We consider the local (single site or bond) perturbations, e.g., local spin flips, and probe the scrambling with similar local (single site or bond) operators.
These shortest wave-length perturbations allow   probing the entire dispersion band of  elementary excitations and the direction-dependent maximal group velocities in the underlying system.
We show that topologically nontrivial quantum phases, such as chiral or helical spin ordering, allows for electric-field controlled anisotropic scrambling and a direction-dependent buildup of quantum correlations.
We analyze the left-right asymmetric scrambling and determine the directional dependent butterfly velocities in the cases with conserving $SU(2)$ or $U(1)$ symmetries.
Assisted with exact numerical results, we assess the sensitivity of OTOC to the \text{(non-)integrability} of the studied models. Since the spin ordering induces ferroelectricity, it is possible to act indirectly on the spin via an external electric field (that couples to the ferroelectric polarization) and modify the dynamic of OTOC, as will be demonstrated below.

A complementary approach to the investigated short-wavelength limit will be OTOC with the low-energy large wave-length excitations, where the probing is done with similar low-energy large wave-length detectors.
%

The  paper is organized as follows: Sec.~\ref{sec:model} specifies the mathematical model; Sec.~\ref{sec:analytical_results} introduces left-right-asymmetry measures for OTOC and by virtue of symmetry the set of non-trivial cases is identified. Also, analytical results for the early time regime and ${L=4}$-spin model are presented.
Sec.~\ref{sec:otoc_num_res} contains a discussion of the numerical results for spin chains of ${L=22}$ and ${L=102}$ sites, including the implications of chirality on the directional asymmetry of scrambling. We also discuss the short- and long-time limits for ${L=22}$ sites and verify the conjectured universal form Eq.~\ref{eq:OTOC1_alpha_vb} around the ballistic wavefront for spin chains of ${L=102}$ sites. We also present results for the directional dependence of the butterfly velocity, and determine the wavefront shape parameter $\alpha$.
A summary in Sec.~\ref{sec:conclusion} concludes the paper.
Technicalities and detailed calculations are deferred to Appendices.

\section{Theory and Model Hamiltonian}
\label{sec:model}

The Hamiltonian of the studied helical system with spin-driven ferroelectricity reads 
\begin{align}
 \label{eq:Hamiltonian}
  \begin{split}
  \hat{H}
  &= J_1 \sum^{L}_{i=1}\hat{\vec{S}}_i \cdot \hat{\vec{S}}_{i+1}
   + J_2 \sum^{L}_{i=1}\hat{\vec{S}}_i \cdot \hat{\vec{S}}_{i+2}
   - \sum^L_{i=1} B^z_i\spin{z}{i}                               \\
  &+ D \sum^{L}_{i=1} \left(\hat{\vec{S}}_{i} \times \hat{\vec{S}}_{i+1} \right)\ZZ,
   \qquad\qquad
   D=E_y g\PHDG_{\mathrm{ME}}.
 \end{split}
\end{align}
The {$L$} quantum {spins} positioned at sites $i$ along the $x$-axis are described by spin ($1/2$) operators $\hat{\vec{S}}_i$.
The nearest-neighbor exchange interaction is ferromagnetic (${J_1<0}$) whereas the next-nearest-neighbor is antiferromagnetic (${J_2>0}$) resulting generally in a frustrated spin order.
Typical values, e.g., for $\text{LiCu}_2\text{O}_2$ are ${J_1 = -11 \pm 3}$~meV and ${J_2\approx 7 \pm 1}$~meV~\cite{Gippius2004,*Drechsler2005,Park2007}.
External electric and magnetic (in general site dependent) fields are applied along $y$ ($E_y$) and $z$ ($B^z_i$) axes, respectively.
The emergent spin-driven ferroelectric polarization follows from ${\hat{\vec{P}}=g\PHDG_{\mathrm{ME}}\hat{\vec{e}}\PHDG_{x} \times \hat{\vvv{\upkappa}}}$, 
where $\hat{\vec{e}}_x$ is the unit vector along the chain, ${\hat{\vvv{\upkappa}} = \sum^L_{i=1}\hat{\vvv{\upkappa}}\PHDG_{i} = {\sum^L_{i=1} (\hat{\vec{S}}_i \times \hat{\vec{S}}_{i+1})}}$ is the vector spin chirality, and $g\PHDG_{\mathrm{ME}}$ is the magnetoelectric coupling constant.
An external electric field (in our case along the $y$ axis), which can be generated via dielectric or liquid ion gating \cite{Leighton2019}, couples to $ {\hat{\vec{P}}}$ as ${-{\hat{\vec{P}}}\cdot{\bf E}\PHDG_y}={D \sum^{L}_{i=1} (\hat{\vec{S}}_{i} \times \hat{\vec{S}}_{i+1} )\PHDG_z}$.
The coefficient $D$ encompasses the electric field strength and the magnetoelectric coupling constant and mimics an electric-field tunable inverse Dzyaloshinskii Moriya (DM) interaction term. Hence, depending on the strength and direction of the external fields the system can be driven to the chiral or nonchiral phase.
This term breaks explicitly the Dihedral group $\mathbb{D}_2$ symmetry:
It is symmetric with respect to time inversion, $\hat{\mathcal{T}}$, but antisymmetric with respect to spin-flip, $\hat{\mathcal{Z}}$ (e.g., ${\hat{\mathcal{Z}}=\prod^L_{i=1} 2\spin{x}{i}}$ or ${\hat{\mathcal{Z}}=\exp(\im \pi\sum^L_{i=1} \spin{x}{i})}$), or a spatial inversion, $\hat{\mathcal{P}}$.
The vector-spin-chirality order parameter, ${\vvv{\upkappa}=\brkt{\hat{\vvv{\upkappa}}}}$, is nonzero in the helical phase and disappears for collinear spin ordering \cite{Katsura2005,Mostovoy2006,Sergienko2006,Tokura2014}.
Of interest here is the impact of the chirality on the delocalization of quantum information, that signifies the loss of information under time evolution, meaning scrambling.

A unitary local rotation of the spins about the $z$ axis by the angle ${\vartheta =-\arctan\left(D/J_{1}\right)}$, ${\spin{+}{j} \rightarrow \spin{+}{j}e^{-\im j\vartheta}}$, converts the Hamiltonian \eqref{eq:Hamiltonian} to
\begin{align}
 \label{eq:Hamiltonian2}
\begin{split}
  \hat{H}_T
  &=
  \frac{J^\prime_1}{2} \sum^L_{i=1}\left(\spin{+}{i} \spin{-}{i+1}+\spin{-}{i} \spin{+}{i+1}\right)
  +
  J_1 \sum^L_{i=1} \spin{z}{i} \spin{z}{i+1}
  \\
  &
  + \frac{J^\prime_2}{2} \sum^L_{i=1} \left(\spin{+}{i} \spin{-}{i+2}+\spin{-}{i} \spin{+}{i+2}\right)
  + J_2 \sum^L_{i=1}\spin{z}{i} \spin{z}{i+2}
  \\
  &
  - D^\prime \sum^L_{i=1}\left(\hat{\vec{S}}_{i} \times \hat{\vec{S}}_{i+2} \right)\ZZ
  - \sum^L_{i=1}  B^z_i\spin{z}{i}.
\end{split}
\end{align}
Here ${\spin{\pm}{i}=\spin{x}{i}\pm\im \spin{y}{i}}$ are spin raising/lowering operators on site $i$, ${J^\prime_1 = \sqrt{J_1^{2}+D^{2}}}$, $J^\prime_{2}=J_{2}\left(J_{1}^{2}-D^{2}\right)/\left(J_{1}^{2}+D^{2}\right)$, and $D^\prime=DJ_{1}J_{2}/\left(J_{1}^{2}+D^{2}\right)$. 
Depending on the values of parameters, in the case of a homogeneous magnetic field (${B_i^z=B_z}$), the ground state of Hamiltonians \eqref{eq:Hamiltonian}/\eqref{eq:Hamiltonian2} can be either ferromagnetic, chiral, or nematic \cite{Azimi2014,Azimi2016}. For ${J_{2}=0}$, the DM interaction term is absorbed by the transverse-exchange term. The Hamiltonian is equivalent to the ferromagnetic {\em easy-plane} XXZ model with renormalized exchange $J^\prime_1$, exchange anisotropy $J_1/J^\prime_1=1/\sqrt{1+(D/J_1)^2}<1$, and admits an exact solution through the Bethe Ansatz. The chirality in the case of open boundary conditions (OBC) is $\kappa^z \sim J_1D/(J_1^2+D^2)$.
For ${J_{2}\neq0}$, ${D\neq0}$, the system is not integrable and displays mixed GOE/GUE level statistics in the case of randomly distributed $B_i^z$ \cite{Stagraczynski2017}.

In the case of a zero magnetic field (${B_i^z=0}$) and vanishing DM interaction (${D=0}$), the total spin $\hat{\vec{S}}_{\mathrm{tot}}=\sum_{i}{\hat{\vec{S}}_{i}}$ is conserved, $[\hat{\vec{S}}_{\mathrm{tot}},\hat{H}]=0$, and the system is $SU(2)$ symmetric.
Otherwise, for a homogeneous magnetic field (${B^z_i=B_z}$), the $z$-component of the total spin $\spin{z}{\mathrm{tot}}=\sum_{i}{\spin{z}{i}}$ is a conserved quantity, $[\spin{z}{\mathrm{tot}},\hat{H}]=0$, and our model is $U(1)$ symmetric.
Therefore, the total number of ``down'' (``up'') spins can be used to characterize any eigenstate of the system. Also, each $S^z_{\mathrm{tot}}$-subsector can be solved independently.
The magnetic field $B_z$ only causes a constant shift in the energy within each subsector and does not affect eigenstates.
Besides, the spectrum of Hamiltonian~\eqref{eq:Hamiltonian} is symmetric with respect to 
a spin-flip~\footnote{${S_i^x \stackrel{\hat{\mathcal{Z}}}{\rightarrow} S_i^x}$, ${S_i^y \stackrel{\hat{\mathcal{Z}}}{\rightarrow} -S_i^y}$, ${S_i^z \stackrel{\hat{\mathcal{Z}}}{\rightarrow} -S_i^z}$, and $\hat{H}(J_1,J_2,D,B_z) \stackrel{\hat{\mathcal{Z}}}{\rightarrow} \hat{H}(J_1,J_2,-D,-B_z)=\hat{H}(J_1,J_2,-D,B_z) + 2 B_z\sum^L_{i=1} \spin{z}{i}$.}, $\hat{\mathcal{Z}}$,
in combination with a spatial inversion~\footnote{$\hat{H}(J_1,J_2,D,B_z) \stackrel{\hat{\mathcal{P}}}{\rightarrow} \hat{H}(J_1,J_2,-D,B_z)$.}, $\hat{\mathcal{P}}$,
\begin{equation}
 \label{eq:zp_mapping}
 \hat{H}(J_1,J_2,D,B_z)
 \!\stackrel{\hat{\mathcal{P}}\hat{\mathcal{Z}}}{\longrightarrow}\!
 \hat{H}(J_1,J_2,D,B_z)+2 B_z\sum^L_{i=1} \spin{z}{i}.
\end{equation}
The fully {\em saturated} state, all spins either up or down, is a trivial the eigenstate of the Hamiltonian.
We choose the state $|0\rangle\equiv{|\!\uparrow\uparrow\dots\uparrow\rangle}$ as a ferromagnetic reference state (vacuum state) and consider ${S^z_{\mathrm{tot}} \geqslant 0}$ sectors only.
We call $M$-excitation state ($M$-magnon state) the state with $M$ spins flipped down with respect to the ferromagnetic reference state.
These states comprise the $M$-excitation/magnon sector with ${S^z_{\mathrm{tot}}=L/2-M}$.

\begin{figure}[!t]
 \includegraphics[width=\columnwidth]{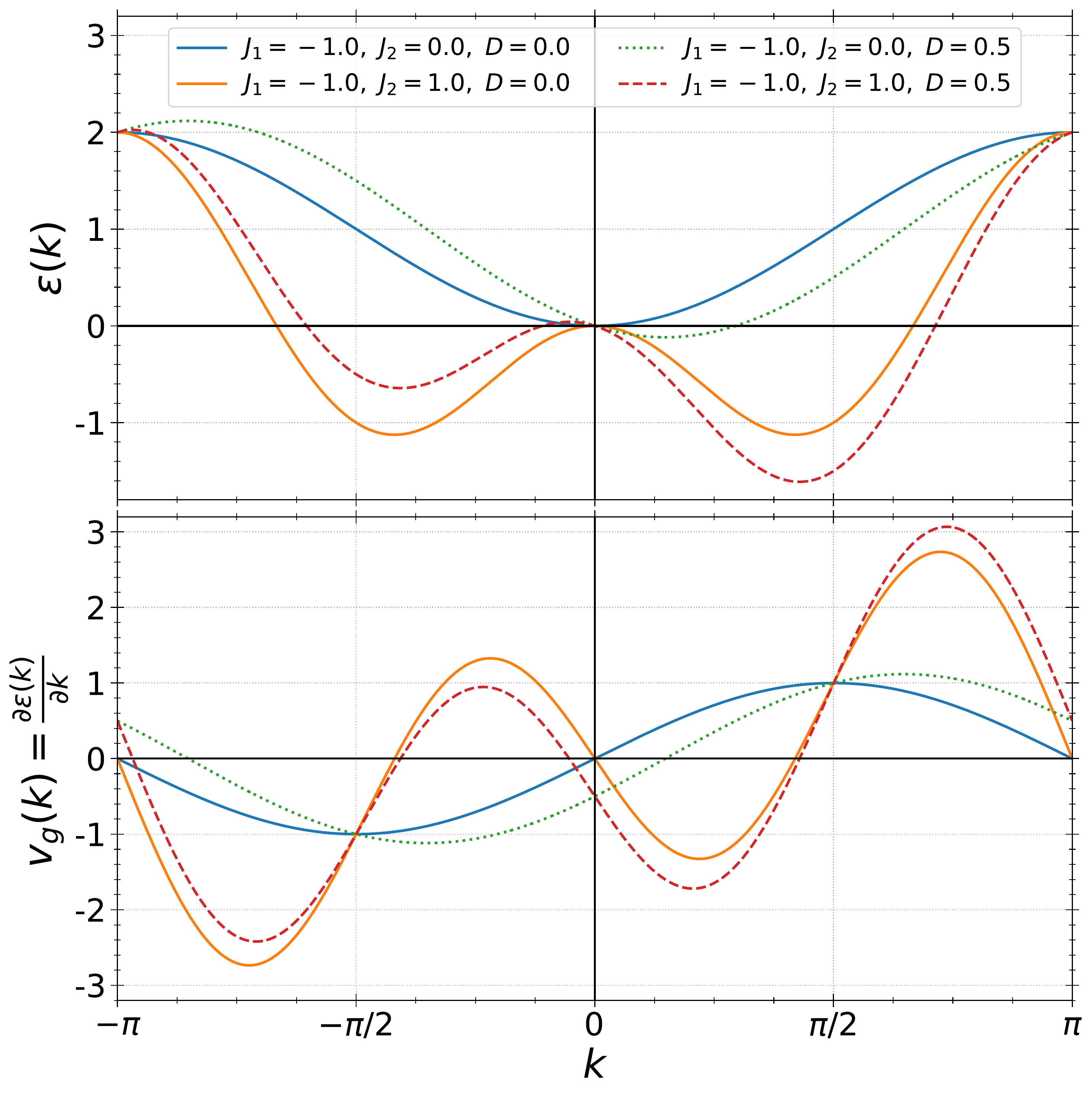}
 \caption{The dispersion relation (upper) and a corresponding group velocity (lower) for different $J_1$, $J_2$, and $D$, parameters. Bands are shifted for ${D\neq0}$ (green dotted and red dashed curves) and ${k\leftrightarrow -k}$ reflection symmetry is explicitly broken. The maximal values of the right (${v_g(k) > 0}$) and left (${-v_g(k)>0}$) group velocities differ only for ${J_2,D\neq 0}$ (red dashed curves), whereas they are equal for any $J_1,\,D$, but ${J_2=0}$ (blue and orange solid- and green dotted-curves).
 }
 \label{fig:dispersion}
\end{figure}

One can solve the one-magnon sector exactly by taking the eigenstates of the total momentum (lattice translation) operator,
$|\Psi_k\rangle=\frac{1}{\sqrt{L}}\sum_j \e^{-\im k j}|j\rangle$ with ${|j\rangle=S^{-}_j|\!\uparrow\uparrow\dots\uparrow\rangle}$, as an ansatz.
The one-magnon dispersion relation for this case reads
\begin{equation}
 \label{eq:dispersionrelations}
  \varepsilon(k) = -(J_1+J_2) + J_1 \cos k + J_2 \cos 2 k - D \sin k\,,
\end{equation}
with the one-magnon energies $\varepsilon(k_n)$ and wavevectors
${k_n={2\pi n}/{L}}$, ${n=0,\dots,L-1}$.
For finite DM interaction (${D\neq 0}$), ${\varepsilon(k) \neq \varepsilon(-k)}$ applies, resulting in a mismatch between the group velocities ${v_{g}(k)=\partial \varepsilon(k)/\partial k}$ of magnons with wavevectors $k$ and $-k$, namely ${|v_{g}(k)|-|v_{g}(-k)|=2D\cos(k)}$ (see also Fig.~\ref{fig:dispersion}, where we show the dispersion \eqref{eq:dispersionrelations} and the corresponding group velocities for the different values of $J_2$ and $D$).
For the one-magnon case, the spreading of information occurs through magnon propagation only.
A probe of scrambling with OTOC of local operators samples the entire magnon band.
Therefore the Lieb-Robinson bound for a one-magnon sector can be expressed in terms of the maximum group velocities of ``right'' ($r$) and ``left'' ($\ell$) moving magnons
\begin{equation}
 \label{eq:v_lr_LR}
  v_{\mathrm{LR}}^r=\max_{k\in[-\pi,\pi]} \frac{\partial \epsilon(k)}{\partial k},
  \ \ \ 
  v_{\mathrm{LR}}^\ell=\max_{k\in[-\pi,\pi]} \left(-\frac{\partial \epsilon(k)}{\partial k}\right)\!.
\end{equation}

A positive next-nearest-neighbor term (${J_{2} > 0}$), generates an additional peak in the dispersion relation Eq.~\eqref{eq:dispersionrelations}. Thus generally there are four local extrema in the group velocity (see Fig.~\ref{fig:dispersion}). Two positive-valued ones correspond to the ``right'' moving branch of magnons, and two negative-valued ones associated with the ``left'' moving branch. In the case of a finite $D\neq 0$, ${v_{\mathrm{LR}}^r\neq v_{\mathrm{LR}}^\ell}$. The second isolated-extremum value in each branch will generate the second wavefront behind the main one (see below Sec.~\ref{sec:otoc_num_res}).
Because the spin-flip operator $\hat{\mathcal{Z}}$ not only flips the spin but also the sign of $D$-term, the dispersion relation for the ${(L-1)}$-magnon sector is reflected (${k \leftrightarrow -k}$) as compared to the $1$-magnon sector, and $v_{\mathrm{LR}}^r$ and $v_{\mathrm{LR}}^\ell$ are exchanged.

For $J_2=0$, the dispersion relation is expressible as 
\begin{equation}
 \label{eq:dispersionrelations_J2_0}
  \varepsilon(k) = -J_1 + \sqrt{J_1^2 + D^2_{\vphantom{1}}}\, \cos(k + q) \,,
\end{equation}
which is re-scaled by ${\sqrt{J_1^2 + D^2_{\vphantom{1}}}/J_1}$ and rigidly shifted by a momentum ${q=\arctan(D/J_1)}$ (the twisted boundary conditions), as compared to the case with a vanishing DM interaction, ${\varepsilon(k) = -J_1 + J_1 \cos(k)}$ for ${D=0}$.
Therefore, the maximal left and right group velocities are equal, ${v_{\mathrm{LR}}^r= v_{\mathrm{LR}}^\ell}=\sqrt{J_1^2 + D^2_{\vphantom{1}}}$, for $J_2=0$ and any $D$ case.
Only the probes with the long-wave excitations --- OTOC with nonlocal operators --- will exhibit the dominant direction of the information spreading,
because of a finite value of the group velocity ${\partial \varepsilon(k)/\partial k \approx D}$ at ${k \rightarrow 0}$.
Those will not be considered here.

Different ${J_2 > 0}$ and ${D}$ parameters renormalize the maximal group velocities without affecting the essence of the results and conclusions.
Therefore, without loss of generality in what follows, we stick with ${J_1=-1}$, ${J_2=0,1}$, and ${D=0,\pm0.5}$. Time will be measured in units of ${1/|J_1|}$.

\section{OTOC Analytical results}
\label{sec:analytical_results}

For simplicity let us set ${g\PHDG_{\mathrm{ME}}=1}$ and study the spreading of the quantum information with pairs of unitary operators ${\hat{W}=\exp(\im \spin{\alpha}{n})}$, ${\hat{V}=\exp(\im\spin{\alpha}{m})}$, and ${\hat{W}=\exp\left(\im\hat{\kappa}^{z}_{n}\right)}$, ${\hat{V}=\exp\left(\im\hat{\kappa}^{z}_{m}\right)}$,
where $\alpha$ is any of ${\{x,y,z\}}$ 
and $\hat{\kappa}^z_n$ is the longitudinal component of the vector spin chirality operator, $\hat{\vvv{\upkappa}}_n$,
\begin{equation}
 \label{eq:chirality_z}
 \hat{\kappa}^z_n = \left(\hat{\vec{S}}_{n} \times \hat{\vec{S}}_{n+1} \right)\ZZ\,.
\end{equation}
With the operator identity
\begin{equation}
 \label{eq:exp_pauli}
 \e^{\im b \sig{\alpha}{}}
 = \cos b\,\mathbb{I} + \im\sin b\,\sig{\alpha}{}\,,
\end{equation}
where $\sig{\alpha}{}$ ($\alpha \in \{x,y,z\}$) is one of the Pauli operators,
the commutator of $\exp(\im \spin{\alpha}{n})$ and $\exp(\im \spin{\beta}{m})$ is expressed in terms of commutators of Pauli operators $\hat{\sigma}^{\alpha}_{n}$ and $\hat{\sigma}^{\beta}_{m}$ as:
\begin{equation}
 \left[
   \e^{\im \spin{\alpha}{n}(t)}
   ,
   \e^{\im \spin{\beta}{m}}
 \right]
 =
 \sin^2\!\left(\tfrac{1}{2}\right)
 \left[
   \sig{\alpha}{n}(t), \sig{\beta}{m}
 \right].
\end{equation}
 OTOC given by Eq.~\eqref{eq:OTOC1} then reads
\begin{equation}
 \label{eq:spin_w_sigmas}
 \brkt{\left|\!\left[
   \e^{\im \spin{\alpha}{n}(t)}
   ,
   \e^{\im \spin{\beta}{m}}
 \right]\!\right|^2}
 =
 \sin^4\!\left(\tfrac{1}{2}\right)
 \brkt{\left|\!\left[
   \sig{\alpha}{n}(t), \sig{\beta}{m}
 \right]\!\right|^2}.
\end{equation}
Eq.~\eqref{eq:spin_w_sigmas} already shows that OTOCs given by these two sets of unitary operators saturate to values that differ by a factor $\sin^4(1/2)$. Hence, they cannot saturate to the same finite value, as other reported studies have suggested for any unitary operators. The Pauli operators are Hermitian and unitary as well, $(\sigma^{\alpha})^{\dagger}\sigma^{\alpha}=(\sigma^{\alpha})^2=\mathbb{I}$.

One can also express OTOC, Eqs.~\eqref{eq:OTOC1}/\eqref{eq:OTOC1_FT}, with the second operator pairs (${\hat{W}=\exp\left(\im\hat{\kappa}^{z}_{n}\right)}$, ${\hat{V}=\exp\left(\im\hat{\kappa}^{z}_{m}\right)}$) in terms of the Pauli operators,
but the resulting expressions are not as compact any more, see Appendix~\ref{sec:spin_chirality}~\footnote{Note also that the Pauli operators together with identity operator build the full operator basis on the given site. Any operator can be decomposed in the sum of strings of Pauli operators plus identity.
}.

\subsection{Scrambling anisotropy measures}
\label{sec:anisotropy_measures}

Let us first introduce the notation with the site dependence of the operators made explicit:
\begin{equation}
 \label{eq:OTOC_C_nm}
  C_{nm}(t)
  =
  \brkt{
  \left[\hat{W}_m(t),\hat{V}_n\right]^\dagger\!
  \left[\hat{W}_m(t),\hat{V}_n\right]
  }
  \,.
\end{equation}
To quantify the directional asymmetry in the scrambling, we define the following left-right asymmetry measures:
\begin{align}
  \label{eq:delta_c_n_m_a}
    \Delta C^a_{n,m}(t)&=\frac{1}{2}\left(C\PHDG_{n,m\vphantom{d}}(t)-C\PHDG_{m,n}(t)\right)\,,\\
  \label{eq:delta_c_n_m_b}
    \Delta C^b_{n,d}(t)&=\frac{1}{2}\left(C\PHDG_{n,n+d}(t)-C\PHDG_{n,n-d}(t)\right)\,,
\end{align}
where the operator $\hat{W}$ is acting on the sites either to the left or to the right of $\hat{V}$.
$\Delta C^a_{nm}(t)$ \eqref{eq:delta_c_n_m_a} measures the directional asymmetry for the exchange of indices, whereas $\Delta C^b_n(t)$ \eqref{eq:delta_c_n_m_b} for a reflection of indices on the $n$-th site. $\Delta C^a_{n,n+d}(t)$ becomes equivalent to $\Delta C^b_{n,d}(t)$ in the case of a translationally invariant system.

Yet another alternative measure for the directional asymmetry utilizes the spatial inversion operator applied solely to the Hamiltonian and the state (operators $\hat{W},~\hat{V}$ are not permuted on the lattice).
Formally, this operation results in the change of sign of the DM amplitude in the Hamiltonian~\eqref{eq:Hamiltonian} ($\mathcal{\hat{\mathcal{P}}}:\,H(J_1,J_2,D) \leftrightharpoons H(J_1,J_2,-D)$) and the directional asymmetry can be defined as
\begin{align}
   \label{eq:delta_c_n_m_c}
    \Delta C^c_{n,m}(t)&=\frac{1}{2}\left(C^{D\geqslant 0}_{n,m\vphantom{d}}(t)-C^{D\leqslant 0}_{n,m}(t)\right)\,,
\end{align}
where the expectation value $C^{D\leqslant 0}_{n,m}(t)$ is taken with respect to the spatially inverted state. For a translationally invariant system $\Delta C^c_{n,n+d}(t)$ also corresponds to $C^a_{n,n+d}(t)$.

Generally, the directional anisotropy in the scrambling can be caused by a chiral term (DM interaction) in the Hamiltonian (through the time-evolved operators) or the nonvanishing chiral order in the probed state.
Besides, OTOC, given by Eq.~\eqref{eq:OTOC_C_nm}, and the explicit asymmetry measures, i.e. Eqs.~\eqref{eq:delta_c_n_m_a}-\eqref{eq:delta_c_n_m_c}, depend on the employed operators, the distance between them, and the specific lattice sites on which these operators act nontrivially.
From the latter two only the distance $d$ between the operators matters in the case of a translationally invariant system,
\begin{equation}
 C_{n,n+d}(t) \equiv C_d(t)\,.
\end{equation}
Furthermore, by inserting $\hat{\mathcal{P}}^2=\mathbb{I}$ on both sides of the squared commutator (scrambling kernel) in Eq.~\eqref{eq:OTOC_C_nm}, it can be shown that
\begin{equation}
 C^{D \geqslant 0}_d(t)=C^{D \leqslant 0}_{-d}(t)\,.
\end{equation}
Therefore, the directional asymmetry measures Eqs.~\eqref{eq:delta_c_n_m_a}-\eqref{eq:delta_c_n_m_c} fulfill the following equalities:
\begin{equation}
 \Delta C^a_{n,n+d}(t) = \Delta C^b_{n,d}(t) = \Delta C^c_{n,n+d}(t) \equiv \Delta C\PHDG_d(t)\,,
\end{equation}
\begin{equation}
 \Delta C\PHDG_d(t) = -\Delta C\PHDG_{-d}(t)\,,
\end{equation}
and
\begin{equation}
 \Delta C^{D \geqslant 0}_d(t) = \Delta C^{D \leqslant 0}_{-d}(t)\,.
\end{equation}

For the sectors with opposite magnetization, ${S^z_{\mathrm{tot}}}={\pm(L/2-M)}$,
the spectrum of the Hamiltonian \eqref{eq:Hamiltonian} is identical, aside of a constant energy shift proportional to an applied magnetic field, $B_z$ (see Sec.~\ref{sec:model}, Eq.~\eqref{eq:zp_mapping}).
This shift is irrelevant when the magnetic field is zero (${B_z=0}$) or scrambling is probed with $\sig{z}{}$-s (``$B_z$'' term commutes with these operators).
The eigenstates in these sectors are mapped onto each other by a combination of a spin-flip ($\hat{\mathcal{Z}}$) and a spatial-inversion ($\hat{\mathcal{P}}$) operators:
\begin{align}
 \hat{H}\ket{\psi}  &= (E + B_z(L-M))\ket{\psi},\\
 \hat{H}\,\hat{\mathcal{P}}\hat{\mathcal{Z}}\ket{\psi} &= (E - B_z(L-M))\hat{\mathcal{P}}\hat{\mathcal{Z}}\ket{\psi}.
\end{align}
Each state $\ket{\phi}$ of a given magnetization sector ($S^z_{\mathrm{tot}}$) has a corresponding state $\hat{\mathcal{P}}\hat{\mathcal{Z}}\ket{\phi}$ in the opposite-magnetization sector ($-S^z_{\mathrm{tot}}$) and for these states
\begin{equation}
 \label{eq:OTOC_C_opos_spin}
  C^{S^z_{\mathrm{tot}} \geqslant 0}_d(t) = C^{S^z_{\mathrm{tot}} \leqslant 0}_{-d}(t)\,.
\end{equation}
Here, the expectation values on the different sides of the equality are taken with respect to the corresponding states $\ket{\phi}$ and $\hat{\mathcal{P}}\hat{\mathcal{Z}}\ket{\phi}$, respectively.
~\footnote{Equality~\eqref{eq:OTOC_C_opos_spin} can be shown by applying the identity $(\hat{\mathcal{P}}\hat{\mathcal{Z}})^2=\mathbb{I}$ on both sides of the squared commutator (scrambling kernel) in Eq.~\eqref{eq:OTOC_C_nm}.}
From Eq.~\eqref{eq:OTOC_C_opos_spin} follows that the asymmetry $\Delta C_{d}(t)$ measured for the corresponding states has an opposite sign, i.e.
\begin{equation}
 \label{eq:delta_C_opos_spin}
  \Delta C^{S^z_{\mathrm{tot}} \geqslant 0}_d(t) = - \Delta C^{S^z_{\mathrm{tot}} \leqslant 0}_d(t)\,.
\end{equation}
As a result, for the $\hat{\mathcal{P}}\hat{\mathcal{Z}}$ symmetric state, like any eigenstate of the Hamiltonian in the zero-magnetization sector ($S^z_{\mathrm{tot}}=0$, the so-called half-filled case), the OTOC is symmetric.
The sum of OTOCs is symmetric for a pair of states that are mapped on each other by $\hat{\mathcal{P}}\hat{\mathcal{Z}}$ such as in the case of equal participation rates of the opposite-magnetization sectors, for example, at finite temperatures (${\beta^{-1}<\infty}$) for a zero magnetic field (${B_z=0}$).
At infinite temperature (${\beta=0}$)~\footnote{All eigenstates are present with the equal weights.}, the scrambling is symmetric in the case of vanishing magnetic field, and the directional asymmetry is invisible for the probe with $\sig{z}{}$-s even in the case of a finite magnetic field.

Finally, because ${\hat{\mathcal{Z}}\hat{\mathcal{P}}\hat{\mathcal{T}}}$ where $\hat{\mathcal{T}}$ is a time inversion operator leaves the system Hamiltonian unchanged,
\begin{align}
 \label{eq:opposz_assym_zpt}
 \begin{split}
   \Delta C^{a,c}_{n,n+d}(t)
  &=
  -\Delta C^{a,c}_{n,n+d}(-t)\,,
  \\
   \Delta C^{b\vphantom{,}}_{n,d}(t)
  &=
  -\Delta C^{b\vphantom{,}}_{n,d}(-t)\,,
 \end{split}
\end{align}
and only odd in $t$ terms are contributing to the directional asymmetry in scrambling.

\subsection{Short-Time Limit}
\label{sec:short_time_limit}

In this section, we determine the leading and subleading contributions to the OTOC kernel (squared commutator) in the short-time limit ${t \ll 1}$.
By expanding the Heisenberg representation of ${\hat{W}}$, ${\hat{W}}(t)=\exp(-\im \hat{H} t)\hat{W}\exp(\im \hat{H} t)$, in time $t$
\begin{align}
  \label{eq:time_expansion_W}
  \hat{W}(t)
  &=
  \hat{W}
  + \im t
  [
    \hat{H}
    ,
    \hat{W}
  ]
  + \frac{(\im t)^2}{2!}
  [
    \hat{H}
    ,
    [
      \hat{H}
      ,
      \hat{W}
    ]
  ]
  \nonumber
  \\
  &\quad
  + \frac{(\im t)^3}{3!}
  [
    \hat{H}
    ,
    [
      \hat{H}
      ,
      [
        \hat{H}
        ,
        \hat{W}
      ]
    ]
  ]
  + \cdots
  \nonumber
  \\
  &=
  \sum\limits_{n=0}^{\infty}
  \frac{\left(\im t\right)^n}{n!}
  [
    \hat{H},\hat{W}
  ]\PHDG_n
  \,,
\end{align}
where
\begin{equation}
 \label{eq:nested_commutators}
  [
    \hat{A},\hat{B}
  ]\PHDG_n
  =
  \underbrace{
  [\hat{A},[\hat{A},\ldots[\hat{A}}_n,\hat{B} ]\ldots ]]
\end{equation}
denotes the nested commutator,
the kernel of OTOC \eqref{eq:OTOC1} can be rewritten in the following form
\begin{align}
 \label{eq:time_expansion_OTOC}
  |[\hat{W}(t),\hat{V}]|^2
  &=
  \left|
    \left(
      \sum\limits_{n=1}^{\infty}
      \frac{\left(\im t\right)^n}{n!}
      \left[
        [
          \hat{H},\hat{W}
        ]\PHDG_n
        ,
        \hat{V}
      \right]
    \right)
  \right|^2
  .
\end{align}
Accordingly, the first nonvanishing commutator $[[\hat{H},\hat{W}]\PHDG_{n=\nu},\hat{V}]\neq 0$ (whereas $[[\hat{H},\hat{W}]\PHDG_{n<\nu},\hat{V}]=0$) determines the leading contribution in $t$ in the short-time limit:
\begin{align}
 \begin{split}
  &
  |[
  \hat{W}(t),\hat{V}
  ]|^2
  \approx
  \frac{t^{2\nu}}{(\nu!)^2}
  \left(
    \!
    \left|
      \left[
        [\hat{H},\hat{W}]\PHDG_\nu,
        \hat{V}
      \right]
    \right|^2
  \right.
  \\
  &
  \qquad + 
  \left.
    \frac{\im t}{\nu+1}
    2
    \,\mathrm{Im}
    \!
    \left(
      \!
      \left[
        [\hat{H},\hat{W}]\PHDG_\nu,
        \hat{V}
      \right]^\dagger
      \!
      \left[
        [\hat{H},\hat{W}]\PHDG_{\nu+1},
        \hat{V}
      \right]
    \!
    \right)
    \!
    \!
  \right).
 \end{split}
 \label{eq:time_expansion_OTOC_leaing}
\end{align}
Here, in the second line, we kept only the leading and the subleading contributions in $t$.
For operators that commute initially ${[\hat{W},\hat{V}] = 0}$ (not commute ${[\hat{W},\hat{V}]\neq 0}$), ${\nu>1}$ ($\nu =0$), the leading term is of order $t^{2\nu}$ and subleading correction to it of order $t^{2\nu+1}$.
For a translationally-invariant case, according to Eq.~\eqref{eq:opposz_assym_zpt}, the leading contribution to the scrambling is always symmetric, and only the subleading one causes the directional asymmetry. For operators that do not commute initially, this leads to a constant ($t$-independent) shift and the linear in $t$ left-right asymmetric contribution.

Early time power-law behavior, instead of initially speculated exponential one, were also reported for chains with local Hamiltonians (see Refs.~\cite{LinMotrunich2018,LinMotrunich2018_2,Riddell2019,Smith2019}). In earlier works~\cite{Dora2017} based on the results for the Luttinger-liquid, this type of behavior was viewed as a distinct feature of integrable models.

\subsubsection{Scrambling Measurements with Pauli operators}
\label{sec:sigma}

Here we only consider the cases where the scrambling is measured solely with Pauli operators,
$\hat{W}=\sig{\alpha}{m}$ and $\hat{V}=\sig{\alpha}{n}$. Both operators have single site support, the sites $m$ and $n$, respectively.
The Hamiltonian $\hat{H}(J_1,J_2,D,B_z)$ \eqref{eq:Hamiltonian} contains quadratic terms in $\sig{}{}$-s, acting on the nearest- (terms ``$J_1$'' and ``$D$'') and the next-nearest-neighbor sites (term ``$J_2$''), 
and the linear one (term ``$B_z$''). 
The commutation relations between the Pauli operators
$ \left[
    \sig{\alpha}{m},\sig{\beta}{n}
  \right]
  = 2\im\,\del{m}{n}\,\lev{\alpha}{\beta}{\gamma}\,\sig{\gamma}{n}
$,
where $\del{m}{n}$ is the Kronecker delta and $\lev{\alpha}{\beta}{\gamma}$ the completely antisymmetric tensor (Levi-Civita symbol) with ${\lev{x}{y}{z}=1}$,
yield:
\begin{align}
\label{eq:comm_H_W}
\begin{split}
  [
    \hat{H}
    ,
    \sig{\alpha}{m}
  ]
  &=
  \frac{\im J_1}{2}
      \sigcroscomp[+]{m}{m-1}{m+1}{\alpha}
  \\
  &+
  \frac{\im J_2}{2}
      \sigcroscomp[+]{m}{m-2}{m+2}{\alpha}
  \\
  &+
  \frac{\im D}{2}
  \big(
    \codel{z}{\alpha}
    \sigdotcomp[-]{m}{m-1}{m+1}
    \\
    & \qquad\quad
    -  
    \sig{z}{m}
    \left(
      \sig{\alpha}{m-1}
      -
      \sig{\alpha}{m+1}
    \right)
  \big)
  \\
  &+
  \im B_z\,
  \lev{z}{\alpha}{\beta}\sig{\beta}{n}
  \,.
\end{split}
\end{align}
The magnetic field ($B_z$) contributions (the last line) are only relevant for $\sig{x,y}{}$-s and they vanish for $\sig{z}{}$-s.
In Eq.~\eqref{eq:comm_H_W}, and in what follows, the {\em Greek superscripts} $\alpha$, $\beta$, and $\gamma$ denote any of ${\{x,y,z\}}$, the {\em Latin subscripts} $j$, $m$, $n$, are the lattice site indices,
and the summation over repeating indices is assumed except for the explicitly given $x$, $y$, or $z$.
For $\sig{\alpha}{m}$, which has a support only on a single site $m$,
the commutator with $\hat{H}(J_1,J_2,D,B_z)$ (Eq.~\eqref{eq:comm_H_W}) stretches over the nearest-neighbor sites to $m$, namely ${m\pm 1}$ (due to ``$J_1$'' and ``$D$'' terms), and the next-nearest-neighbor sites ${m\pm 2}$ (due to ``$J_2$'' term).
Therefore, the support of the resulting operator \eqref{eq:comm_H_W} expands over two additional sites in each direction.
Contributions due to ``$J_2$'' term of the Hamiltonian are similar to those from ``$J_1$'' term, with ${m \pm 2}$ instead of $m \pm 1$.
They also act only on two sites, the width of the support, however, increases by four sites in this case.
Consequently, for the nested commutators in Eq.~\eqref{eq:time_expansion_W},
the width of the support increases by four sites (two in each direction) after each iteration.

One can probe this operator spreading, for example, with yet another Pauli operator, ${\hat{V}=\sig{\alpha}{n}}$, by employing OTOC Eq.~\eqref{eq:OTOC_C_nm}.
In this case, the index of the first nonvanishing term in the expansion of the OTOC kernel, Eq.~\eqref{eq:time_expansion_OTOC}, is ${\nu = \max(1,(|n-m| + 1)\ \mathbf{div}\ 2)}$ when ${J_2\neq0}$ or ${\nu = \max(1,|n-m|)}$ when ${J_2=0}$ but ${J_1\neq0}$ or ${D\neq0}$.
Accordingly, the leading and subleading contributions in OTOC are
\begin{widetext}
\begin{align}
 \label{eq:sigma_leading}
  Q^{(0)}_{mn}(t)
  &=
  \frac{t^{2\nu}}{(\nu !)^2}
  \left|\!\left[
    [\hat{H},\sig{\alpha}{m}]\PHDG_\nu
    ,
    \sig{\alpha}{n}
  \right]\!\right|^2
  ,
  \\
  \label{eq:sigma_subleading}
  Q^{(1)}_{mn}(t)
  &=
   \frac{\im\,t^{2\nu+1}}{\nu!(\nu+1)!}
  \left(
    \left[
      [\hat{H},\sig{\alpha}{m}]\PHDG_\nu
      ,
      \sig{\alpha}{n}
    \right]^\dagger
    \!
    \left[
      [\hat{H},\sig{\alpha}{m}]\PHDG_{\nu+1}
      ,
      \sig{\alpha}{n}
    \right]
    -
    \left[
      [\hat{H},\sig{\alpha}{m}]\PHDG_{\nu+1}
      ,
      \sig{\alpha}{n}
    \right]^\dagger
    \!
    \left[
      [\hat{H},\sig{\alpha}{m}]\PHDG_\nu
      ,
      \sig{\alpha}{n}
    \right]
  \right)
  ,
  \\
  \text{with}\quad
  \nu &=
  \left\{
    \begin{array}{lcl}
      \max(1,(|n-m|+1)\ \mathbf{div}\ 2) & \qquad \qquad \text{for} \quad & J_2\neq 0 \\
      \max(1,|n-m|)                      & \qquad \qquad \text{for} \quad & J_2= 0    \\
    \end{array}
  \right.
  . \nonumber
\end{align}
\end{widetext}

In the Appendix~\ref{sec:sigma_2nd_and_3rd_order}, we explicitly evaluate these two terms for $d=|n-m|\leqslant 2$.

Spin chains allow us to define $\lambda(v)$ for arbitrary large $v$ in contrast to local quantum circuits and relativistic field theories, where there is a strict ``light cone'' beyond which even exponentially weak signaling is impossible.
For rays at fixed velocity $v$ with
 ${vt \gg 1}$ and ${t \ll 1}$ such that ${\nu=vt/2 \in \mathbb{Z}}$ (${J_2\neq 0}$) or ${\nu=vt \in \mathbb{Z}}$ (${J_2 = 0}$),
Eq.~\eqref{eq:sigma_leading} leads to the following estimate for the decay exponent $\lambda(v)$~\eqref{eq:OTOC1_LV} (see Apendix~\ref{sec:decay_exp_small_t}, Eq.~\eqref{eq:v_ln_v_appx})
\begin{equation}
 \label{eq:v_ln_v}
  \lambda(v) \approx -2 v\ln v\,.
\end{equation}
Therefore, ${|\lambda(v)|}$ grows slower than ${2 v^{\alpha}}$ where ${\alpha>1}$ at (extremely) early times.

\subsection{Exact Expressions for $L=4$ site system}

For illustrative purposes, we present analytical solutions for the system of four spins (${L=4}$) for the particular choice of operator pairs ${W=\sig{z}{n\pm 1}}$, ${V=\sig{z}{n}}$, and ${W=\exp (\im\hat{\kappa}^{z}_{n\pm 1})}$, ${V=\exp (\im\hat{\kappa}^{z}_{n})}$.

In the case of fully polarized states, 
${|\!\uparrow\uparrow\uparrow\uparrow\rangle}$ and ${|\!\downarrow\downarrow\downarrow\downarrow\rangle}$,
the corresponding Hilbert sub-spaces are one-dimensional and there is no scrambling in the system.
The two-magnon sector has a zero magnetization, ${S^z_{\mathrm{tot}}=0}$ (half-filled case), hence all asymmetric contributions vanish (see Sec.~\ref{sec:anisotropy_measures}).
The remaining one- and three-magnon sectors correspond to the case with opposite magnetization, and the asymmetric contributions only differ by a sign (see Sec.~\ref{sec:anisotropy_measures}). Therefore, we only consider the one-magnon sector and evaluate OTOC~\eqref{eq:OTOC_C_nm} for the system eigenstates with finite or vanishing chirality.

We consider the eigenstates of the Hamiltonian \eqref{eq:Hamiltonian}, 
with eigenenergies ${-(D+J_2)}$ and ${(J_1+J_2)}$,
\begin{align}
 \label{eq:one_magnon_gs_L4_Ch}
  \ket{\mathrm{Ch}}&=
    \frac{1}{2}
    \left(
         {|\!\downarrow\uparrow\uparrow\uparrow\rangle}
     -\im{|\!\uparrow\downarrow\uparrow\uparrow\rangle}
     -   {|\!\uparrow\uparrow\downarrow\uparrow\rangle}
     +\im{|\!\uparrow\uparrow\uparrow\downarrow\rangle}
    \right)
   \,,
   \\
 \label{eq:one_magnon_gs_L4_W}
  \ket{\mathrm{W}}&=
    \frac{1}{2}
    \left(
         {|\!\downarrow\uparrow\uparrow\uparrow\rangle}
     +   {|\!\uparrow\downarrow\uparrow\uparrow\rangle}
     +   {|\!\uparrow\uparrow\downarrow\uparrow\rangle}
     +   {|\!\uparrow\uparrow\uparrow\downarrow\rangle}
    \right)
   \,,
\end{align}
which have a finite ($\braket{\mathrm{Ch}}{\hat{\kappa}^z}{\mathrm{Ch}}=1/4$) and zero chirality ($\braket{\mathrm{W}}{\hat{\kappa}^z}{\mathrm{W}} = 0$).
They are also the eigenstates of the translation operator with the crystal momentum $\pi/2$ and $0$, respectively.
The states $\ket{\mathrm{W}}$ and $\ket{\mathrm{Ch}}$ are also well known from quantum information theory as {$W$-} and twisted {$W$-state}~\cite{Wilde2017}.

For small $t$, expanding OTOC up to subleading contribution in $t$, for the directional asymmetries we have
(detailed calculations are given in Appendix~\ref{sec:L4_app_one_three_magnon}):
\begin{align}
 \begin{split}
 \label{eq:scrambling4spins_Assym_Ch}
  \left|\Delta C^{\mathrm{Ch}}_{d=1}\right|
  &=
  2\,t^3 \left(J_1^3 + D^2 J_1 + 8 D J_1 J_2 \right)
  +
  \mathcal{O}\left(t^5 \right),
 \end{split}
\\
 \begin{split}
 \label{eq:scrambling4spins_Assym_W}
  \left|\Delta C^{\mathrm{W}}_{d=1}\right|
  &=
  2\,t^3 \left(D^3 + J_1^2 D + 8 D J_1 J_2 \right) 
  +
  \mathcal{O}\left(t^5 \right),
 \end{split}
\end{align}
respectively.
Expressions for $\ket{\mathrm{Ch}}$ and $\ket{\mathrm{W}}$ are similar, with only $D$ and $J_1$ exchanged (${D \leftrightarrow J_1}$).
The leading term in $t$ is the cubic one.
As expected, the scrambling is symmetric 
for vanishing DM interaction, ${D=0}$, in the case of nonchiral eigenstate $\ket{\mathrm{W}}$.

Obviously, for ${|d|=2}$, the asymmetric contribution vanishes, because ${d=2}$ and ${d=-2}$ are equivalent --- one can reach the site on ${|d|=2}$ distance from both ends of the chain with PBC.

To better quantify the directional asymmetry in scrambling, we also check OTOC with the vector spin chirality operators, ${\hat{V}_n=\exp(\im\kappa^z_n)}$ and ${\hat{W}_m=\exp(\im\kappa^z_m)}$. In this case, again, a nonchiral state exhibits the directional asymmetry only if the Hamiltonian has a nonvanishing DM interaction (the chiral term, ${D\neq 0}$).
In the short-time limit, the directional asymmetries in OTOC are:
\begin{widetext}
\begin{align}
 \label{eq:scrambling4spins_Assym_Ch_chiral_series}
  \left|\Delta C^{\mathrm{Ch}}_{d=1}\right|
  &=
  8|J_1|
  \left|
  t-\frac{1}{6}\left(7 J_1^2 + 3 \left(D+2 J_2\right)^2\right)\,t^3 + O\left(t^5\right)
  \right|
  \sin^4\!\tfrac{1}{4}\cos^2\!\tfrac{1}{4}
  \,,\\
  \left|\Delta C^{\mathrm{W}}_{d=1}\right|
  &=
 \label{eq:scrambling4spins_Assym_W_chiral_series}
  8|D\,|
  \left|
   t-\frac{1}{6}\left(D^2 + 3 \left(
                                    2J_1^2 + \left(J_1 - 2J_2\right)^2
                                    -4 J_1 \left(J_1+2 J_2\right)\cos\tfrac{1}{2} 
                              \right)
                \right)t^3
    + O\left(t^5\right)
  \right|
  \sin^4\!\tfrac{1}{4}\cos^2\!\tfrac{1}{4}
  \,,
\end{align}
\end{widetext}
respectively.
The leading term is linear in $t$ because chiralities on the neighboring bonds share the site and $[\exp(\im\kappa^z_{n\pm 1}),\exp(\im\kappa^z_{n})]\neq 0$.

Comparing Eqs.~\eqref{eq:scrambling4spins_Assym_Ch}-\eqref{eq:scrambling4spins_Assym_W} with Eqs.~\eqref{eq:scrambling4spins_Assym_Ch_chiral_series}-\eqref{eq:scrambling4spins_Assym_W_chiral_series}, it is evident that the different observables are not equally sensitive to the asymmetric spreading of quantum information.

\section{OTOC Numeric results}
\label{sec:otoc_num_res}

We performed systematic numerical studies of OTOC by employing exact techniques.
The ground state for the given parameters is obtained by generalized block Davidson exact-diagonalization methods and
unitary time evolution is carried out by the Krylov's subspace Arnoldi method.
The latter iteratively computes the product of the matrix exponential with a given vector without explicitly constructing the matrix exponential.

\begin{figure*}[!htbp]
  \begin{minipage}[b]{0.6325\linewidth}
  \subfloat{\includegraphics[width=0.495\textwidth]{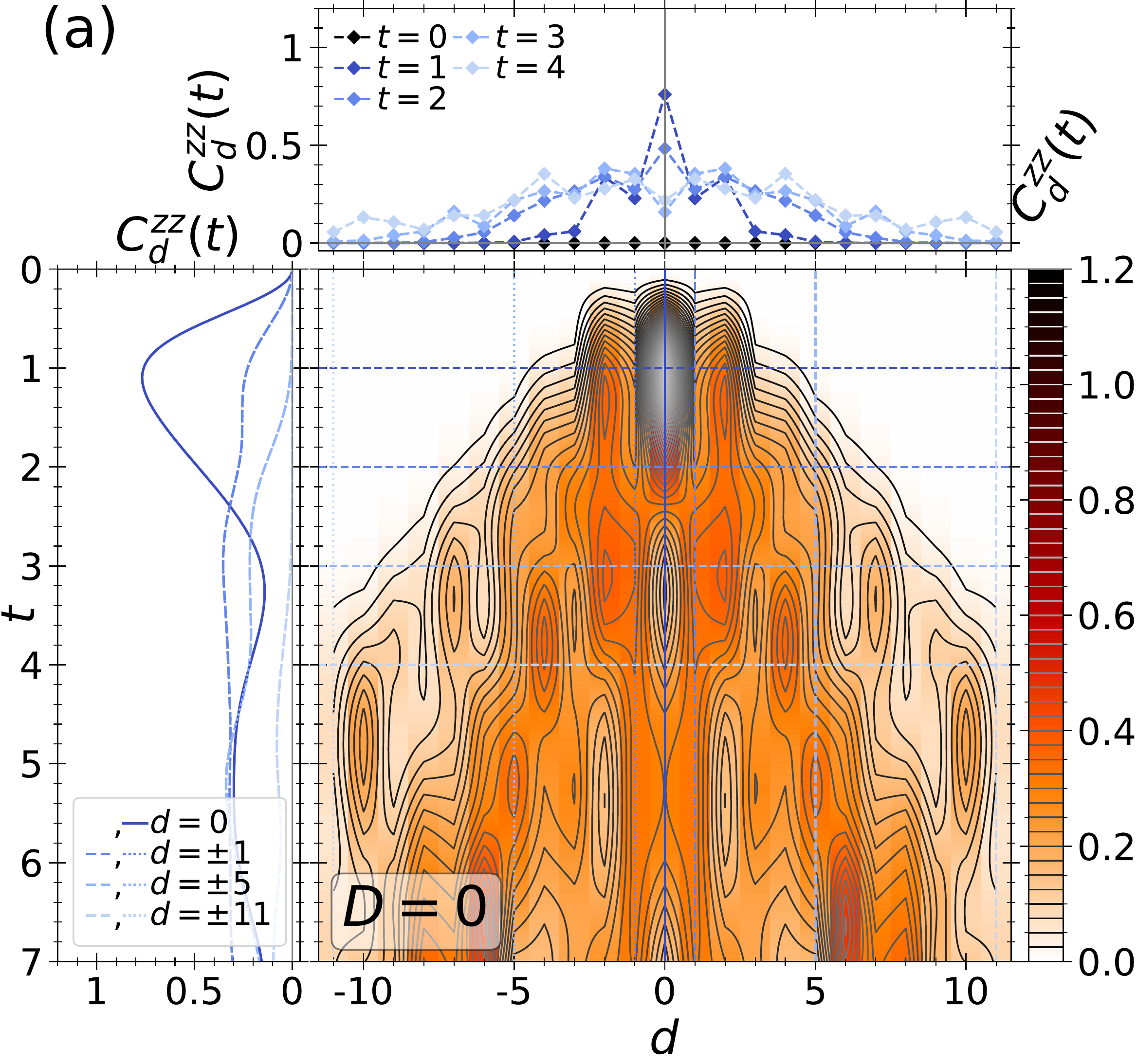}}
  \hfill
  \subfloat{\includegraphics[width=0.495\textwidth]{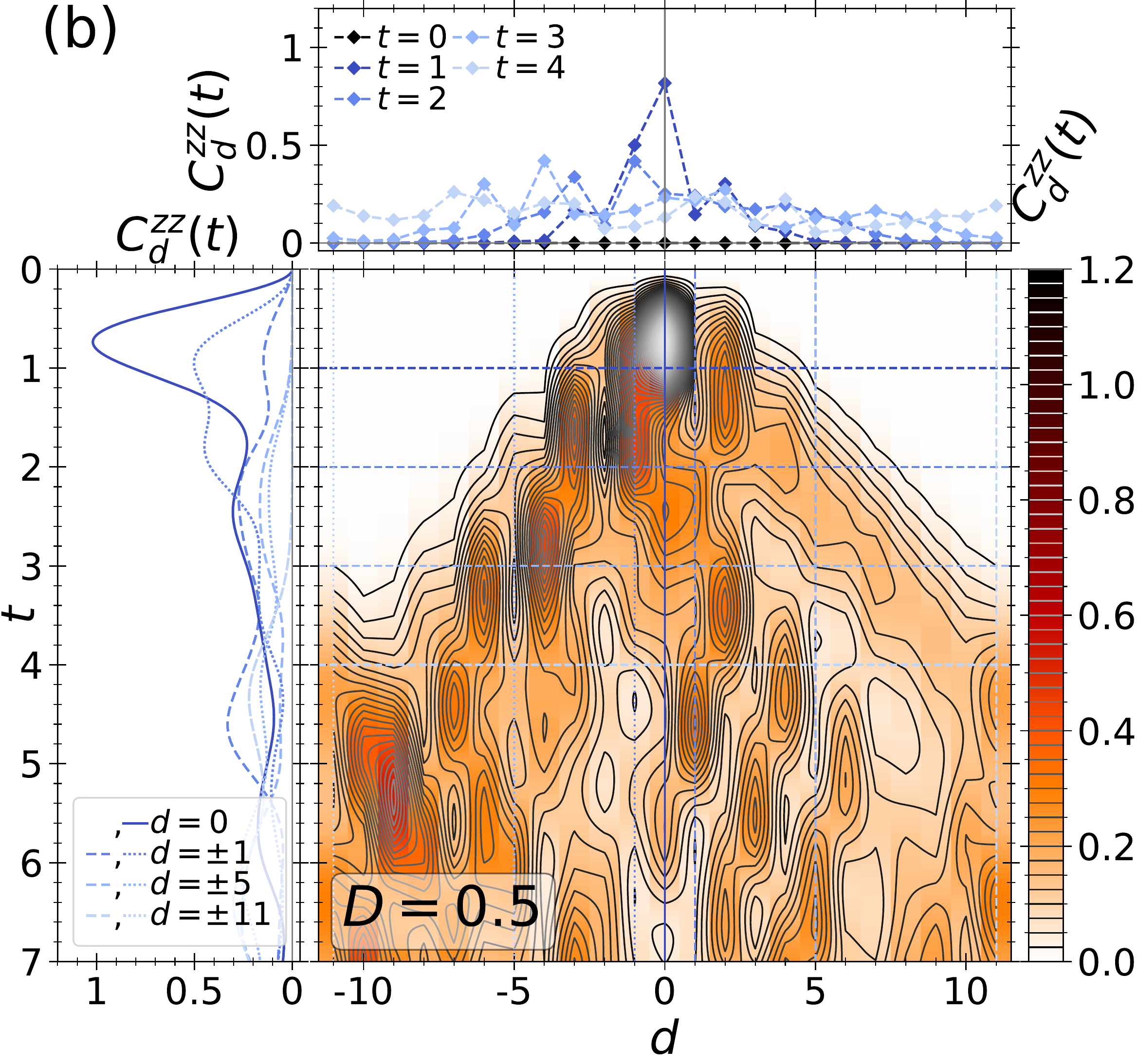}}\\
  \vfill
  \subfloat{\includegraphics[width=0.495\textwidth]{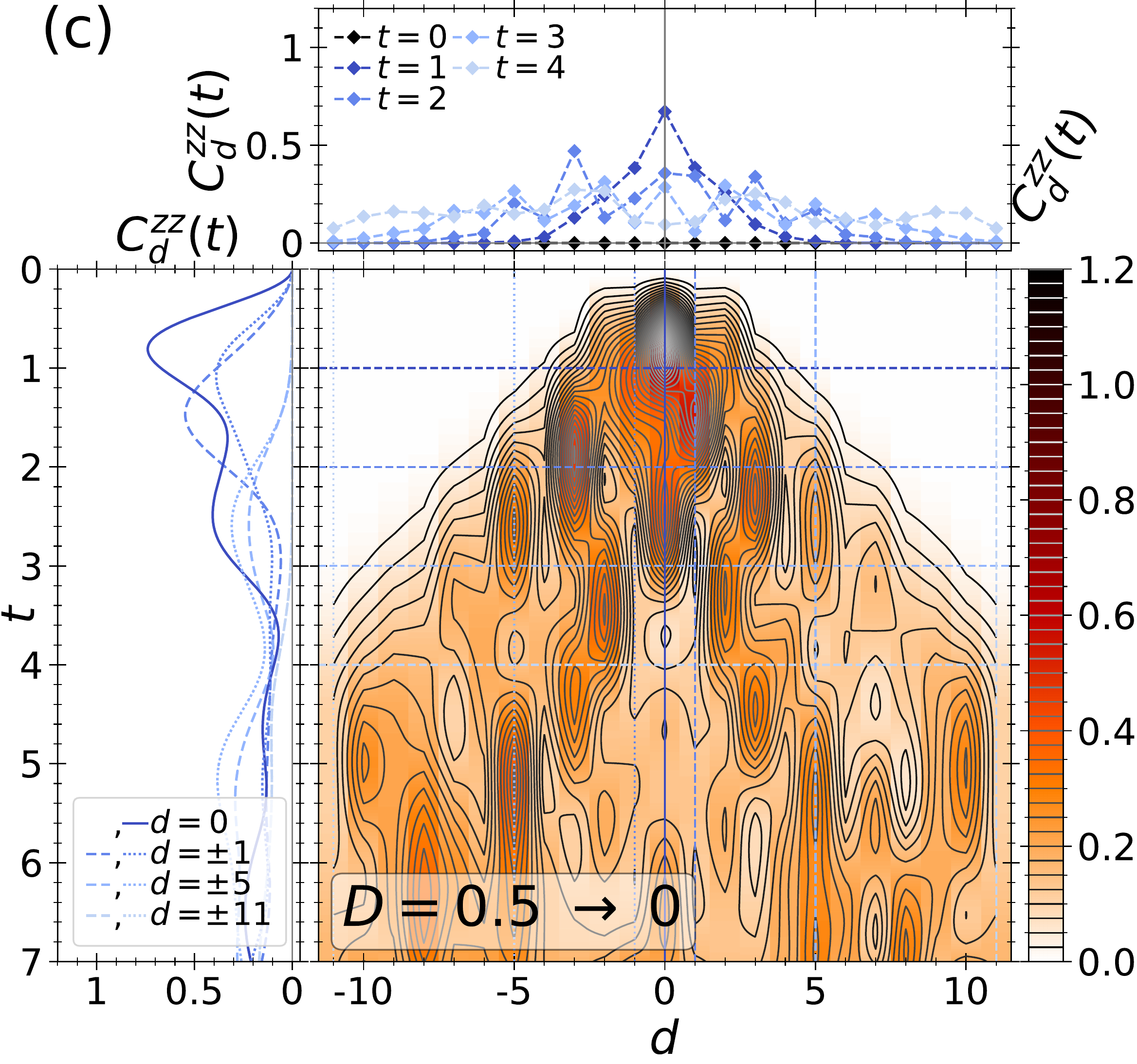}}
  \hfill
  \subfloat{\includegraphics[width=0.495\textwidth]{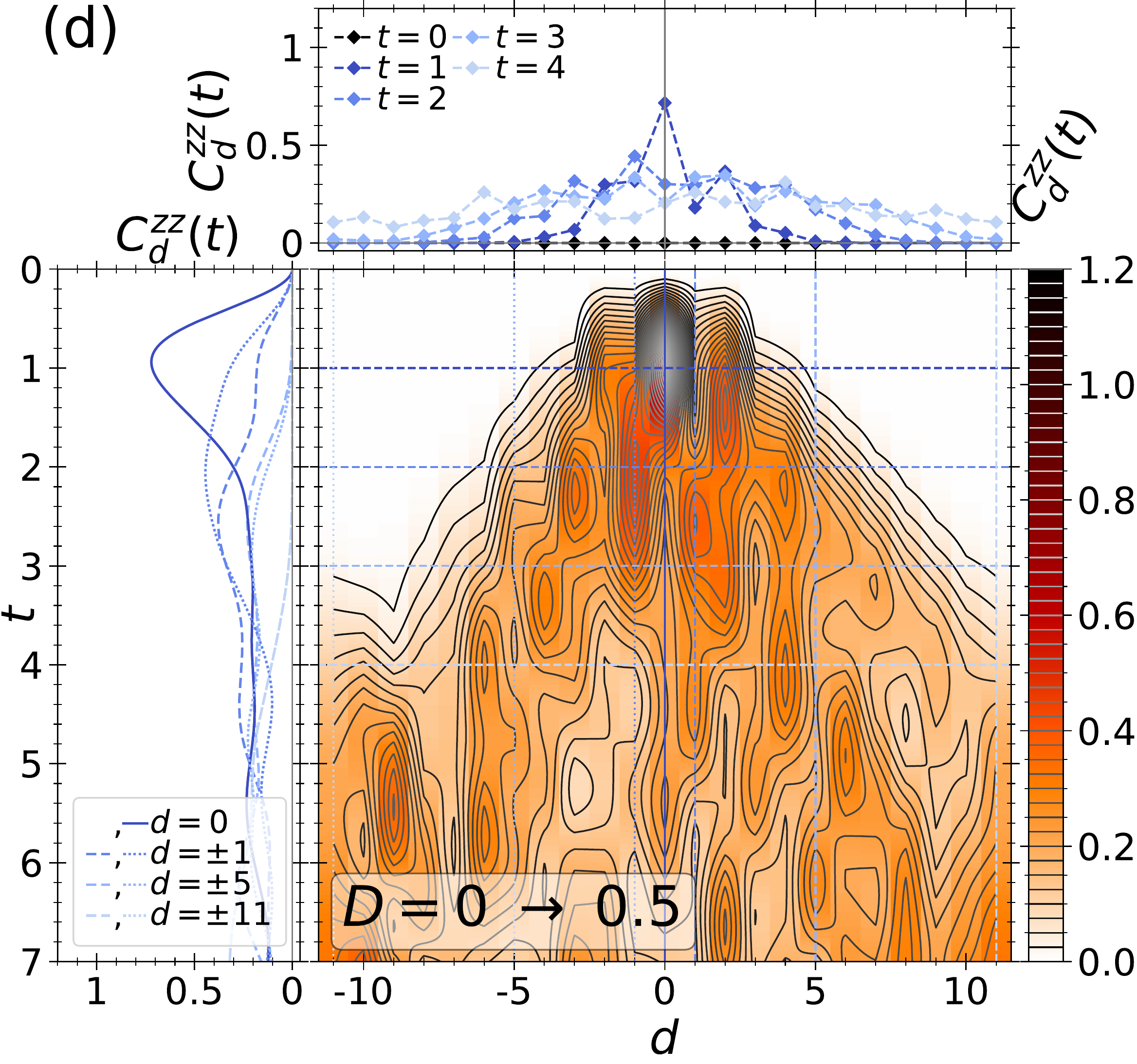}}
  \end{minipage}
  \hfill
  \begin{minipage}[b]{0.3575\linewidth}
  \subfloat{\includegraphics[width=\textwidth]{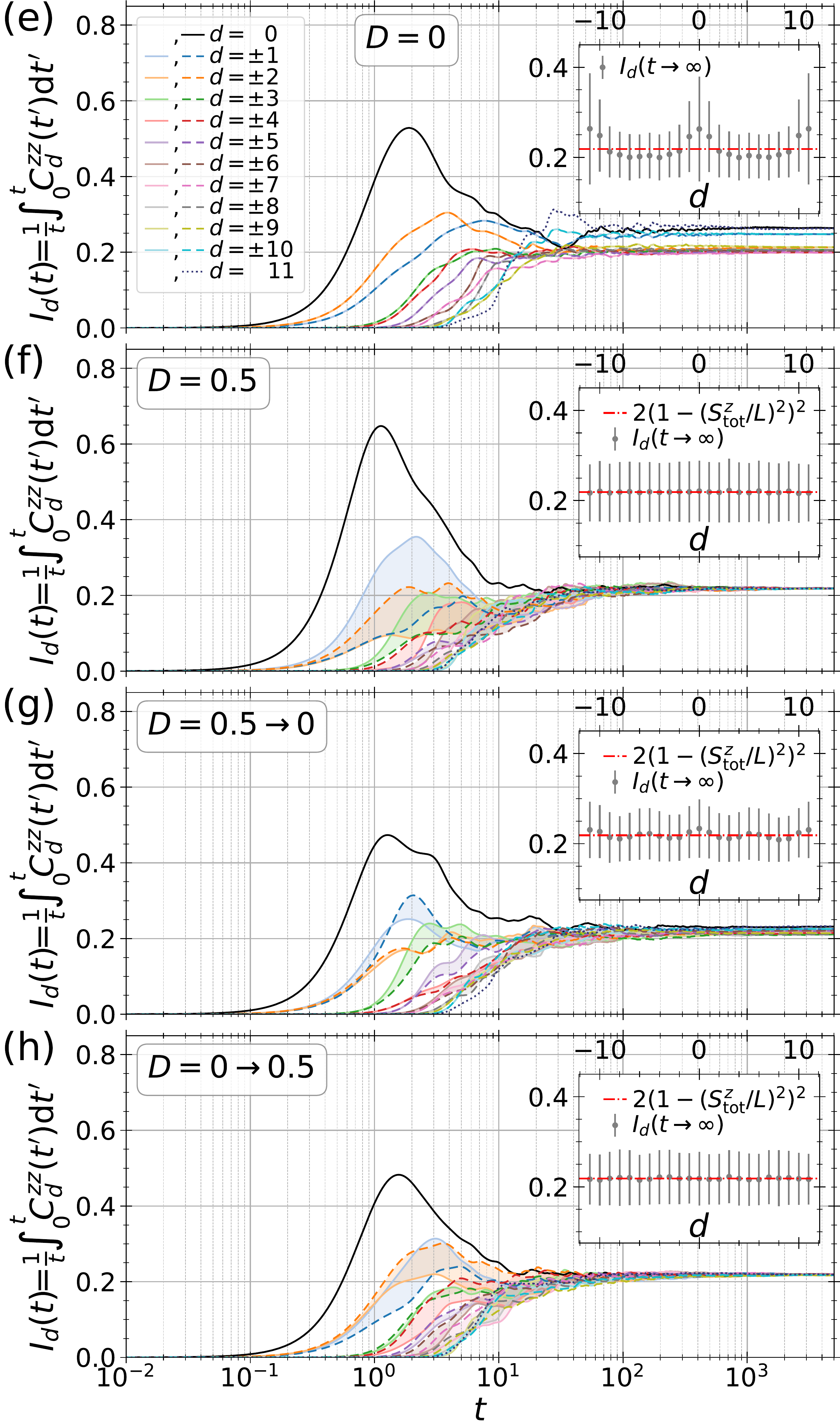}}
  \end{minipage}
 \caption{Spatiotemporal evolution of the OTOC $C^{zz}_d(t)$ ((a)-(d)) and time-average of it ((e)-(h)) for nonchiral- ((a),(e) and (d),(h)) and chiral-states ((b),(f) and (c),(g)),
          in $2$-excitation sector (${S^z_{\mathrm{tot}}=(L/2-2)}$) of ${L=22}$ spin chain with PBC; ${J_2=-J_1=1}$; time is measured in units of $|J_1|^{-1}$.
          The initial chiral- or nonchiral-state is prepared as the ground state of the system with a vanishing (${D=0}$) or finite (${D=0.5}$) DM interaction, respectively.
          Plots (c),(g) and (d),(h): DM interaction is quenched at ${t=0}$ ${D=0.5\rightarrow0}$ (c),(g) and ${D=0\rightarrow0.5}$ (d),(h).
          Plots (a)-(d): Contour lines are interpolated to non integer $d$; the left and the top subplots correspond to vertical and horizontal cross-sections of the main plot at distances $d=0,\pm1,\pm5,\pm11$ and at times $t=0,1,2,3,4$, respectively; a white spot around ${d=0}$ ${t=0.75}$ is due to contour lines.
          Insets in plots (e)-(g) show the time-averaged values with the standard deviation for $t\geqslant 400$ values; red dash-dotted line corresponds to the value expected in the case of complete scrambling.
         }
\label{fig:czzd}
\label{fig:longtime_czzd}
\end{figure*}

\begin{figure*}[!htbp]
  \begin{minipage}[b]{0.6325\linewidth}
  \subfloat{\includegraphics[width=0.495\textwidth]{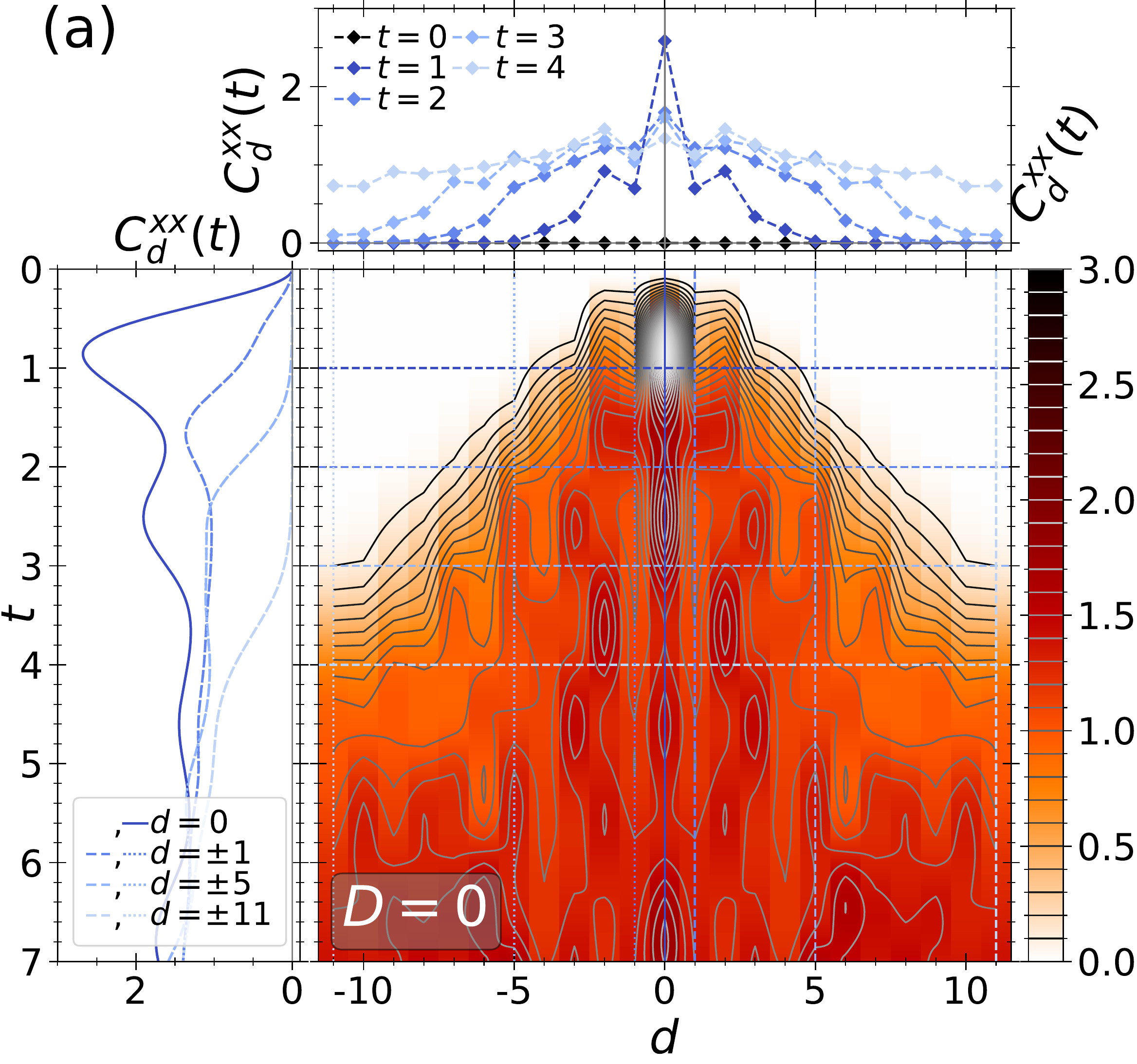}}
  \hfill
  \subfloat{\includegraphics[width=0.495\textwidth]{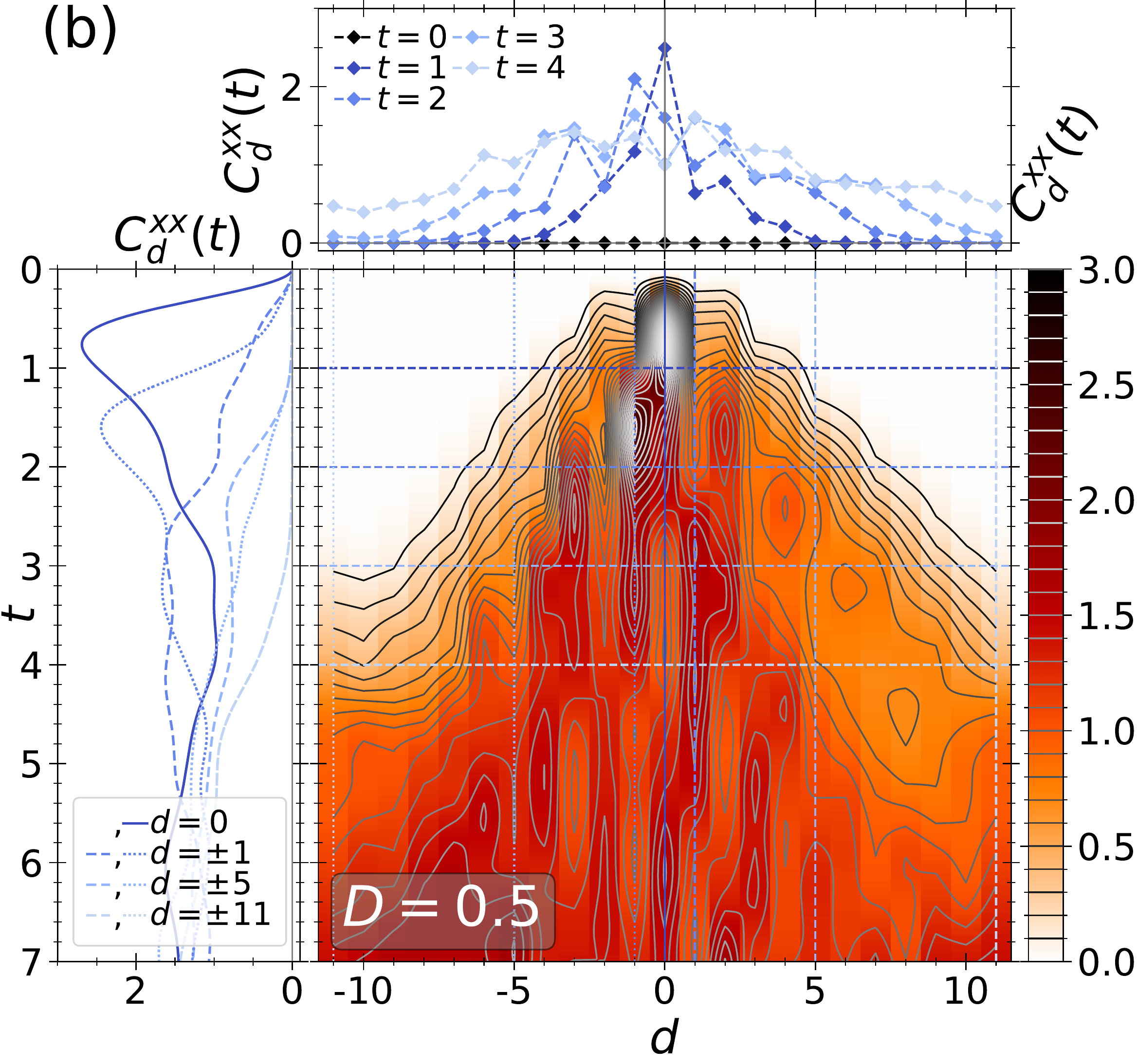}}\\
  \vfill
  \subfloat{\includegraphics[width=0.495\textwidth]{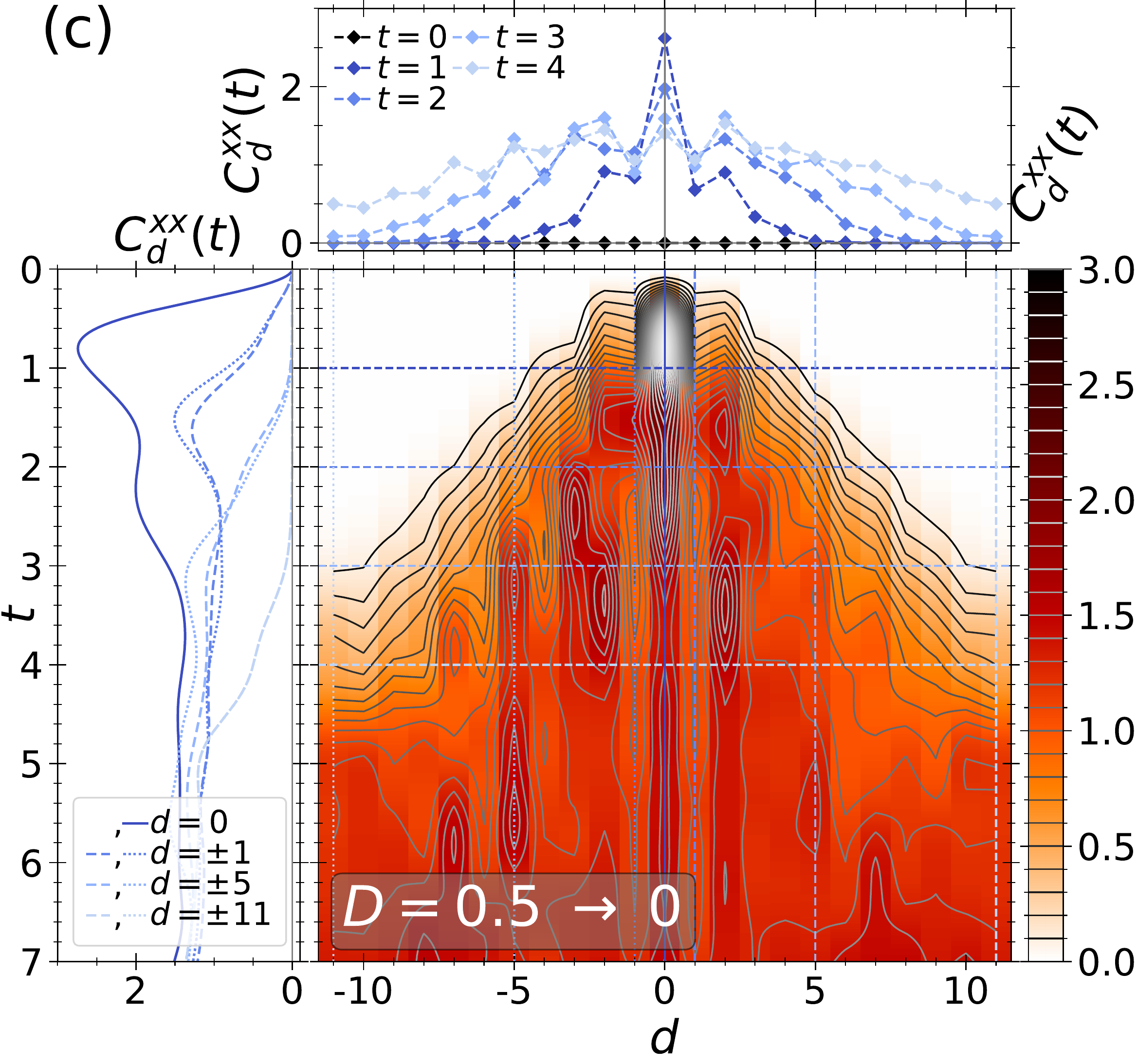}}
  \hfill
  \subfloat{\includegraphics[width=0.495\textwidth]{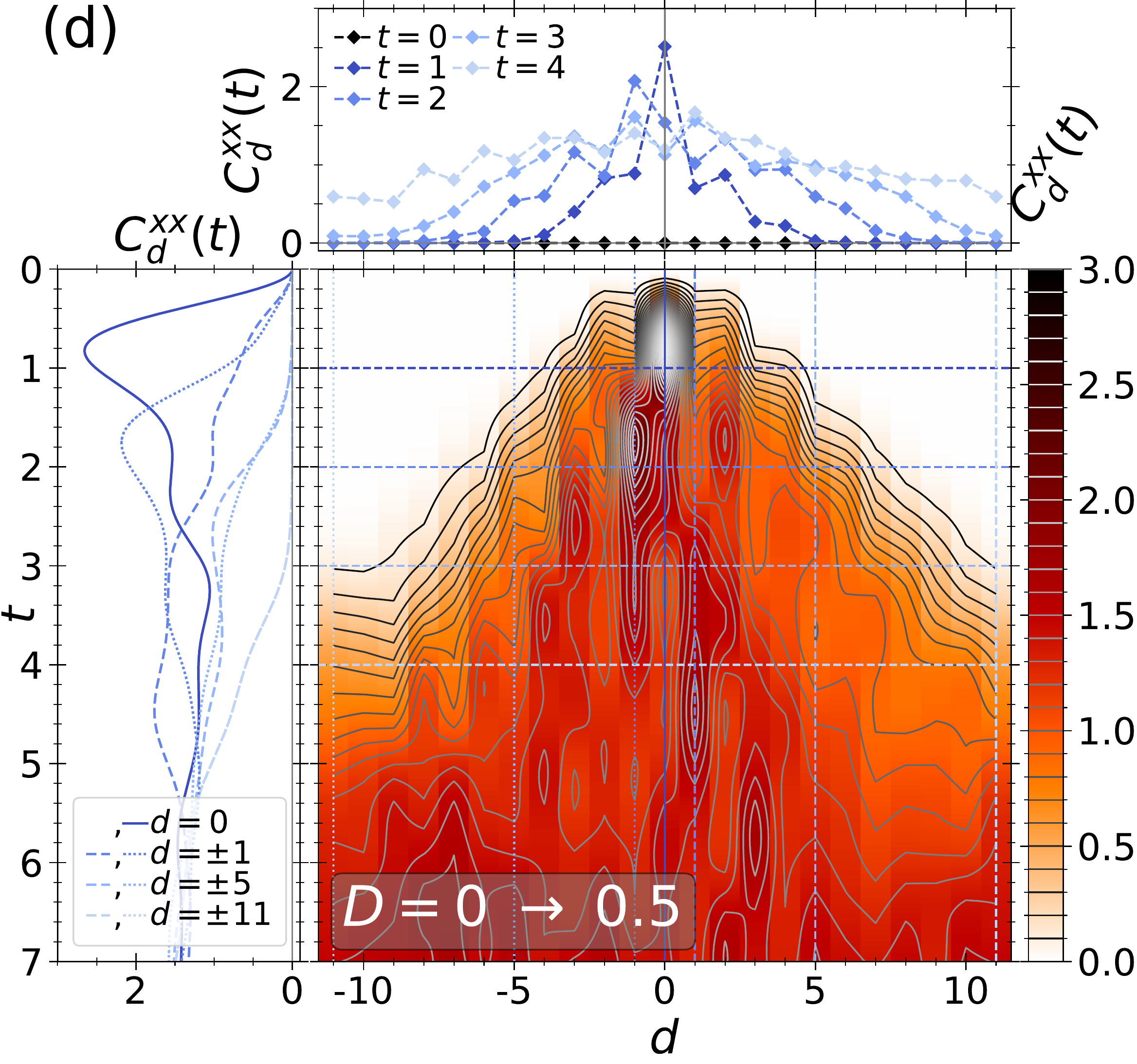}}
  \end{minipage}
  \hfill
  \begin{minipage}[b]{0.3575\linewidth}
  \subfloat{\includegraphics[width=\textwidth]{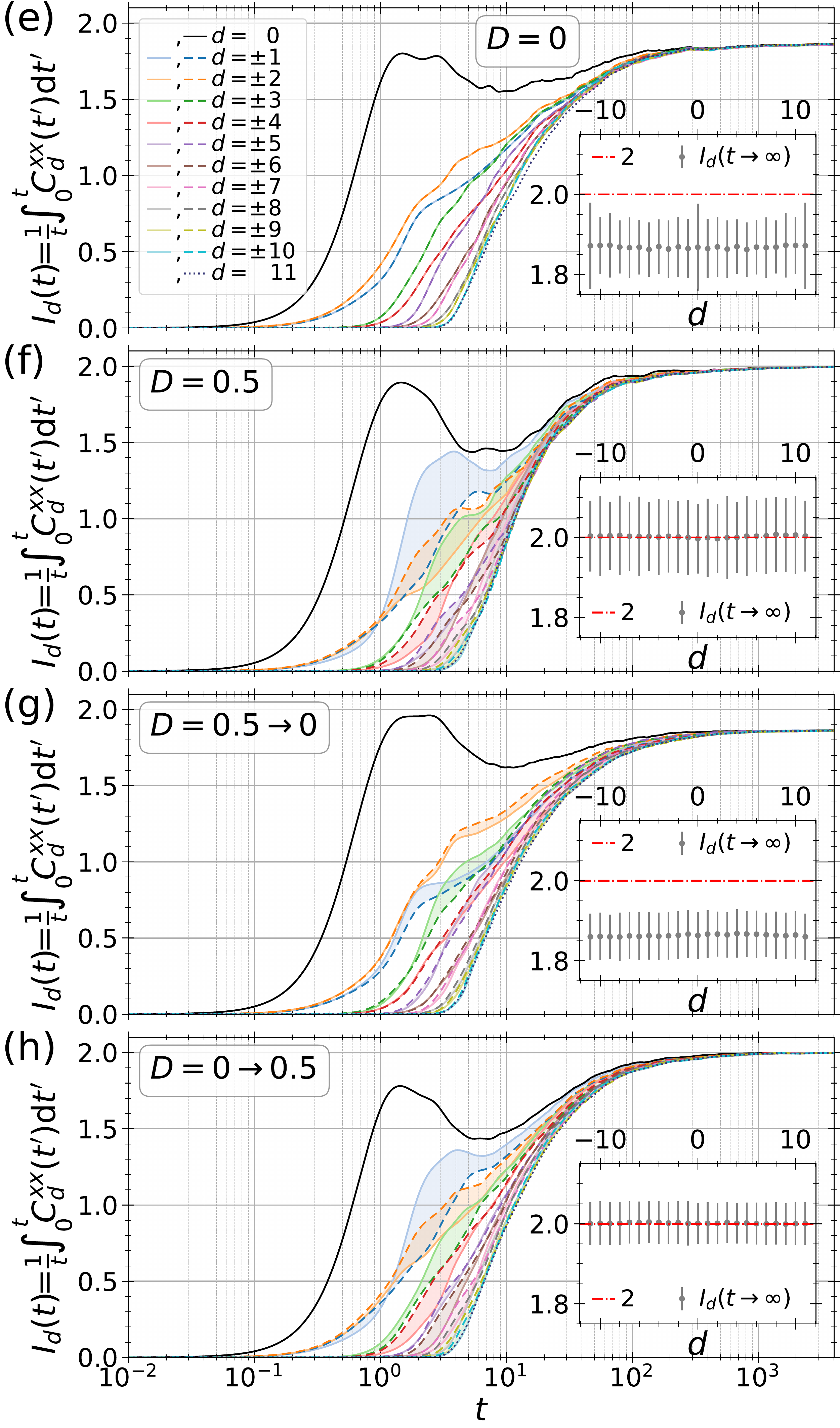}}
  \end{minipage}
 \caption{Spatiotemporal evolution of the OTOC $C^{xx}_d(t)$ ((a)-(d)) and time-average of it ((e)-(h)) for nonchiral- ((a),(e) and (d),(h)) and chiral-states ((b),(f) and (c),(g)),
          in $2$-excitation sector (${S^z_{\mathrm{tot}}=(L/2-2)}$) of ${L=22}$ spin chain with PBC; ${J_2=-J_1=1}$; time is measured in units of $|J_1|^{-1}$.
          The initial chiral- or nonchiral-state is prepared as the ground state of the system with a vanishing (${D=0}$) or finite (${D=0.5}$) DM interaction, and ${B_z=1.235}$ and ${B_z=1.51}$, respectively. Finite magnetic field (${B_z\neq 0}$) does not change behavior of $C^{xx}_d(t)$ qualitatively.
          Plots (c),(g) and (d),(h): DM interaction is quenched at ${t=0}$ ${D=0.5\rightarrow0}$ (c),(g) and ${D=0\rightarrow0.5}$ (d),(h).
          Plots (a)-(d): Contour lines are interpolated to non integer $d$; the left and the top subplots correspond to vertical and horizontal cross-sections of the main plot at distances $d=0,\pm1,\pm5,\pm11$ and at times $t=0,1,2,3,4$, respectively; a white spot around ${d=0}$ ${t=0.75}$ is due to contour lines.
          Insets in plots (e)-(g) show the time-averaged values with the standard deviation for $t\geqslant 400$ values; red dash-dotted line corresponds to the value expected in the case of complete scrambling.
         }
\label{fig:cxxd}
\label{fig:longtime_cxxd}
\end{figure*}

As an initial state, we consider the ground state of the system in the two-magnon sector (${S^z_{\mathrm{tot}}=L/2 - 2}$) for finite, ${D=0.5}$, or zero, ${D=0}$, DM interaction.
The former is also the chiral state, $\kappa^z \neq 0$, whereas the latter one has a vanishing chirality order, $\kappa^z= 0$.
We refer to them as chiral and nonchiral states, respectively.
We consider two scenarios, one with unchanged system Hamiltonian,
and another with DM interaction quenched at ${t = 0}$ from ${D=0.5}$ to ${D=0}$ or from ${D=0}$ to ${D=0.5}$, respectively.
With these setups, we examine:
\begin{itemize}
 \item[a)] nonchiral initial state and time evolution with Hamiltonian without DM interaction, the symmetric case;
 \item[b)] chiral initial state and time evolution with the Hamiltonian with a finite DM interaction (chiral Hamiltonian), the asymmetric case;
 \item[c)] the role of chirality at the level of state  ---  chiral initial state and time evolution with Hamiltonian with vanishing DM interaction;
 \item[d)] the role of chirality at the level of Hamiltonian (time-evolved operators) level  ---  nonchiral initial state and time evolution with the chiral Hamiltonian.
\end{itemize}
We consider both integrable and non-integrable cases.
We always use the PBC.
For non-integrable case with ${J_2=-J_1=1}$, the considered quench in DM interaction is across the dynamical (as well as static) phase-transition line between the chiral and nonchiral phases \cite{Azimi2016}. In the case of quenched Hamiltonian, the initial state would correspond to non-trivial excitations in the same ${S^z_{\mathrm{tot}}}$-sector (all considered Hamiltonians preserve ${S^z_{\mathrm{tot}}}$). 

We investigate the quantum information scrambling by employing OTOC for three different pairs of operators:
\begin{align}
 \label{eq:czzd}
  C^{zz}_{d}(t)
  &=
  \brkt{[\sig{z}{n + d}(t),\sig{z}{n}]^\dagger[\sig{z}{n + d}(t),\sig{z}{n}]},
  \\ 
 \label{eq:cxxd}
  C^{xx}_{d}(t)
  &=
  \brkt{[\sig{x}{n + d}(t),\sig{x}{n}]^\dagger[\sig{x}{n + d}(t),\sig{x}{n}]},
  \\
 \label{eq:ckkd}
  C^{\kappa\kappa}_{d}(t)
  &=
  \brkt{[\e^{\im\kappa^{z}_{n + d}(t)},\e^{\im\kappa^{z}_{n}}
        ]^\dagger
        [\e^{\im\kappa^{z}_{n + d}(t)},\e^{\im\kappa^{z}_{n}}
        ]}.
\end{align}
Taking $\spin{x/z}{n}$ or $\exp(\im\spin{x/z}{n})$ instead of $\sig{x/z}{n}$ will only lead to a constant multiplicative factor (see Sec.~\ref{sec:analytical_results}).
The results for $\sig{y}{}$-s will be the same, as in the case of $\sig{x}{}$-s, because of rotational symmetry about $z$-axis.
The first Eq.~\eqref{eq:czzd} and the third Eq.~\eqref{eq:ckkd} act in the same ${S^z_{\mathrm{tot}}}$-sector (recall that Hamiltonian \eqref{eq:Hamiltonian}, as well as $\sig{z}{j}$ and $\kappa^z_j$ preserve the $z$-component of the total spin). Hence, the entire calculations can be performed in the same ${S^z_{\mathrm{tot}}}$-sector.
The second measure given by Eq.~\eqref{eq:cxxd} acts across the ${S^z_{\mathrm{tot}}}$-sectors (${\sig{x}{j}=\spin{+}{j}+\spin{-}{j}}$).
Nonetheless, only $S^z_{\mathrm{tot}}$, ${S^z_{\mathrm{tot}}\pm 1}$, and ${S^z_{\mathrm{tot}} \pm 2}$ sectors are involved in the time evolution in this case, reducing the necessary computational complexity considerably.
We only compute ${\ket{\psi(t)}=\hat{W}(t)\hat{V}\ket{\psi_0}}$ and ${\ket{\phi(t)}=\hat{V}\hat{W}(t)\ket{\psi_0}}$ and recover OTOC $C(t)$ as
\begin{equation}
  C(t)=\brktprod{\psi(t)}{\psi(t)} + \brktprod{\phi(t)}{\phi(t)} - 2\mathrm{Re}(\brktprod{\phi(t)}{\psi(t)})\,.
\end{equation}
With this approach, we need to propagate appropriate states only once forward and once backward in time.

We will distinguish three time regimes, early- (${t\ll 1}$, when the time series expansion is still valid), intermediate- (near and around the approaching wavefront, ${t\approx d/\VB(\vec{\hat{n}})}$), and long-time (${t \rightarrow \infty}$) regimes. We investigate the intermediate-time regime more accurately at the end of this section.

\begin{figure}[!htbp]
 \includegraphics[width=\columnwidth]{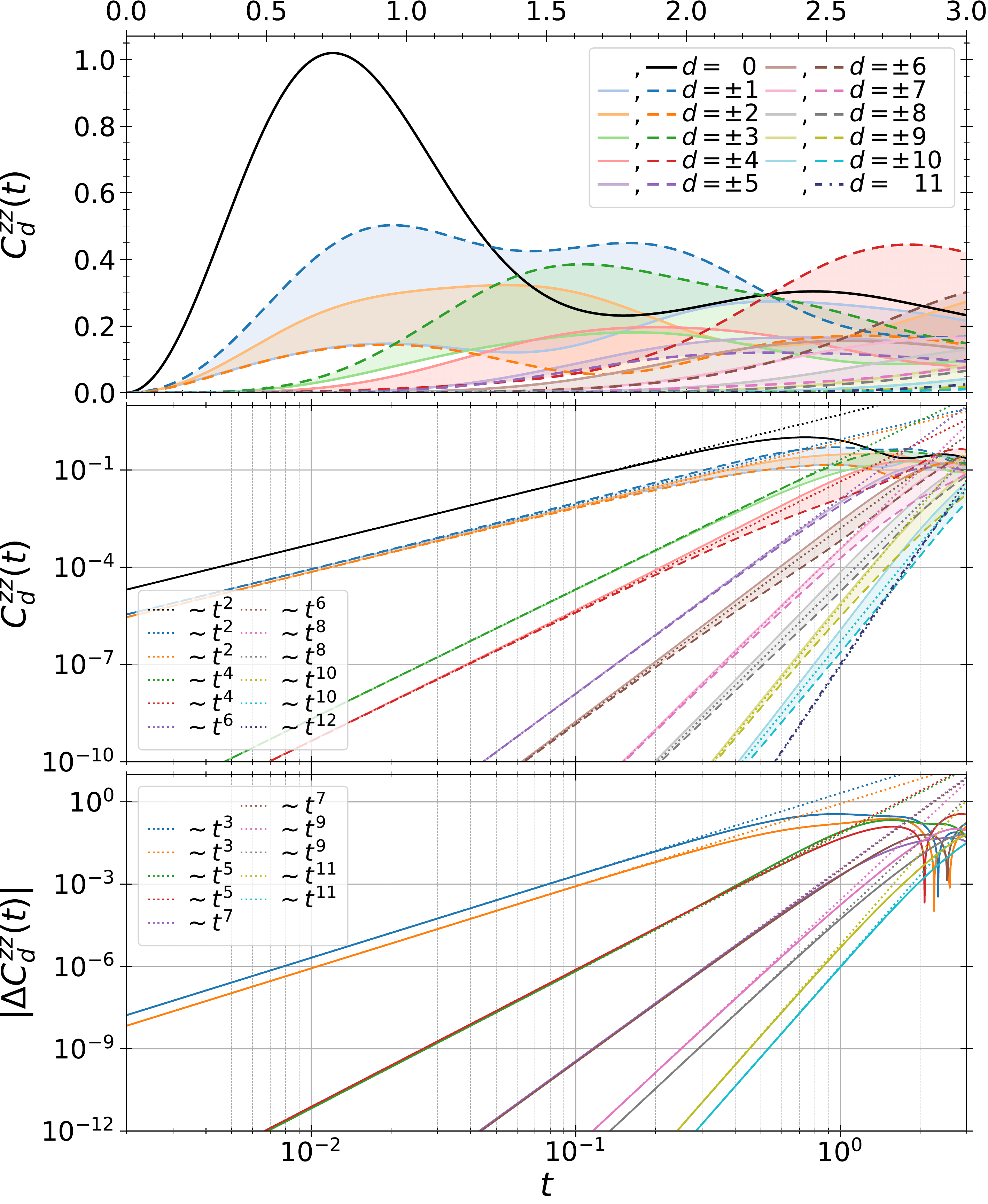}
 \caption{Short-time behavior of OTOC $C^{zz}_d(t)$ from Fig.~\ref{fig:czzd}~d) (${D=0.5}$)
          and a corresponding directional asymmetry $|\Delta C^{zz}_d(t)|$.
          Dotted lines in the center and bottom log-log plots correspond to power-law fits for ${t \ll 1}$ .
         }
\label{fig:shorttime_czzd}
\end{figure}

\begin{figure}[!htbp]
 \includegraphics[width=\columnwidth]{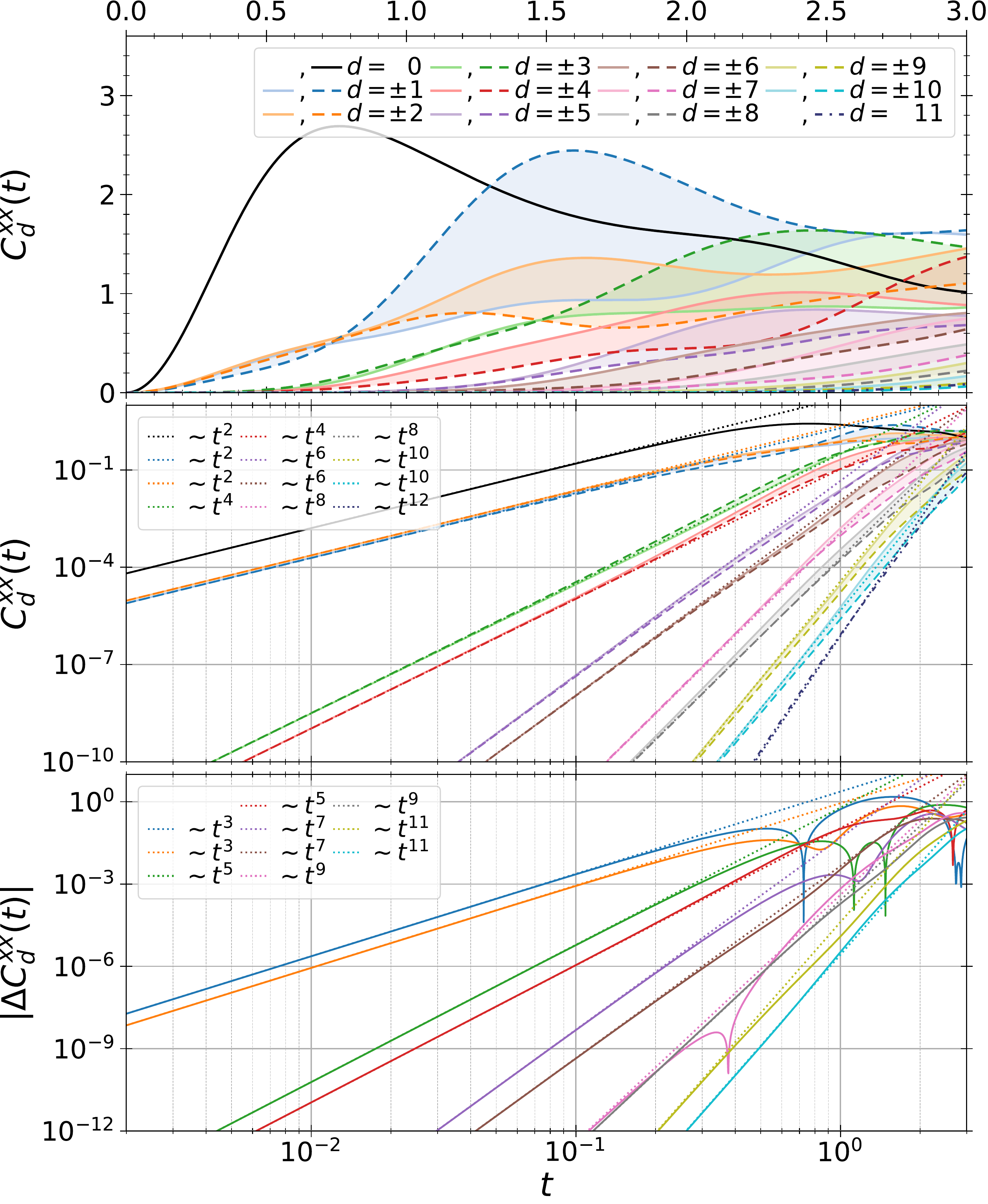}
 \caption{Short-time behavior of OTOC $C^{xx}_d(t)$ from Fig.~\ref{fig:cxxd}~d) (${D=0.5}$)
          and a corresponding directional asymmetry $|\Delta C^{xx}_d(t)|$.
          Dotted lines in the center and bottom log-log plots correspond to power-law fits for ${t \ll 1}$ .
         }
\label{fig:shorttime_cxxd}
\end{figure}

\begin{figure}[!htbp]
 \includegraphics[width=\columnwidth]{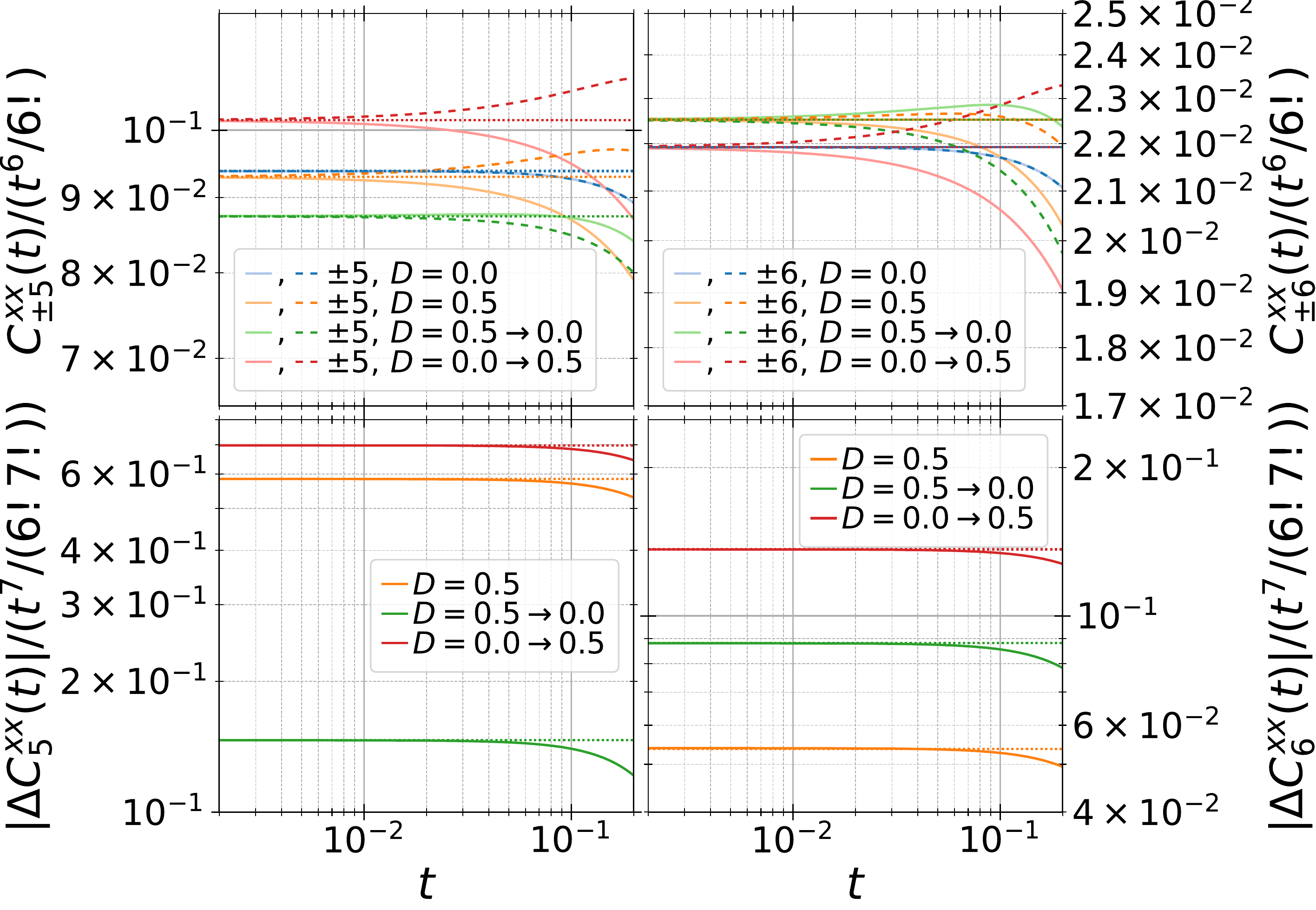}
 \caption{Short-time behavior of the scaled OTOC $C^{xx}_d(t)$ (top row) from Fig.~\ref{fig:cxxd}
          and a corresponding directional asymmetry $|\Delta C^{xx}_d(t)|$ (bottom row) for fixed distance $d=5$ (left column) and $d=6$ (right column).
          Dotted lines correspond to power-law fits for ${t \ll 1}$.
         }
\label{fig:shorttime_cxxd_quartet}
\end{figure}

\begin{figure*}[!htbp]
  \begin{minipage}[b]{0.635\linewidth}
  \subfloat{\includegraphics[width=0.495\textwidth]{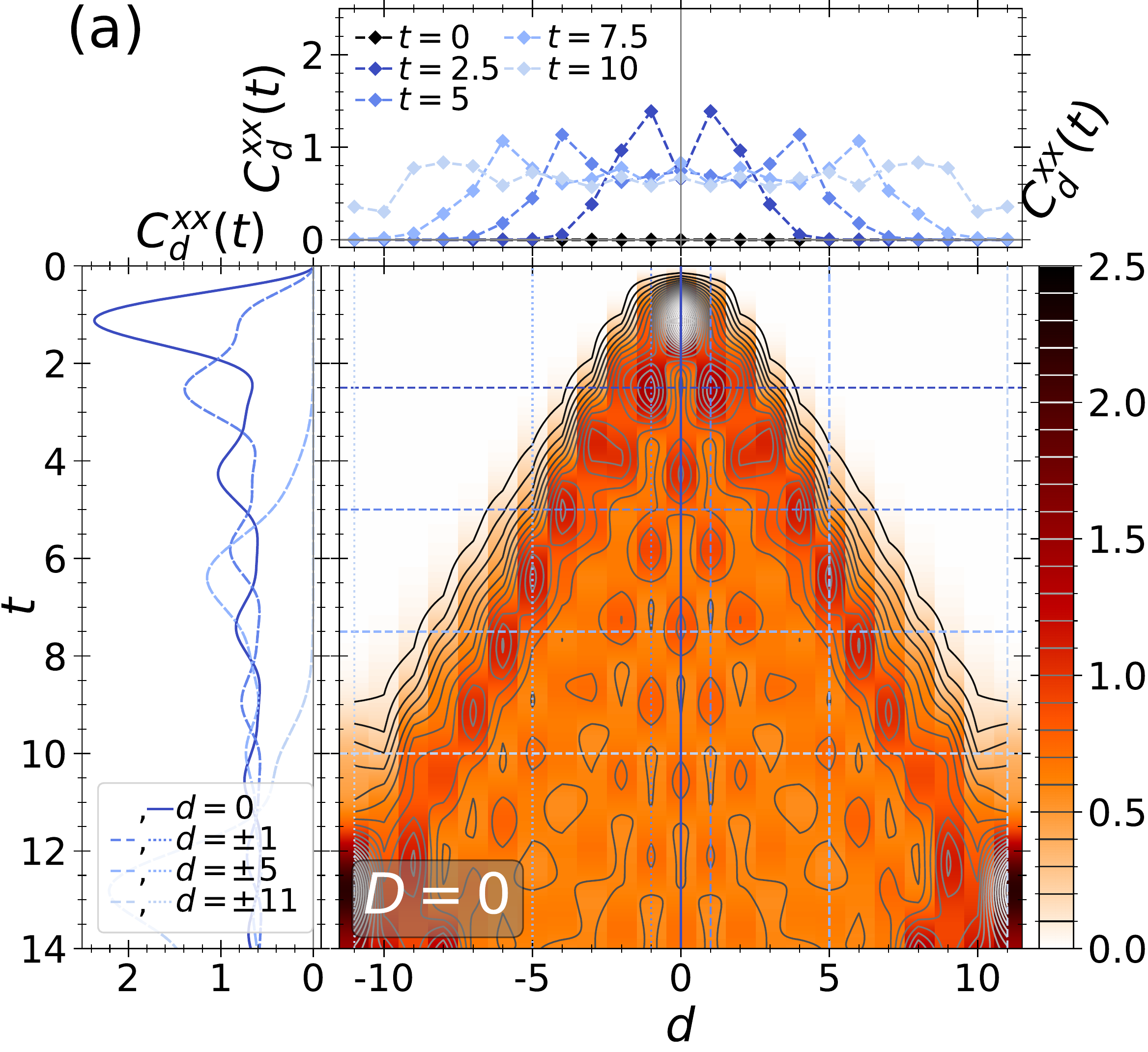}}
  \hfill
  \subfloat{\includegraphics[width=0.495\textwidth]{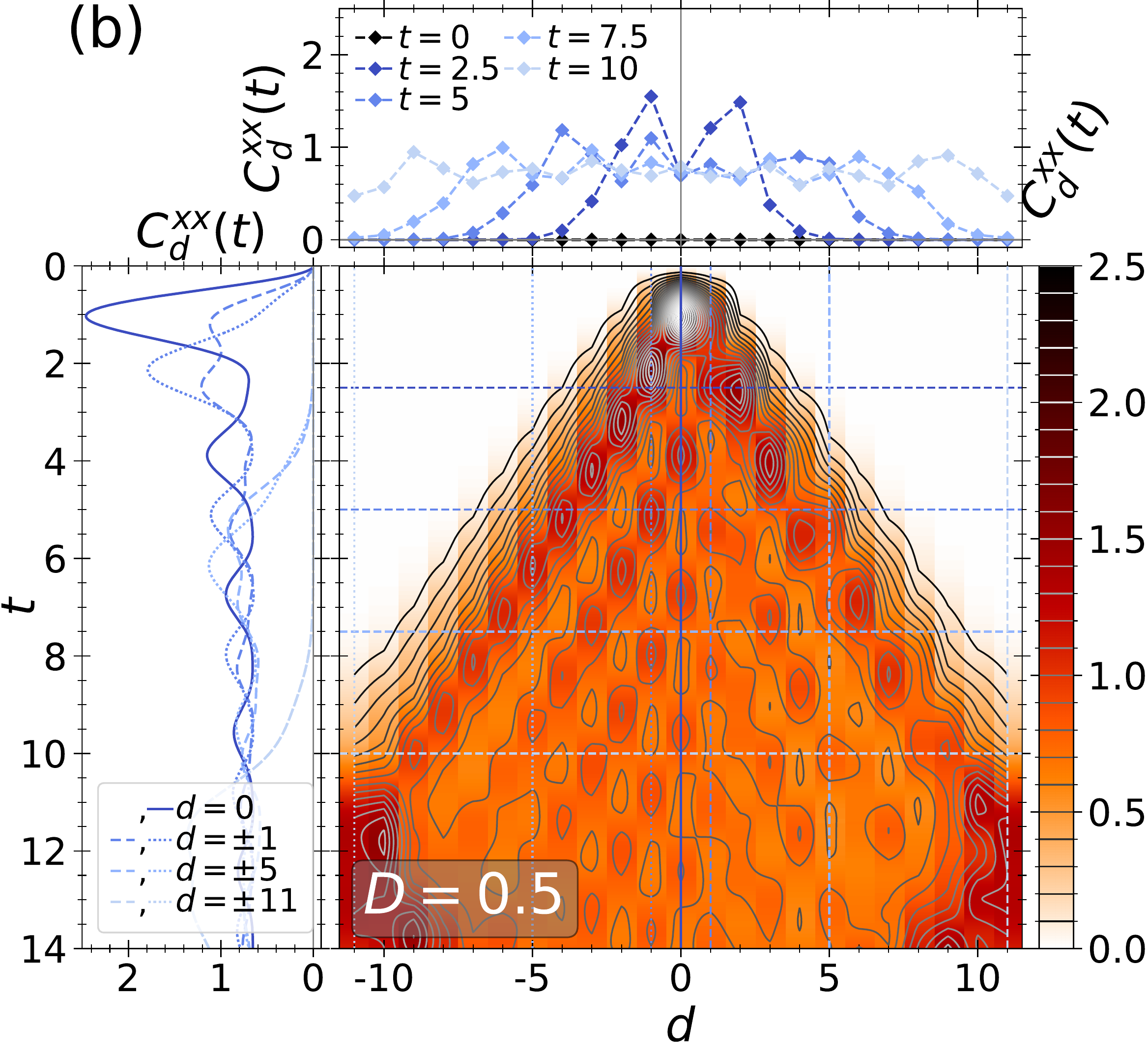}}\\
  \vfill
  \subfloat{\includegraphics[width=0.495\textwidth]{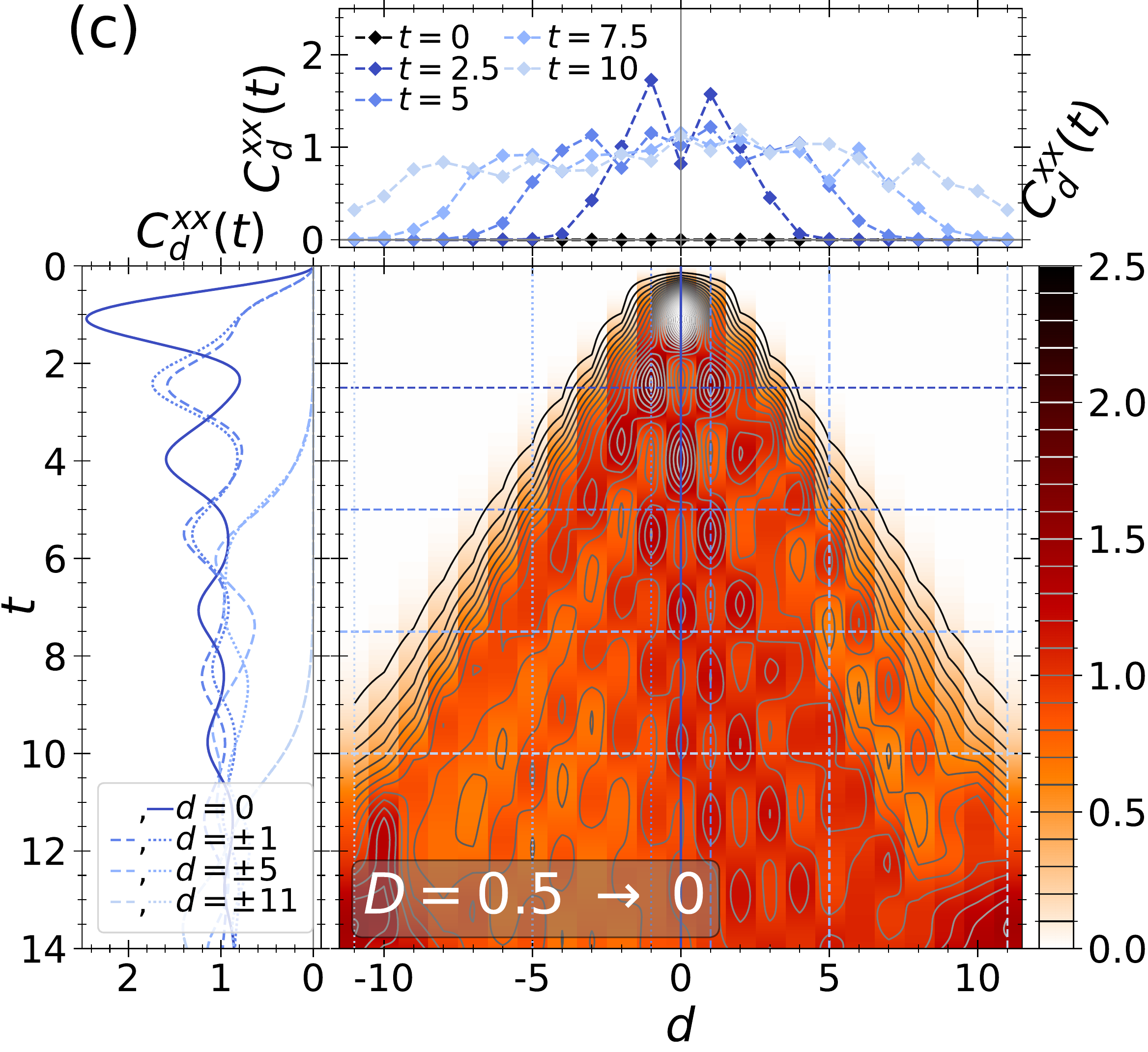}}
  \hfill
  \subfloat{\includegraphics[width=0.495\textwidth]{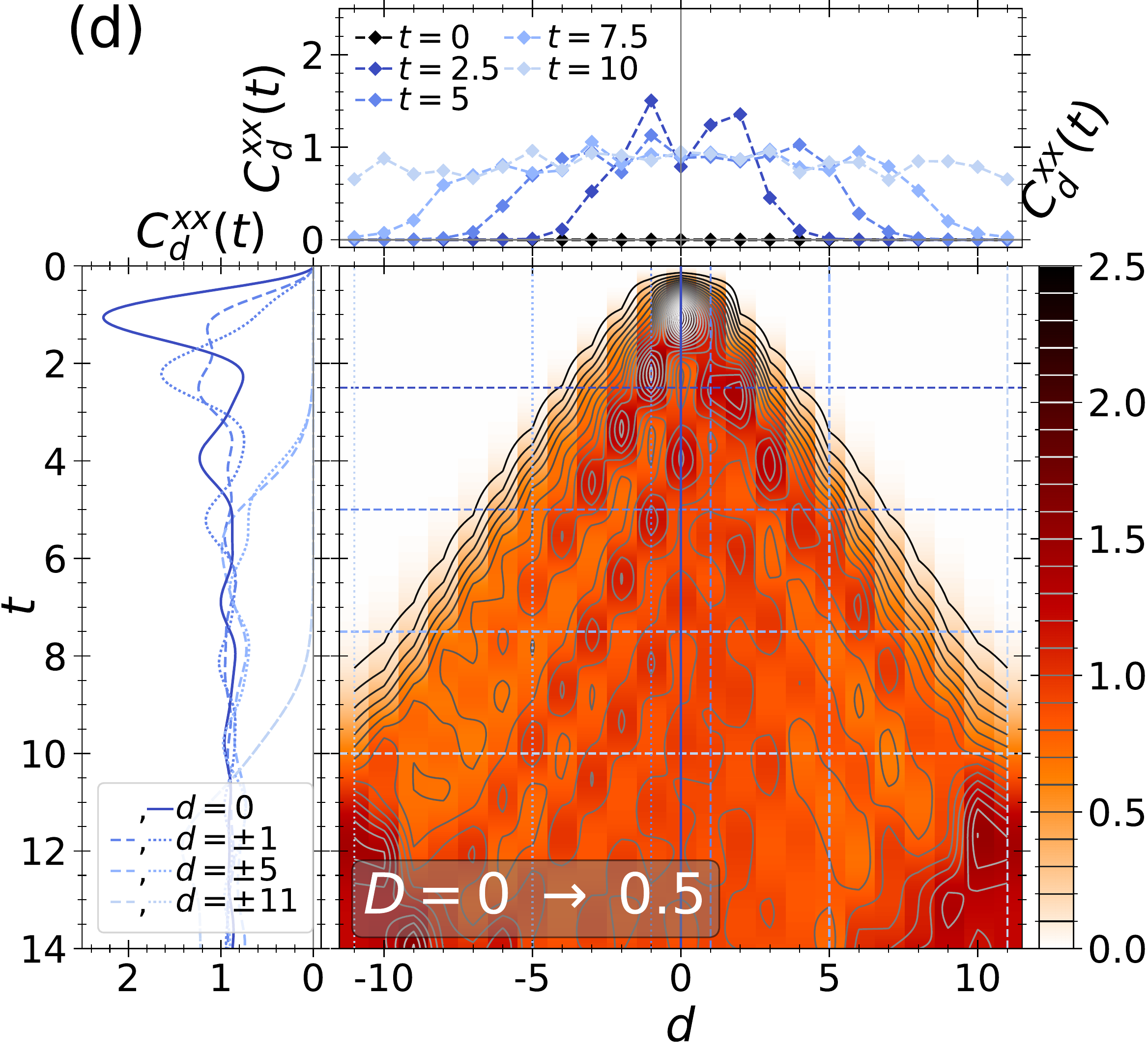}}
  \end{minipage}
  \hfill
  \begin{minipage}[b]{0.355\linewidth}
  \subfloat{\includegraphics[width=\textwidth]{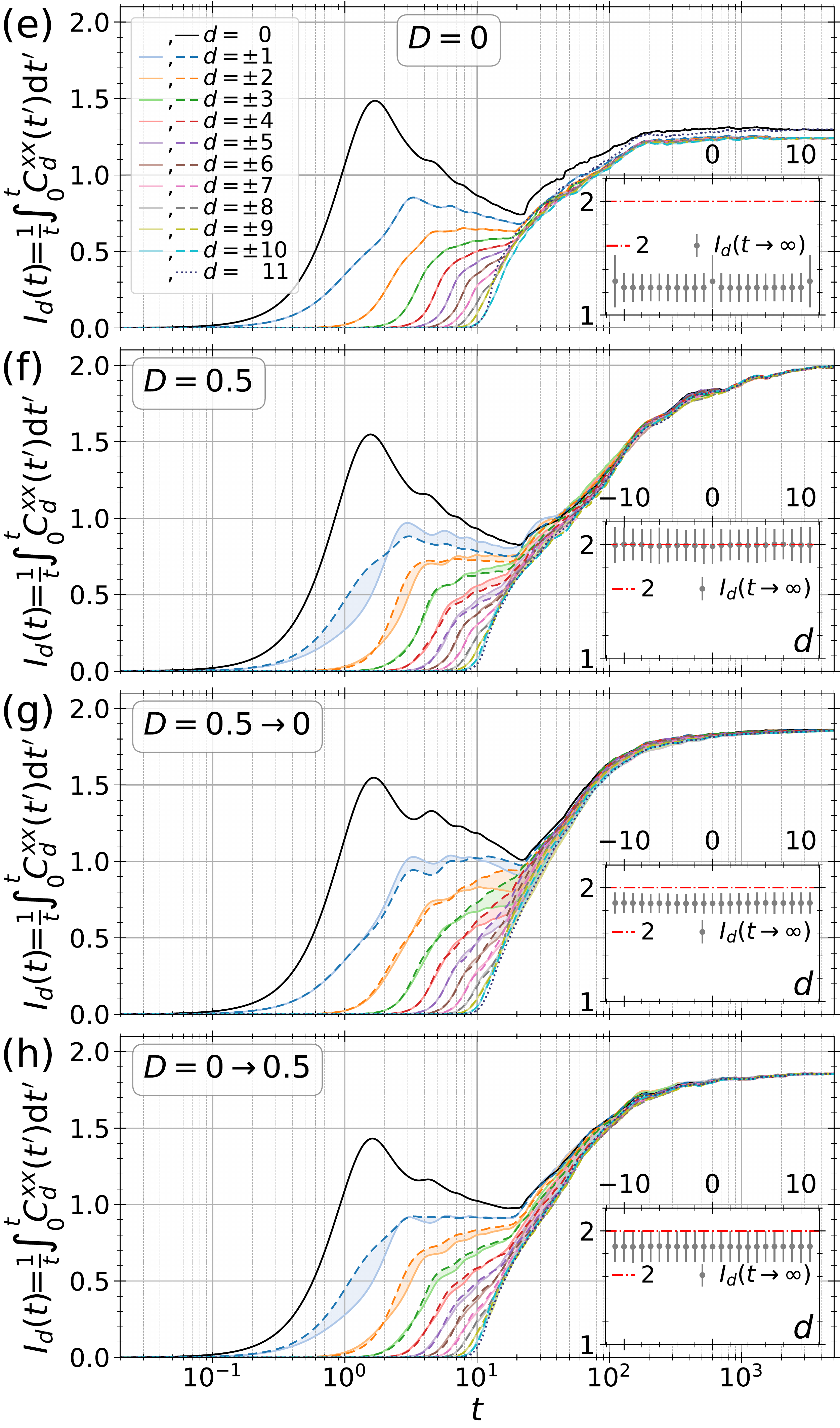}}
  \end{minipage}
 \caption{Spatiotemporal evolution of the OTOC $C^{xx}_d(t)$ ((a)-(d)) and time-average of it ((e)-(h)) for nonchiral- ((a),(e) and (d),(h)) and chiral-states ((b),(f) and (c),(g)),
          in $2$-excitation sector (${S^z_{\mathrm{tot}}=(L/2-2)}$) of ${L=22}$ spin chain with PBC; ${J_1=-1}$; ${J_2=0}$; time is measured in units of $|J_1|^{-1}$.
          The initial chiral- or nonchiral-state is prepared as the ground state of the system with a vanishing (${D=0}$, XXX-Heisenberg) or finite (${D=0.5}$ easy-plane XXZ-Heisenberg) DM interaction, and ${B_z=0}$ and ${B_z=0.099}$, respectively. Finite magnetic field (${B_z\neq 0}$) does not change behavior of $C^{xx}_d(t)$ qualitatively.
          Plots (c),(g) and (d),(h): DM interaction is quenched at ${t=0}$ ${D=0.5\rightarrow0}$ (c),(g) and ${D=0\rightarrow0.5}$ (d),(h).
          Plots (a)-(d): Contour lines are interpolated to non integer $d$; the left and the top subplots correspond to vertical and horizontal cross-sections of the main plot;
          a white spot around ${d=0}$ ${t=1}$ is due to contour lines.
          Insets in plots (e)-(g) show the time-averaged values with the standard deviation for $t\geqslant 800$ values; red dash-dotted line corresponds to the value expected in the case of complete scrambling.
         }
\label{fig:cxxd_J2_0}
\end{figure*}

\begin{figure}[!thpb]
 \includegraphics[width=\columnwidth]{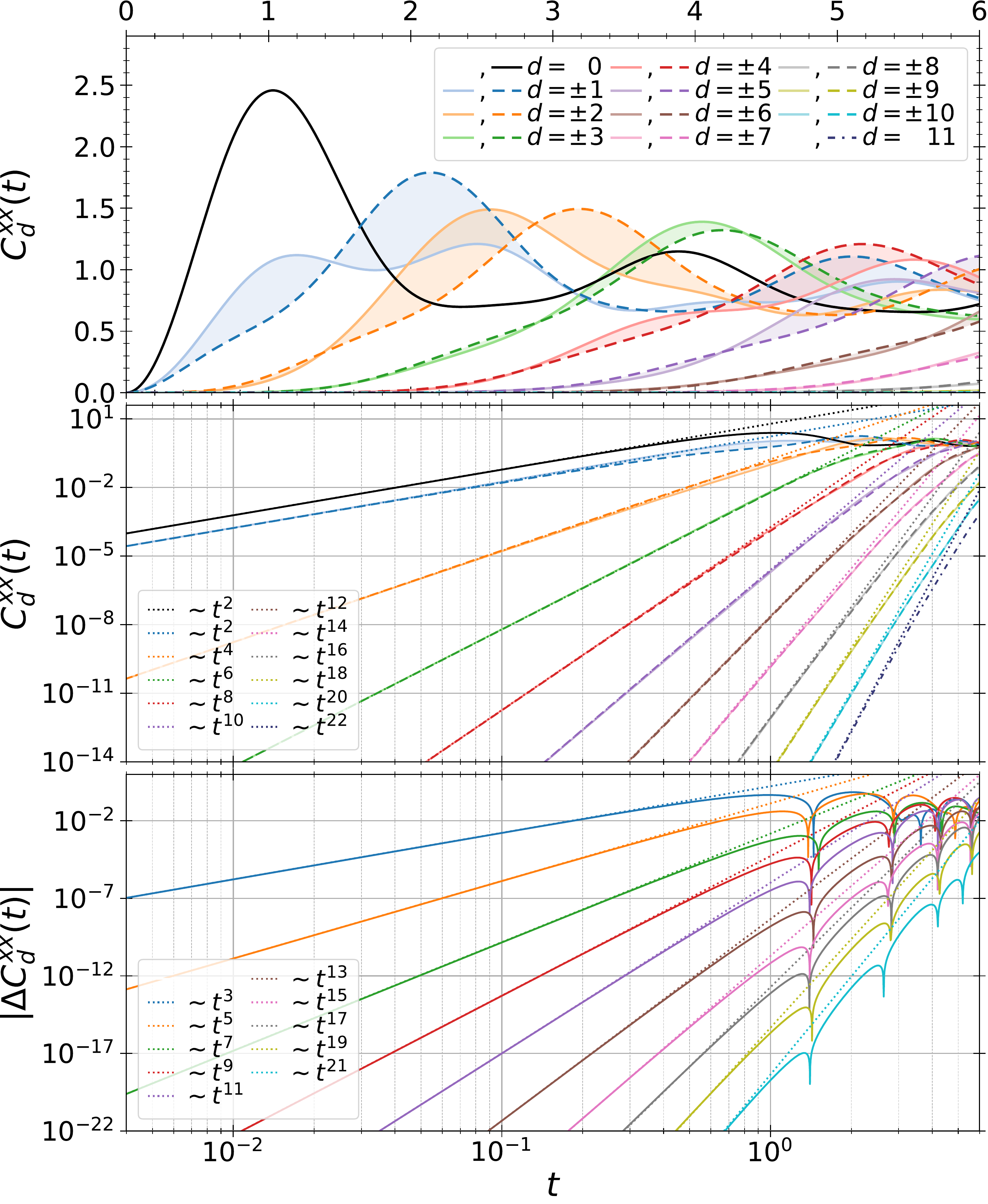}
 \caption{Short-time behavior of OTOC $C^{xx}_d(t)$ from Fig.~\ref{fig:cxxd_J2_0}~d) (${D=0.5}$)
          and the corresponding directional asymmetry $|\Delta C^{xx}_d(t)|$.
          Dotted lines in the center and bottom log-log plots correspond to power-law fits for ${t \ll 1}$ .
         }
\label{fig:shorttime_cxxd_J2_0}
\end{figure}

\begin{figure}[!b]
 \includegraphics[width=\columnwidth]{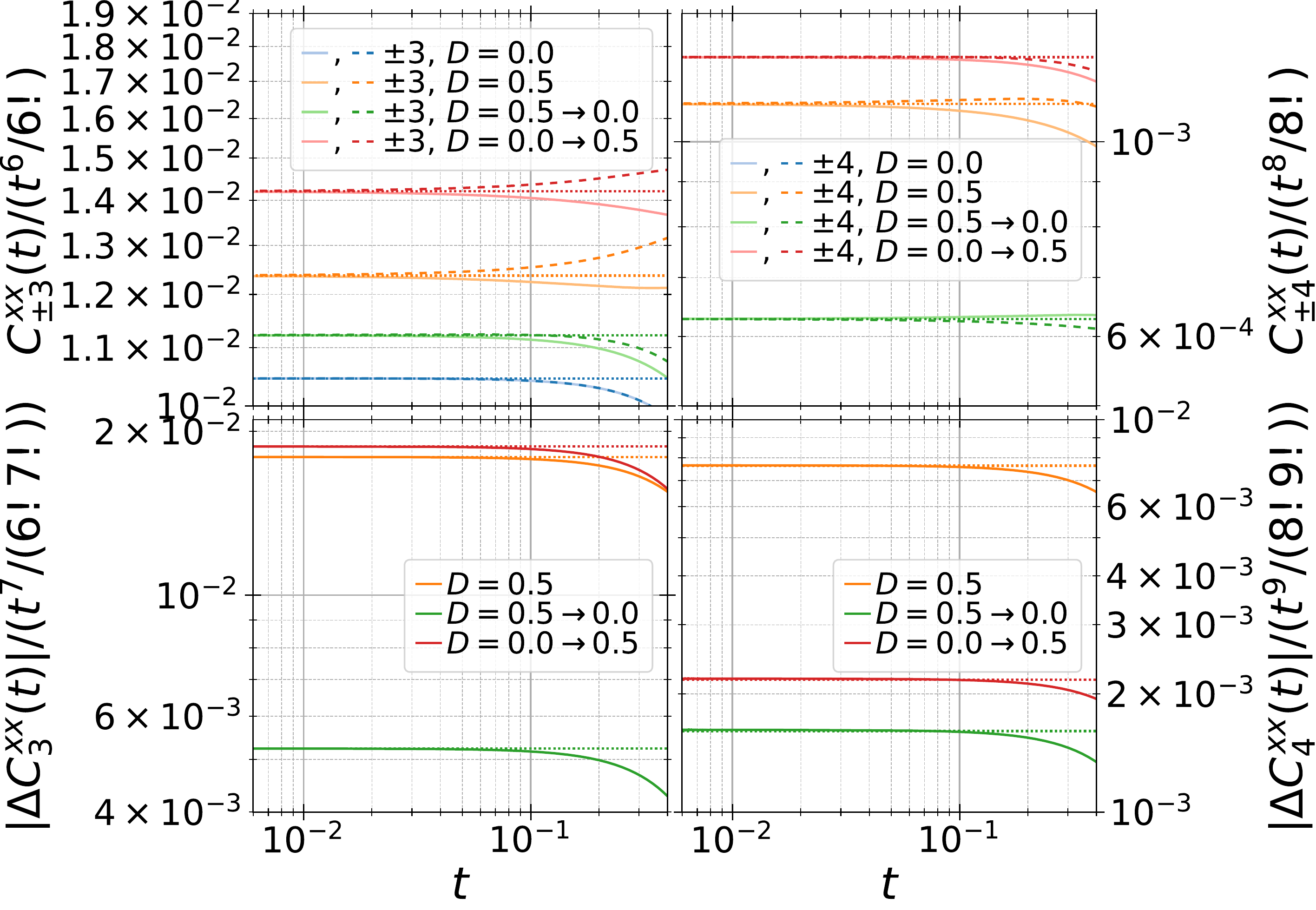}
 \caption{Short-time behavior of the scaled OTOC $C^{xx}_d(t)$ (top row) from Fig.~\ref{fig:cxxd_J2_0}
          and the associated directional asymmetry $|\Delta C^{xx}_d(t)|$ (bottom row) for fixed distance $d=3$ (left column) and $d=4$ (right column).
          Dotted lines correspond to power-law fits for ${t \ll 1}$.
         }
\label{fig:shorttime_cxxd_quartet_J2_0}
\end{figure}

\begin{figure}[!htbp]
 \includegraphics[width=0.49\columnwidth]{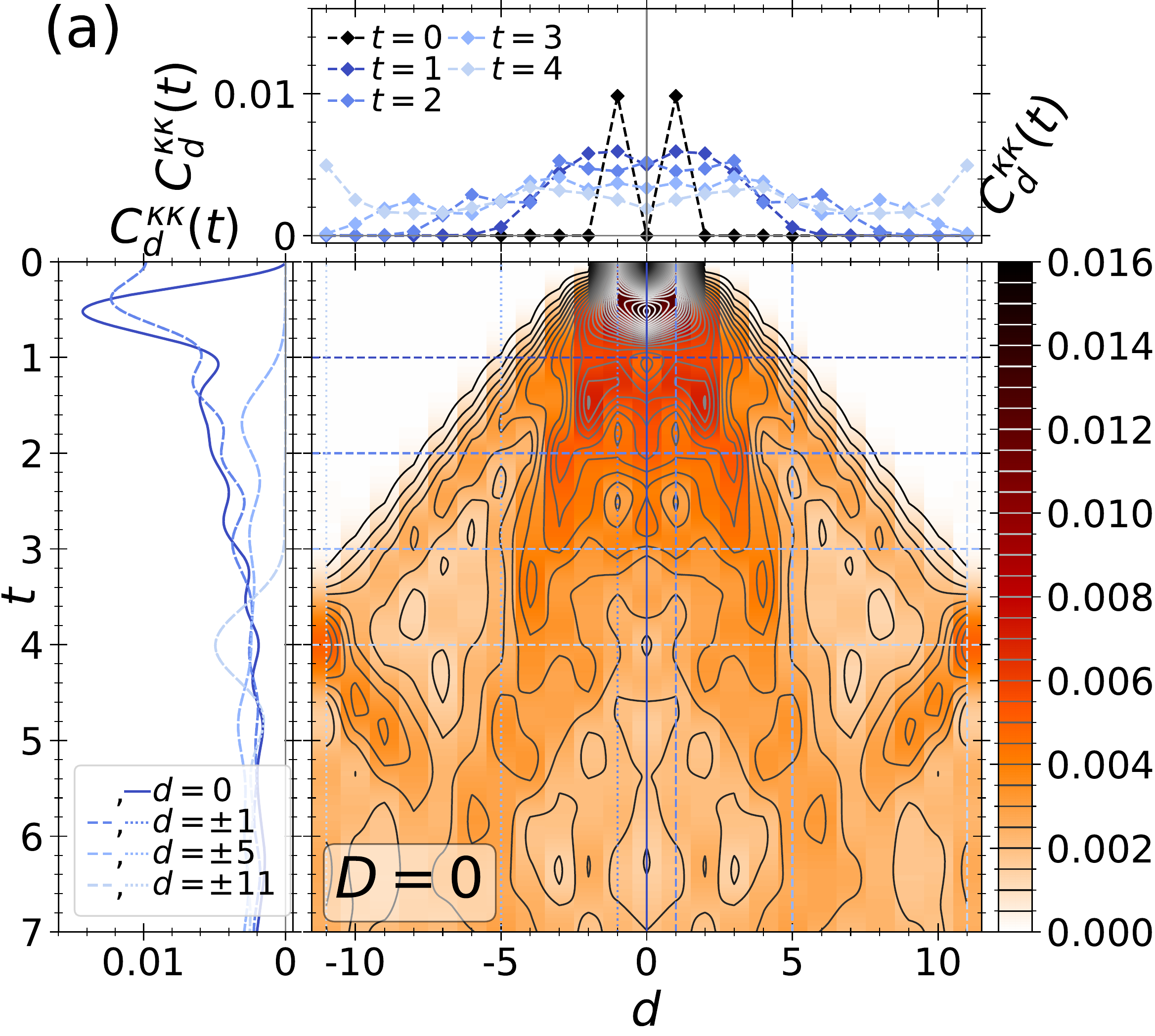}
 \includegraphics[width=0.49\columnwidth]{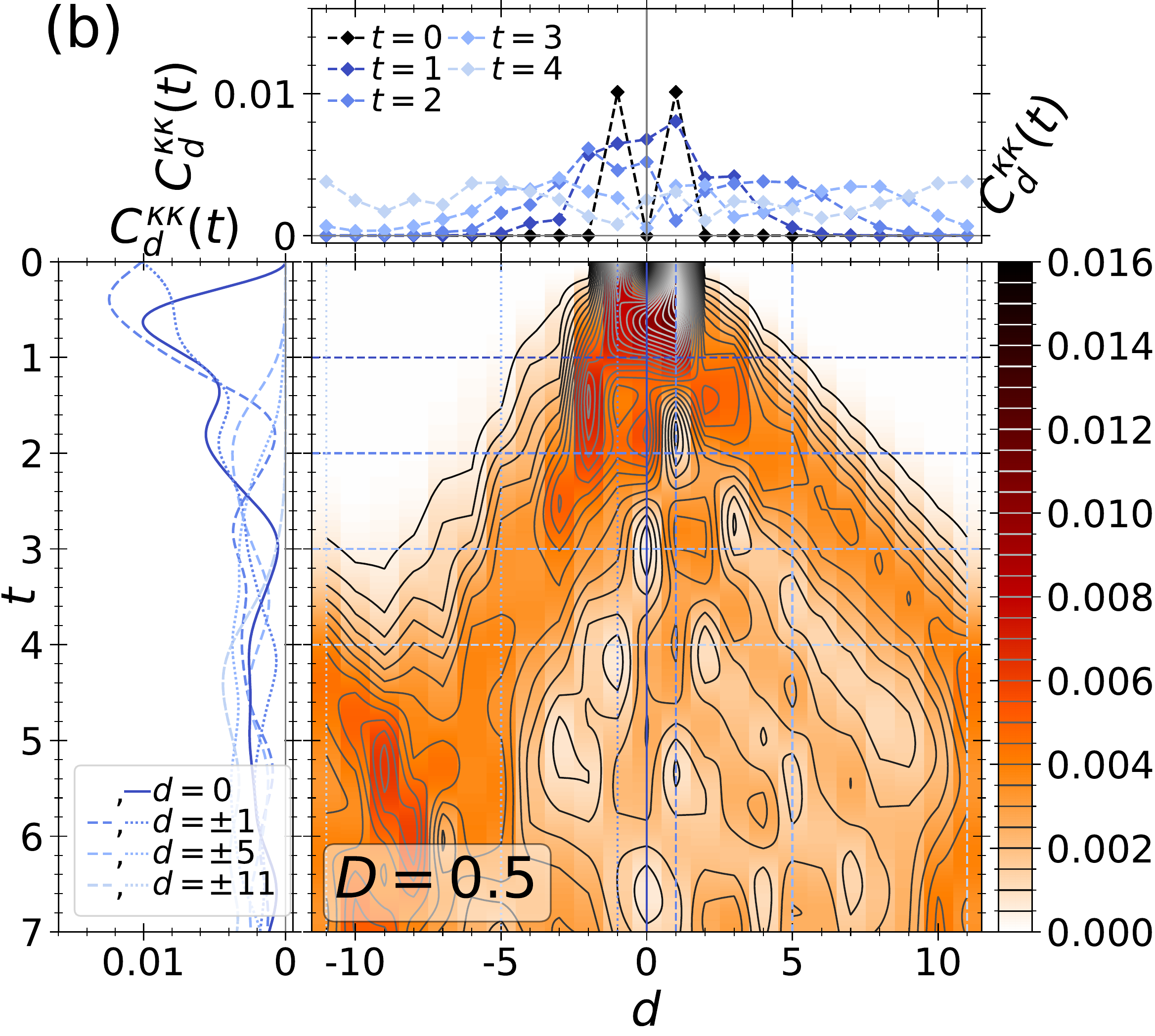}\\
 \includegraphics[width=0.49\columnwidth]{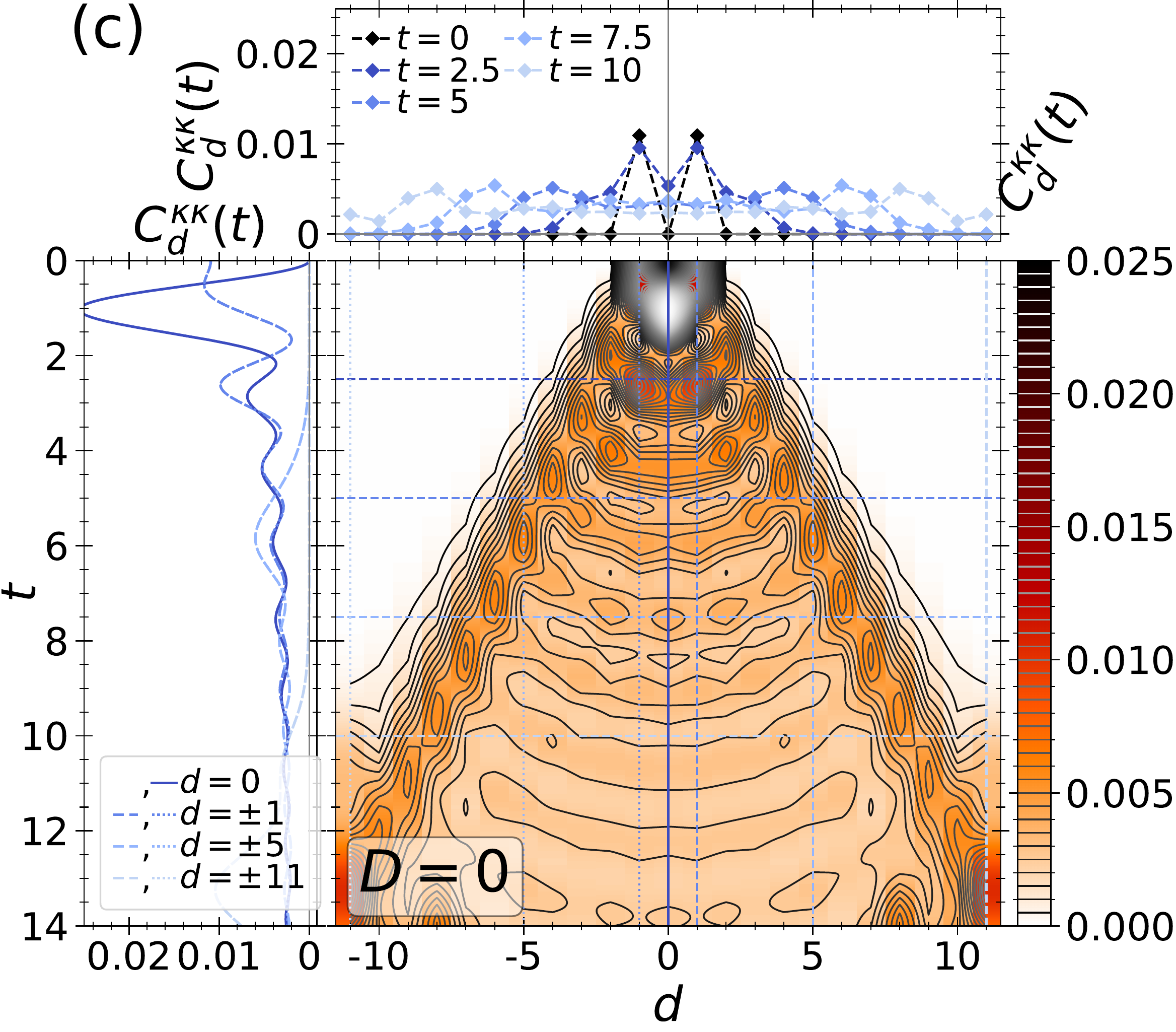}
 \includegraphics[width=0.49\columnwidth]{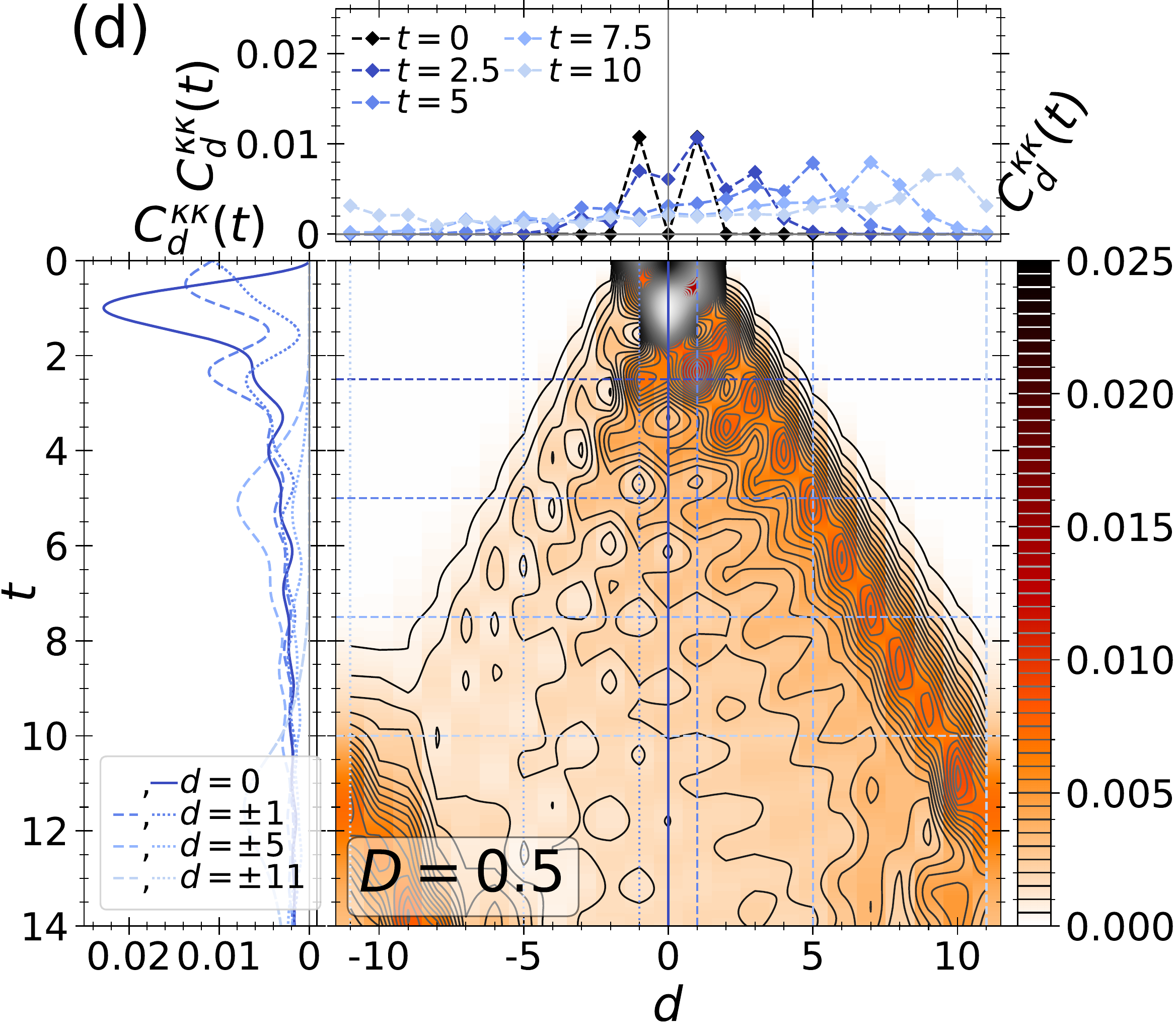}
 \caption{Spatiotemporal evolution of the OTOC $C^{\kappa\kappa}_d(t)$~\eqref{eq:ckkd} for fully nonchiral- ((a) and (c)) and chiral-cases ((b) and (d)),
          in $2$-excitation sector (${S^z_{\mathrm{tot}}=(L/2-2)}$) of ${L=22}$ spin chain with PBC;
          The initial chiral- and nonchiral-state is prepared as the ground state of the system with a vanishing
          (${D=0}$) or finite (${D=0.5}$) DM interaction, respectively;
          Time is measured in units of $|J_1|^{-1}$.
          Plots (a)-(b): ${J_2=-J_1=1}$.
          Plots (c)-(d): integrable case, ${J_1=-1}$, ${J_2=0}$.
          Contour lines are interpolated to non integer $d$.
          For each panel, the left and the top subplots correspond to vertical and horizontal cross-sections of the main plot.
          Black spot around ${d=0}$ ${t=0}$ is due to contour lines (cf. ${t=0}$ cut).
         }
\label{fig:ckkd_ckkd_J2_0}
\end{figure}

\subsection{Early- and long-time behavior}

We start with the non-integrable case (${J_2=-J_1=1}$). Figures~\ref{fig:czzd}-\ref{fig:longtime_cxxd} show the spatiotemporal evolution of OTOC for the ${L=22}$ site system. The results correspond to the four mentioned combinations of the setup.
In all studied cases, we found that OTOC spreads ballistically in both directions, falling sharply outside a light cone, before the left and right fronts collide (recall that we use PBC), see Fig.~\ref{fig:czzd} and Fig.~\ref{fig:cxxd}.
As expected, the OTOC for the nonchiral state is fully symmetric when DM interaction is also zero (${D=0}$) --- the symmetric case (see Fig.~\ref{fig:czzd}(a) and Fig.~\ref{fig:cxxd}(a)).
For a chiral initial state and time evolution with the system Hamiltonian with a finite DM interaction --- the asymmetric (chiral) case, the speeds of the left and right propagating wavefronts do differ, see Fig.~\ref{fig:czzd}(b) and Fig.~\ref{fig:cxxd}(b).
Comparing the cases with chiral state vs.~chiral Hamiltonian, one sees that (Fig.~\ref{fig:czzd}(c) and Fig.~\ref{fig:cxxd}(c)) the contour lines corresponding to ${C^{zz}_d(t)\leqslant 0.05}$ and ${C^{xx}_d(t)\leqslant 0.25}$ thresholds remain symmetric even for the chiral state but the time evolution under the nonchiral Hamiltonian.
The lines matching to higher than the threshold values are asymmetric.
In the case of a chiral Hamiltonian (finite DM interaction), the wavefronts spread asymmetrically even in the case of a nonchiral initial state, see Fig.~\ref{fig:czzd}(d) and Fig.~\ref{fig:cxxd}(d).
The asymmetry is most apparent for the chiral state and the time-evolution with the chiral Hamiltonian (${D=0.5}$).
Barely, but one can still distinguish the second wavefront due to the second isolated maximum in the left and right branches of the group velocity (see plots (a)-(c) in Figs.~\ref{fig:czzd} and \ref{fig:cxxd}).
The scrambling around and behind the wavefronts is considerably stronger for $\sig{x}{}$- as compared to $\sig{z}{}$-operators.
This can be also connected to the nonlocal character of the effective Fermionic representation of the corresponding $\sig{x}{}$ operators, similar to the findings made for quantum Ising chains~\cite{LinMotrunich2018} and larger dimensions of the total ${\spin{z}{\mathrm{tot}}}$ sectors involved in the scrambling:
${S^z_{\mathrm{tot}} \pm 1}$ and ${S^z_{\mathrm{tot}}\pm2}$ sectors for $\sig{x}{}$ as compared to only ${S^z_{\mathrm{tot}}}$ in the case of $\sig{z}{}$.

In the long-time regime, the saturation of OTOC to its maximum value $2$ (equivalently decay of OTOC Eq.~\eqref{eq:OTOC1_FT} to 0) at long times for all subsystems (i.e., for all separations) and for all operators $\hat{W}$ and $\hat{V}$ implies a complete quantum information scrambling \cite{Hosur2016}. At longer times, in the case of scrambling, $C^{zz}_d(t)$ should converge to
\begin{equation}
 \label{eq:czzd_inf_value}
  C^{zz}_d(t\rightarrow \infty)
  = 2 \left(1-\left(\frac{2 S^z_{\mathrm{tot}}}{L}\right)^2 \right)^2,
\end{equation}
(see Appendix~\ref{sec:apendix_sat_val_OTOC}), which certainly differs from $2$ when ${S^z_{\mathrm{tot}} \neq 0}$. It will be different from value $2$ even if one takes the infinite temperature ensemble instead of a pure state. Typically, one expects $2$ for the Hermitian and at the same time unitary operators, like $\sig{z}{}$-s, but because the $z$-component of the total spin is conserved, ${\brkt{\sig{z}{n}} \neq 0}$, ${\brkt{\sig{z}{n+d}(t)}\neq 0}$, hence, the commutator in Eq.~\eqref{eq:czzd} is not a connected one \cite{Aleiner2016}, and the result deviates from the value $2$.
For $\sig{x}{}$-s on the other hand, ${\brkt{\sig{x}{n+d}(t)} = 0}$ and ${\brkt{\sig{x}{n}} = 0}$, because of the rotational symmetry about $z$-axis.
Therefore, the squared commutator in Eq.~\eqref{eq:cxxd} is the connected one, and in the case of full scrambling, it converges to 
\begin{equation}
 \label{eq:cxxd_inf_value}
 C^{xx}_d(t\rightarrow \infty) = 2\,,
\end{equation}
as it was typically expected for the unitary operators \cite{Aleiner2016} (see also Appendix.~\ref{sec:apendix_sat_val_OTOC}).

In Figure~\ref{fig:longtime_czzd}(e)-(h) and Fig.~\ref{fig:longtime_cxxd}(e)-(h), on a semi-log plot, we show the time-averaged values of OTOC, $I_d(t)=\tfrac{1}{t}\int_0^t C^{zz}_d(t')\ud t'$ and $I_d(t)=\tfrac{1}{t}\int_0^t C^{xx}_d(t')\ud t'$, respectively.
The actual values, Eq.~\eqref{eq:czzd_inf_value} and Eq.~\eqref{eq:cxxd_inf_value}, are only acquired for systems with finite DM interaction (see Fig.~\ref{fig:longtime_czzd}(e)-(h) and Figs.~\ref{fig:longtime_cxxd}(e)-(h)).
The latter also indicates the vanishing long-time average of corresponding $F(t)$.
In the case of a vanishing DM interaction (${D=0}$) for both the chiral and the nonchiral initial state, the values to which the time-averaged values of $C^{xx}_d(t)$ saturate are smaller than 2 (see plots (e) and (g) in Fig.~\ref{fig:longtime_cxxd}).
They are homogeneous, however (do not depend on the distance, $d$).
The long-time limit of the time-averaged values of $C^{zz}_d(t)$ have a clear dependence on the distance $d$ for ${D=0}$ cases, being largest on the same and farthest sites (see plots (e) and (g) in Fig.~\ref{fig:longtime_czzd} and the insets therein).
For the case with the nonchiral initial state, the interior eigenenergies of the system Hamiltonian are also at least doubly degenerate.

For the short-time limit (early-time regime), in all cases that we studied, we observe a power-law growth of the OTOC, as shown, for example, for ${D=0.5}$, in the center panels of Figs.~{\ref{fig:shorttime_czzd}-\ref{fig:shorttime_cxxd}} which is consistent with the discussion in Sec.~\ref{sec:short_time_limit}. At leading order, the OTOC behaves as $t^{2\nu}$, where ${\nu=\max(1,(|d|+1)\ \mathbf{div}\ 2)}$. For the chiral initial state or in a system with a nonvanishing DM interaction, we also observe the asymmetric subleading $t^{2\nu+1}$ corrections.
This is also demonstrated in Fig.~\ref{fig:shorttime_cxxd_quartet}, where we plot the scaled values of $C^{xx}_d(t)$ and the corresponding scaled values of the directional asymmetry ${|\Delta C^{xx}_d(t)|}$ for the fixed ${|d|=5}$ and ${|d|=6}$ distances as a function of time (see also the bottom panels in Figs.~\ref{fig:shorttime_czzd}-\ref{fig:shorttime_cxxd}).
All these are also in line with the discussion in Sec.~\ref{sec:short_time_limit}.
As we see, integrability is not essential for the power-law behavior of OTOC. It is generic for any quantum system with local-Hamiltonian and finite on-site subspaces.

We continue with integrable limit ${J_2=0}$ and ${J_1 = -1}$.
In this case (${J_2=0}$), DM interaction can be gauged out from the Hamiltonian \eqref{eq:Hamiltonian},
leading to the effective easy-plane XXZ-Heisenberg model (see Sec.~\ref{sec:model}, Eq.~\eqref{eq:Hamiltonian2}) with modified, twisted boundary conditions ${\spin{\pm}{L+j}=e^{\pm\im \vartheta L}\spin{\pm}{j}}$, in the case of PBC. For ${\vartheta L = 2\pi \gamma}$ with ${\gamma \in \mathbb{Z}}$ (integer multiples of $2\pi$), corresponding to a ``magical'' ${D=-J_1\tan(\gamma\pi/L)}$, one recovers the PBC in the latter case too. One can avoid these subtleties by considering OBC, but the translation invariance is lost in this case, complicating further the assessments of analytical results. Therefore, we study PBC.

Figs.~\ref{fig:cxxd_J2_0}-\ref{fig:shorttime_cxxd_quartet_J2_0}, show the results for $C^{xx}_d(t)$, Eq.~\eqref{eq:cxxd}. 
In the integrable case, OTOC also spreads ballistically in both directions, falling sharply outside the light cone before the left and right fronts collide. We find a linear light cone behavior which agrees with the Lieb-Robinson bound with velocity ${v^{\mathrm{LR}}=\sqrt{J_1^2+D^2}}$, corresponding to the maximal velocity in easy-plane XXZ-Heisenberg chain with exchange amplitude ${\sqrt{J_1^2+D^2}}$.
The overall amplitude and speed of the scrambling are lower as compared to case with ${J_2\neq 0}$; the velocities of the front propagation to the left and the right are now equal, and directional asymmetry only shows up in the amplitudes of OTOC (cf. the non-integrable case with ${J_2 = 1}$).
For $C^{zz}_d(t)$, we do not expect any directional asymmetry --- except the one, caused by twisted boundary conditions --- because unlike $\sig{x/y}{}$, $\sig{z}{}$ remains unchanged under the mentioned gauge transformation. 

For OTOC with low-energy long-wavelength probes (not studied here) we expect to observe a different velocities for the front propagation to the left and to the right for nonvanishing DM interaction also in the case of ${J_2=0}$.
The velocities for the low-energy long-wavelength probes, however, will not correspond to the Lieb-Robinson velocities for the given parameters.

The OTOC for the nonchiral state is fully symmetric in the case of a vanishing DM interaction, ${D=0}$ (see Fig.~\ref{fig:cxxd_J2_0}(a)).
The contour lines corresponding to ${C^{xx}_d(t)\leqslant 0.5}$ threshold remain symmetric for all considered cases, unlike the case with finite, ${J_2=1}$, where the vanishing of DM interaction was also required (cf. Fig.~\ref{fig:cxxd}).

In the long-time limit, the results shown in Fig.~\ref{fig:cxxd_J2_0} evidence that only the case with the chiral initial state and the chiral Hamiltonian (finite, ${D=0.5}$, DM interaction), saturates the desired value $2$ (see Fig.~\ref{fig:cxxd_J2_0}(d)).
For ${D=0.5}$ and, in general, ${D \neq -J_1\tan(2 k\pi/L)}$ the Hamiltonian eigenspectrum is not degenerate.
For the quenched cases (plots (g) and (h) in Fig.~\ref{fig:cxxd_J2_0}), the values to which $C^{xx}_d(t)$ saturate do not depend on distance $d$, but are still smaller than $2$. The symmetric case (nonchiral state and nonchiral Hamiltonian, ${D=0}$) is different. The values to which the time-averaged OTOC saturate are farther away from expected $2$. There is also a structure in the time-averaged OTOC, namely for the farthest (${d=11}$) and closest (${d=0}$) distance time-averaged OTOC values are the same and larger than the rest. Spectra of the system Hamiltonian and decomposition of the initial state in eigenvectors of the system Hamiltonian exhibit an extra symmetry in this case, containing two- and four-fold degenerate eigenpairs, indicating that OTOC is capable of distinguishing higher symmetric phases (see also Appendix~\ref{sec:apendix_sat_val_OTOC}).


For the short-time limit, also, for the integrable case, we observe a power-law growth of OTOC in all four studied setups, as shown, e.g., for ${D=0.5}$ in the center panel of Fig.~\ref{fig:shorttime_cxxd_J2_0}.
At leading order, $C^{xx}_d(t)$ increases as $t^{2\nu}$, but now with ${\nu=\max(1,|d|)}$ (cf. ${J_2 = 1}$).
For the chiral initial state (${\kappa^z \neq 0}$) or in the system with a nonvanishing DM interaction (${D \neq 0}$) we also observe the asymmetric subleading, $t^{2\nu+1}$, corrections to it. This is also demonstrated in Fig.~\ref{fig:shorttime_cxxd_quartet_J2_0}, where we plot the scaled values of $C^{xx}_d(t)$ and the corresponding scaled values of the directional asymmetry ${|\Delta C^{xx}_d(t)|}$ for the fixed ${|d|=3}$ and ${|d|=4}$ distances (see also the bottom panel in Fig.~\ref{fig:shorttime_cxxd_J2_0}).
All these are also consistent with the discussion in Sec.~\ref{sec:short_time_limit}.
Similar power-law behavior was reported in refs.~\cite{Dora2017,Riddell2019}, where the authors studied the one-dimensional spinless fermions with nearest-neighbor repulsion (which is equivalent to the 1D Heisenberg XXZ) and non-interacting XX model, respectively. 

Finally, in Fig.~\ref{fig:ckkd_ckkd_J2_0} we show the results for the spatiotemporal evolution of $C^{\kappa\kappa}_d(t)$ \eqref{eq:ckkd} for symmetric (${D=0}$, plots (a) and (c)) and asymmetric cases (${D=0.5}$, plots (b) and (d)) for non-integrable ${J_2=-J_1=1}$ (plots (a)-(b)) and integrable $J_2=0$, $J_1=-1$ (plots (c)-(d)) cases.
For this operator pairs, we also found ballistically spreading fronts in both directions, falling sharply outside a light cone, before the left and right fronts collide.
As expected, the OTOC for the nonchiral case is fully symmetric (see plots (a) and (c) in Fig.~\ref{fig:czzd}). One can identify the second light cone for the non-integrable ${J_2=-J_1=1}$ case for a vanishing DM interaction (see Fig.~\ref{fig:ckkd_ckkd_J2_0}(a)) reflecting the two distinct isolated maxima in the single-magnon group velocities.
The directional asymmetry in scrambling for the chiral case is also more pronounced when probed with chirality ${\e^{\im\kappa^{z}_{j}}}$ operators instead of ${\sig{x/z}{j}}$-operators (cf. Fig.~\ref{fig:ckkd_ckkd_J2_0}(b) with Fig.~\ref{fig:czzd}(b) and Fig.~\ref{fig:cxxd}(b) and Fig.~\ref{fig:ckkd_ckkd_J2_0}(d) with Fig.~\ref{fig:cxxd_J2_0}(b)).
The long-time behavior of time-averaged OTOC (not shown here) is less conclusive in this case --- because we do not have exact (analytic) values to which the long-time averaged OTOC should converge --- in order to compare with.
Therefore, we cannot judge whether desired values are reached, even in a non-integrable case. Only for the integrable case and symmetric setup, we can for sure say that the scrambling is incomplete, exhibiting explicit distance dependence in the long-time averaged OTOC values.

For the short-time limit, $C^{\kappa\kappa}_d(t)$ exhibits a power-law growth in an early time regime. However, there are also differences. 
Because the chirality operators on the adjacent links (${d=1}$) do not commute (they share one site), the largest among the leading terms in the directional asymmetry of OTOC is linear in time, $\sim t$, as compared to cubic, $\sim t^3$, for ${\sig{x/z}{j}}$-operators (see Fig.~\ref{fig:shorttime_ckkd} in Appendix~\ref{sec:appendix_early_time_reg}). These show that different observables are not equally sensitive to the asymmetric spreading of quantum information.

\begin{figure*}[!htbp]
 \includegraphics[width=\textwidth]{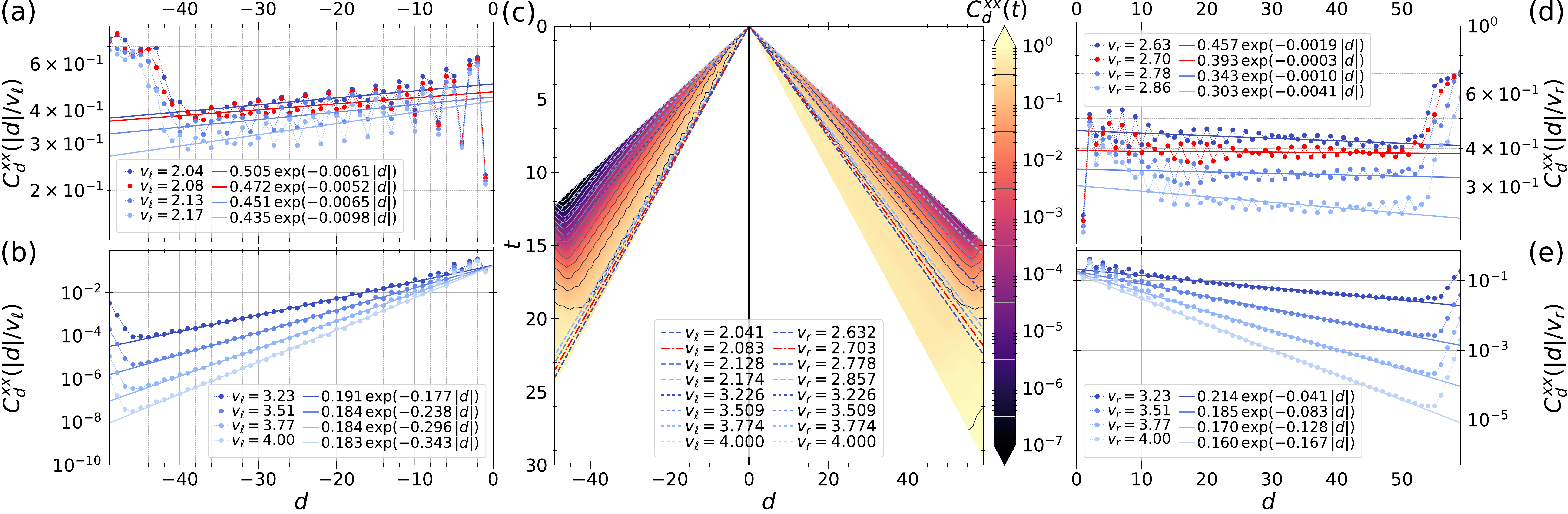}
 \caption{Spatiotemporal evolution of the OTOC $C^{xx}_d(t)$ \eqref{eq:cxxd} along the constant velocity rays for a chain with ${L=102}$ sites with PBC in a two-magnon sector (${S^z_{\mathrm{tot}}=L/2 - 2}$), ${J_2=-J_1=1}$, ${D=0.5}$, ${B_z=1.604}$, the chiral case. A finite magnetic field does not change the behavior of $C^{xx}_d(t)$ qualitatively.
          The central panel, plot (c), a false-color logarithmic plot at fixed velocity rays for $v_{\ell/r}=1/(t_0 + n \Delta t)$, with $t_0 =0.25$, $\Delta t = 0.005$, and ${n=0,\dots,50}$.
          Pairs of semi-log plots (a), (b), and (d), (e), show several cuts at the given velocities for the left and right going rays (also shown on the central panel, plot (c)), together with the corresponding linear regression lines. The linear regressions (exponential fits) are performed for the data in the range ${10\leqslant d \leqslant 40}$ and ${10\leqslant d \leqslant 50}$ for the left and right going rays, respectively; dotted lines are just guides for the eye; the red dots and the red lines represent the data with the minimal inclination.
         }
\label{fig:cxxt_rays}
\end{figure*}

All OTOC probes with local operators show  similar expansion speed which is close to the  maximal group velocities of single-magnon excitations for the given direction~\footnote{Close to the saturation magnetization, the entire single-magnon bandwidth is available and the excitations with maximal group velocities are present in the excitation spectrum.}.
The latter is governed by the model Hamiltonian parameters (see Eq.~\eqref{eq:v_lr_LR} in Sec.~\ref{sec:model}).

\subsection{Intermediate time regime}

In this section, we discuss the intermediate time regimes in the OTOC propagation,
i.e., the times up to the wavefront reaching the location at a given distance or
equivalently the fixed-velocity rays, for speeds larger than butterfly velocity.

\begin{figure}[!thbp]
 \includegraphics[width=\columnwidth]{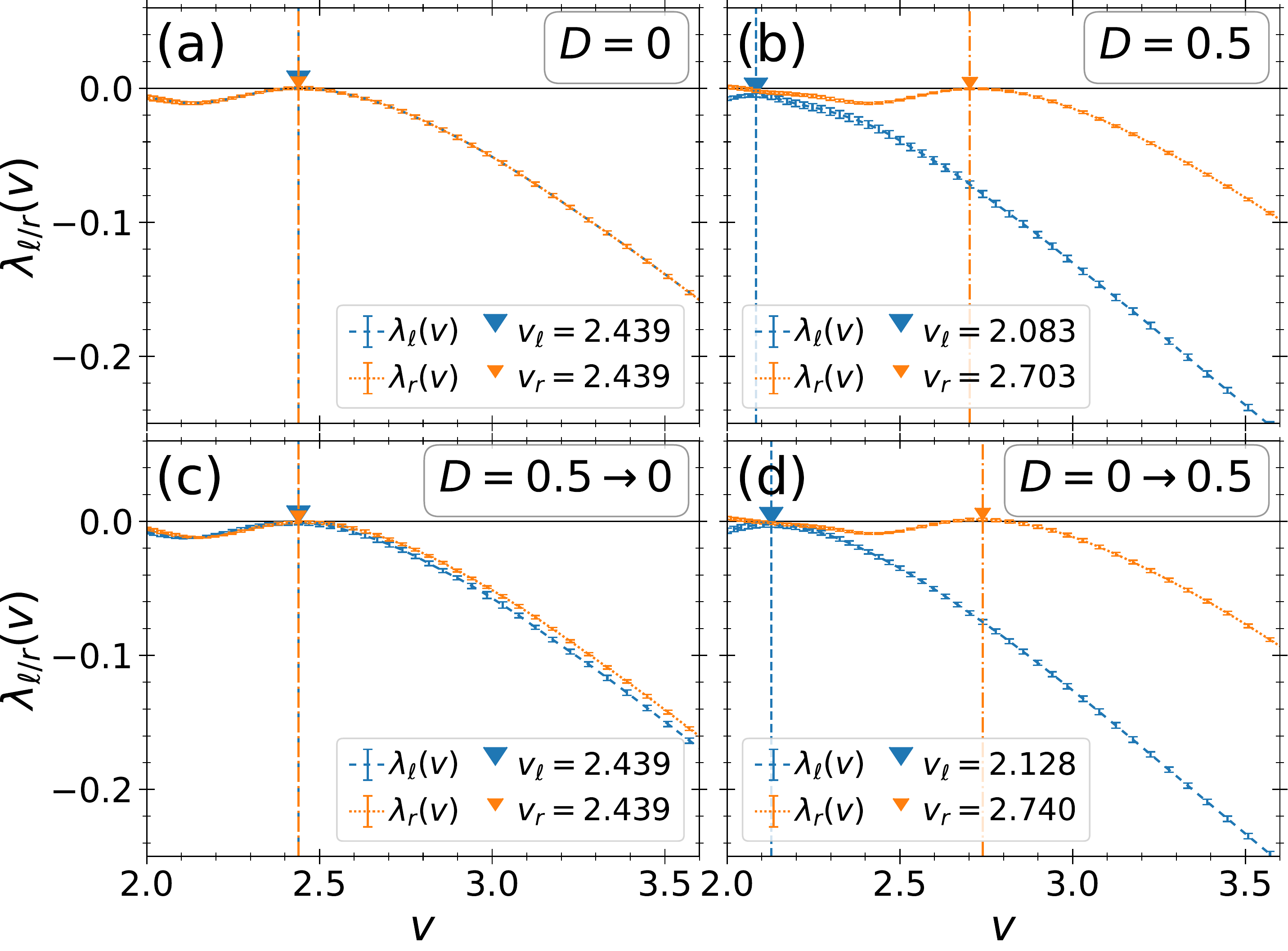}
 \caption{The exponents of the decay $\lambda_\ell(v)$ and $\lambda_r(v)$ of the fitted OTOC for the left and right fixed velocity rays, respectively, for all four studied cases.
          Exponential fits are performed for the data in the ranges ${10 \leqslant d \leqslant 44}$ (a), (c) and ${10 \leqslant d \leqslant 40}$ and ${10 \leqslant d \leqslant 50}$, for the left and right branch, respectively, (b),(d).
          Fitting error bars and the local maximum, corresponding to the highest velocity $v_{\ell/r}$ (triangles) are also shown.
         }
\label{fig:lambda_v}
\end{figure}

\begin{figure}[!thbp]
 \includegraphics[width=\columnwidth]{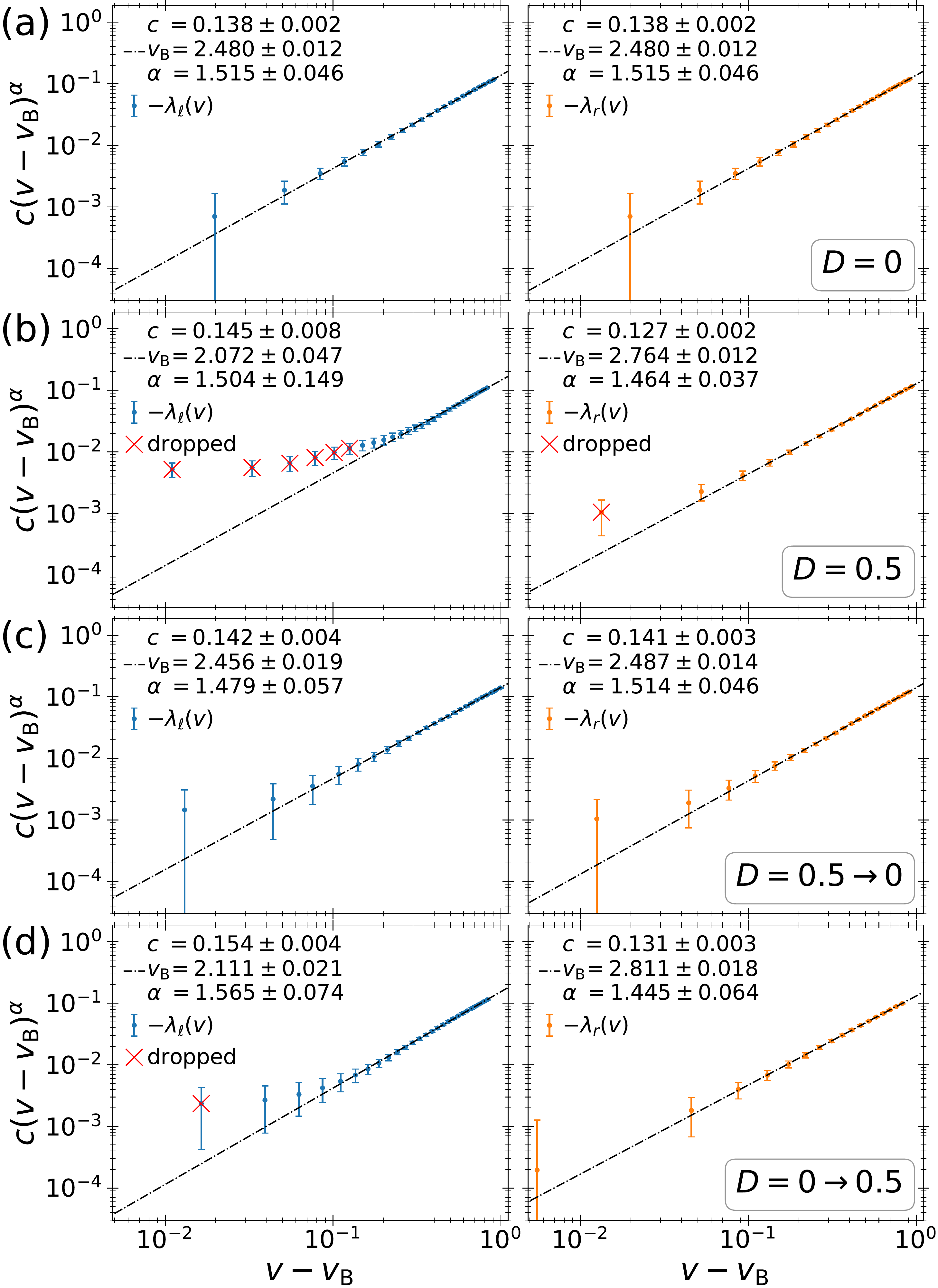}
 \caption{Log-log plots of the $\lambda_{\ell/r}(v)$ from Fig.~\ref{fig:lambda_v} (blue and orange dots with error bars) along with the least-square fits for the range ${v_{\ell/r} < v < 1.4\,v_{\ell/r}}$ to ${c(v-\VB(\vec{\hat{n}}))^\alpha}$ (dash-dotted lines).
         }
\label{fig:lambda_v_fits}
\end{figure}

To verify the proposed universal form, Eq.~\eqref{eq:OTOC1_alpha_vb}, of the space-time evolution of OTOC around and outside of the ballistic light cone, we study the ${L=102}$-spin chain in a two-magnon sector. ${L=102}$ sites give us a possibility to examine distances, $d$, large enough so that the local correlation effects are weak, and longer times for which the propagating wavefronts have not collided yet due to PBC.
Long chains give us also the opportunity to crosscheck whether exponential behavior sets in for some time interval.

\begin{figure*}[!tbp]
 \includegraphics[width=\textwidth]{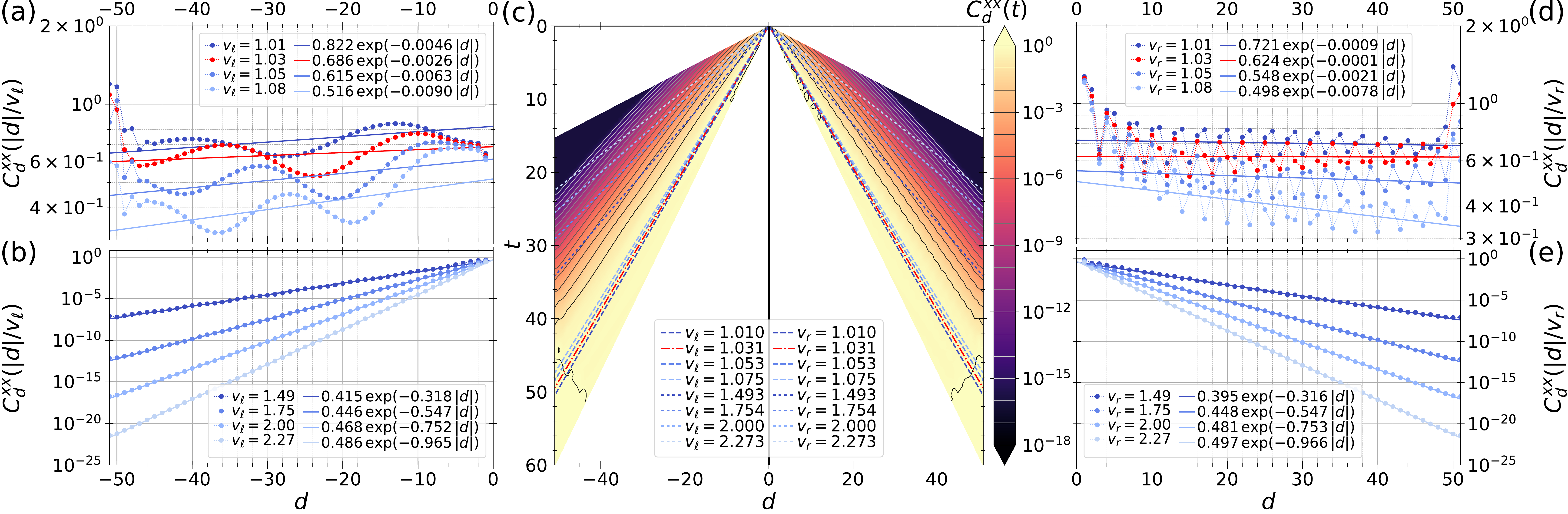}
 \caption{Spatiotemporal evolution of the OTOC $C^{xx}_d(t)$ \eqref{eq:cxxd} along the constant velocity rays for a chain with ${L=102}$ sites with PBC in the two-magnon sector (${S^z_{\mathrm{tot}}=L/2 - 2}$), ${J_1=-1}$, ${J_2=0}$, and ${D=0.5}$, the chiral case.
           The central panel, plot (c), a false-color logarithmic plot at fixed velocity rays for $v_{\ell/r}=1/(t_0 + n \Delta t)$, with $t_0 =0.3$, $\Delta t = 0.01$, and ${n=0,\dots,90}$.
           Pairs of semi-log plots (a), (b), and (d), (e), show several cuts at the given velocities for the left and right going rays (also shown on the central panel, plot (c)), together with the corresponding linear regression lines. The linear regressions (exponential fits) are performed for the data in the range ${9\leqslant d \leqslant 46}$; dotted lines are just guides for the eye; the red dots and the red lines represent the data with the minimal inclination.
          }
\label{fig:cxxd_rays_J2_0}
\end{figure*}

\begin{figure}[!htbp]
 \includegraphics[width=\columnwidth]{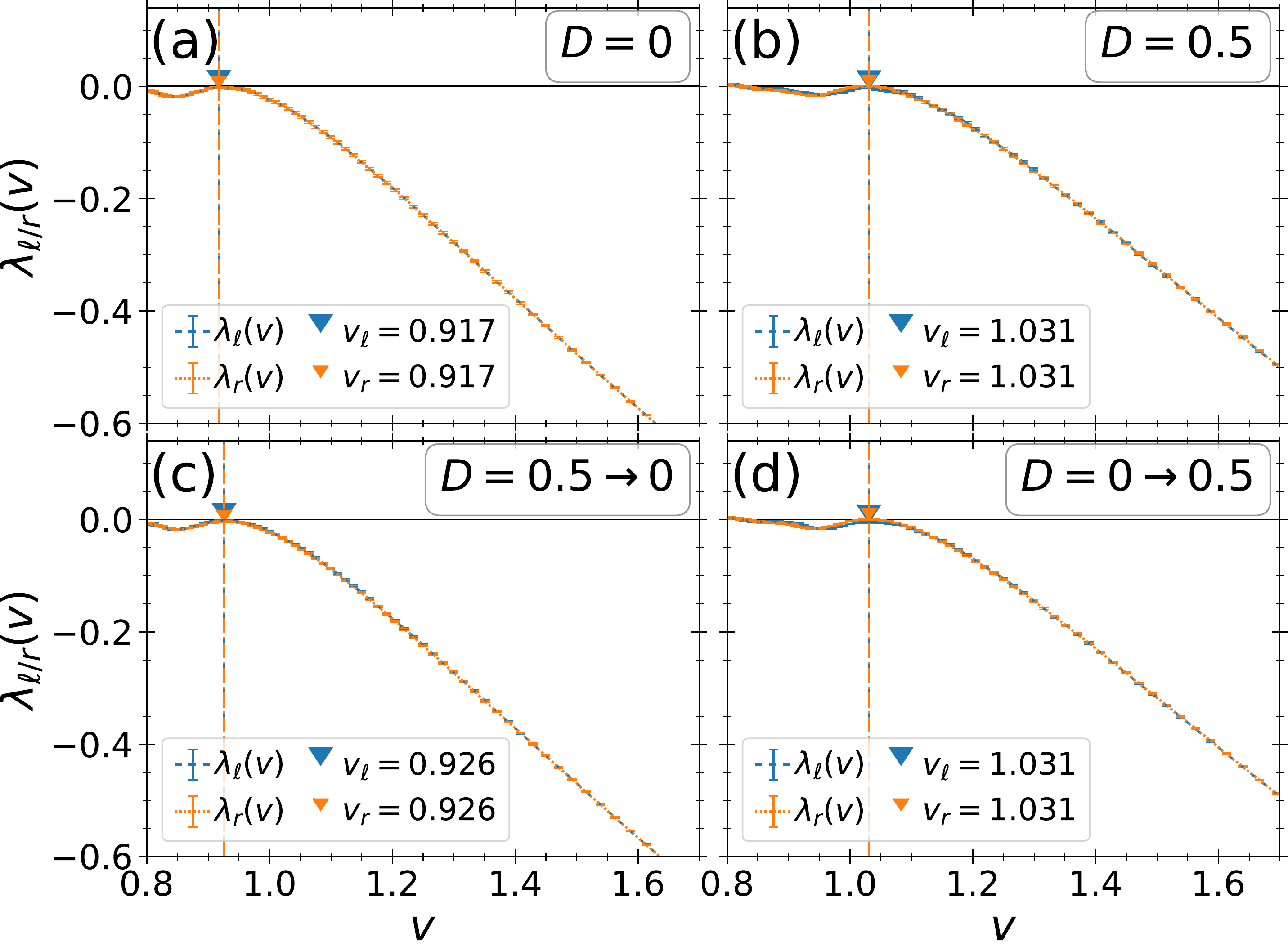}
 \caption{The exponents of decay $\lambda_\ell(v)$ and $\lambda_r(v)$ of the fitted OTOC for the left and right fixed velocity rays, respectively, for all four studied cases.
          Exponential fits are performed for the data in the range ${9 \leqslant d \leqslant 46}$.
          Fitting error bars and the local maximum, corresponding to the highest velocity $v_{\ell/r}$ (triangles) are also shown.
         }
\label{fig:lambda_v_J2_0}
\end{figure}

\begin{figure}[!thbp]
 \includegraphics[width=\columnwidth]{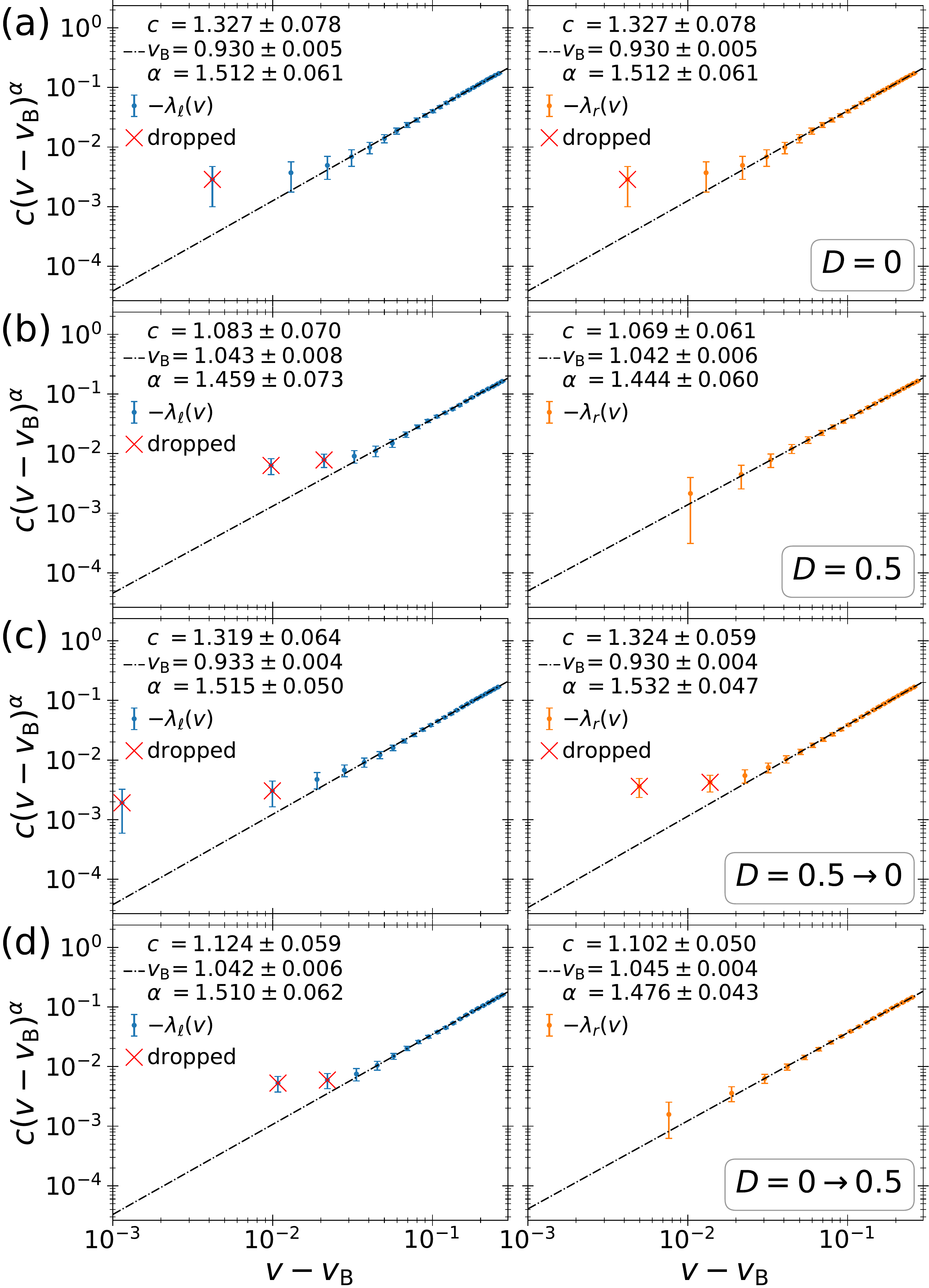}
 \caption{Log-log plots of the $\lambda_{\ell/r}(v)$ from Fig.~\ref{fig:lambda_v_J2_0} (blue and orange dots with error bars) along with the least-square fits for the range ${v_{\ell/r} < v < 1.3\,v_{\ell/r}}$ to ${c(v-\VB(\vec{\hat{n}}))^\alpha}$ (dash-dotted lines).
         }
\label{fig:lambda_v_fits_J2_0}
\end{figure}

To assess the proposed universal functional form, Eq.~\eqref{eq:OTOC1_alpha_vb}, we consider the $C^{xx}_d(t)$ behavior along the fixed-velocity rays.
For this, we sample the data at ${t = (t_0 + n\Delta t)d}$ for ${d=-101,\dots,101}$ with ${n\in\mathbb{N}}$. The results would correspond to rays with ${v=\pm 1/(t_0+n\Delta t)}$ velocities (``$+$'' for the right ($r$) and ``$-$'' for the left ($\ell$) wavefronts).
We fit the obtained numerical results in two steps. First, we identify the exponents of the decay, $\lambda(\vec{v})$, for each of the fixed-velocity rays, employing the linear-regression on a semi-log scale. Then we fit (with a non-linear fit) the obtained exponents, $\lambda(\vec{v})$-s, to ${c(v-\VB(\vec{\hat{n}}))^\alpha}$ function, with $\alpha$, $\VB({\hat{\vec{n}}})$, and $c$ fitting parameters, in order to determine a scaling exponent $\alpha$ and a butterfly velocity $\VB(\vec{\hat{n}})$, for the left and right branches, ${\vec{\hat{n}}=-\vec{e}_x}$ and ${\vec{\hat{n}}=\vec{e}_x}$, respectively.

We start with the non-integrable case ${J_2=-J_1=1}$. As a representative example in Fig.~\ref{fig:cxxt_rays}, we show spatiotemporal evolution of the OTOC $C^{xx}_d(t)$ \eqref{eq:cxxd} along the fixed-velocity rays for ${D=0.5}$ (the asymmetric/chiral case), ${t_0=0.25}$, and ${n=0,\dots,50}$.
Figures~\ref{fig:cxxt_rays}(a),~\ref{fig:cxxt_rays}(b),~\ref{fig:cxxt_rays}(d),~and~\ref{fig:cxxt_rays}(e) show $C^{xx}_d(t)$ along selected fixed-velocity rays (${v_{\ell/r}}$) on a semi-log scale together with the corresponding linear-regression lines (exponential fits). The same rays we also display in Fig.~\ref{fig:cxxt_rays}(c).
We perform fits for ${10\leqslant d \leqslant 40}$ and ${10\leqslant d \leqslant 50}$, for to the left and right rays, respectively, to avoid effects caused by short-range correlations at short distances and collisions of wavefronts for long distances.
In Fig.~\ref{fig:lambda_v}, we show the obtained exponents ${\lambda_{\ell/r}}(v)$ (left and right branches, respectively) of the OTOC decay together with the standard deviation as a function of ray velocity ${v}$ for all four studied cases.
We also display the local maximum of ${\lambda_{\ell/r}}(v)$ corresponding to the largest velocity $v_{\ell/r}$.
For smaller velocities (${v<v_{\ell/r}}$), fitting to the exponential form is less reliable (other functions might fit better) even when the error bars (the standard deviations) of the fitted values are small.
We found that the directional asymmetry in the exponents of the decay is determined by the Hamiltonian under which the system evolves in time, and there is a negligible influence of the finite chirality of the initial state. On the level of ${\lambda(\vec{v})}$, there is no directional asymmetry when the time-evolving Hamiltonian does not contain a chiral term (DM interaction).

Within the accuracy of the fitted data, we do not identify any region with strictly positive values of $\lambda_{\ell/r}$ (see also Fig.~\ref{fig:lambda_v}), which is consistent with the expected behavior that there is no exponential divergence for the quantum system with local Hamiltonians and finite-dimensional sub-spaces of constituent parts, even in the case of well-separated operators \cite{Kukuljan2017,Khemani2018VDLE,XuSwingle2019}.

To determine the butterfly velocity $\VB(\vec{\hat{n}})$ and the exponent $\alpha$, we have performed a non-linear curve fit of the data in the range ${v_{\ell/r}< v < 1.4\,v_{\ell/r}}$ to the function ${c(v-\VB(\vec{\hat{n}}))^\alpha}$, with $\alpha$, $\VB(\vec{\hat{n}})$, and $c$ being fitting parameters.
The fitting results are shown in Fig.~\ref{fig:lambda_v_fits} indicating that
 $\alpha$ depends on the fitting range, decreasing from some ${\alpha \gtrsim 1.5}$ value (for velocities close to $\VB(\vec{\hat{n}})$) towards some ${\alpha > 1}$ with considerably increased fitting range.
Eventually a crossover occurs from ${\sim (v-\VB(\vec{\hat{n}}))^\alpha}$ towards ${\sim v\ln v}$ functional form, which is consistent with expected behavior for ${t\ll 1}$ and ${vt \gg 1}$ case --- the validity region of a small-time expansion, where we expect ${\sim v\ln v}$ behavior \eqref{eq:v_ln_v} (see also Appendix~\ref{sec:decay_exp_small_t}, Eq.~\eqref{eq:v_ln_v_appx}).
The obtained results (within error-bars) reveal that the Hamiltonian under which the system time-evolves, determines entirely the $\VB(\vec{\hat{n}})$ and $\alpha$ values. On the level of ${\lambda(\vec{v})}$, there is no directional asymmetry when the time-evolving Hamiltonian is symmetric (${D=0}$).
It only shows up at the level of different amplitudes in the case of the chiral initial state.
On the other hand, in the case of chiral Hamiltonian (${D\neq 0}$), there is a clear difference between the values obtained for the left and right branches irrespective of the fact whether the chiral or nonchiral initial state were considered.

The determined butterfly velocities are comparable with the maximal group velocities of the single free magnon case (see Sec.~\ref{sec:model}), indicating that the fronts in OTOC are dominantly free-magnon-like.
The same butterfly velocities and exponent $\alpha$ (within error bars) are also obtained for the OTOC with $\sig{z}{}$-s and ${\exp(\im\kappa^z)}$-s.
The fact that $\alpha\approx 3/2$, the specific value for the case of the free particles (see \cite{Xu2020,LinMotrunich2018,Khemani2018VDLE,XuSwingle2019,Riddell2019}), further supports this observation.

All these findings, except one, also hold for the integrable case ($J_2=0$).
For this case, we took $t_0=0.3$, $\Delta t=0.01$, and $n=0,\dots,90$. The results are shown in Figs.~\ref{fig:cxxd_rays_J2_0}-\ref{fig:lambda_v_fits_J2_0}.
The only difference is that the finite chirality is no longer exposed at the level of OTOC decay exponents $\lambda_{\ell/r}$.
Because finite DM interaction merely renormalizes the exchange amplitude ${J_1\rightarrow \sqrt{J_1^2+D^2_{\vphantom{1}}}}$, the maximal group velocities are only increased but are equal for left and right traveling free magnons (see Sec.~\ref{sec:model}), which is manifest in the same left and right decay exponents ${\lambda_\ell(v) \approx \lambda_r(v)}$ (see Fig.~\ref{fig:lambda_v_J2_0}) and the same $\alpha$ and $\VB$ for the left and right branches (see Fig.~\ref{fig:lambda_v_fits_J2_0}).
The directional asymmetry for this case shows up only in the different amplitudes, even in the case of a finite DM interaction and the chiral initial state.

Thus, different left and right butterfly velocities necessitate a finite frustration term (${J_2 \neq 0}$) in addition to a chirality breaking finite DM interaction (${D\neq 0}$) term.

\section{Conclusions}\label{sec:conclusion}

We studied the spreading of quantum spin correlations in frustrated spin chains with spin-current-driven ferroelectricity \cite{Katsura2005}. Such chains are experimentally   realized  in oxide-based single-phase multiferroics \cite{Miyata2021,Schrettle2008}.
The emergent ferroelectric polarization, which allows coupling to an external electric field, is proportional to the spin chirality. Therefore, the latter can also be tuned by external electrical field. The impact of residual chirality on quantum information spreading/delocalization, meaning scrambling, is exposed by analytical and exact numerical results.
To quantify quantum scrambling, we employed OTOC as a useful witness.

The symmetry considerations in Sec.~\ref{sec:anisotropy_measures} showed that scrambling is symmetric for any eigenstate of the Hamiltonian in the zero-magnetization sector ($S^z_{\mathrm{tot}}=0$, i.e., half-filled case) or at finite or infinite temperatures for a zero magnetic field (${B_z=0}$). At infinite temperature in the case of the nonvanishing magnetic field (${B_z \neq 0}$), the directional asymmetry in scrambling escapes the detection by OTOC of $\sig{z}{}$-s.

We found that OTOC exhibits a power-law growth at early-times, irrespective of the integrability of the model. This power-law behavior is extracted by expanding the OTOC kernel (squared commutator) for sufficiently small times ${t \ll 1}$. The found power-law behavior is consistent with the models characterized by local Hamiltonians and finite on-site degrees of freedom \cite{Dora2017,LinMotrunich2018,LinMotrunich2018_2,Riddell2019,Kukuljan2017,Khemani2018VDLE,XuSwingle2019}. 
The leading order is always symmetric and only the subleading corrections exhibit the directional asymmetry.

Exact numerical studies for chains with 22  spins close to saturation magnetization   revealed 
that OTOC   spread ballistically to both (the left and the right) directions, falling sharply outside a light cone before the fronts collide.
The directional asymmetry in scrambling is governed by the chiral coupling, a non-vanishishing dynamical DM interaction.
We verified numerically the conjectured universal form, Eq.~\eqref{eq:OTOC1_alpha_vb}, of OTOC outside and close to the wavefront. For this, we studied the OTOC along fixed velocity rays for a chain with ${L=102}$ sites close to saturation.
We considered the distances on which local correlation effects are weak, and examined longer times for which the wavefronts have not collided yet.
We showed numerically that the proposed universal form Eq.~\eqref{eq:OTOC1_alpha_vb} fits almost perfectly to the obtained data.
Within the accuracy of the fitted data, no regions are found with the simple exponential growth of OTOC. The butterfly velocity is direction dependent only in the cases with chiral Hamiltonians with a finite next-nearest-neighbor exchange, ${J_2 \neq 0}$. For ${J_2=0}$, the chiral term (DM interaction)  symmetrically enhances the butterfly velocity, unlike for the case of low-energy large-wavelength probes where nonvanishing DM would be sufficient.
For spatially local probes   and for  ${J_2=0}$, which is equivalent to the easy-plane XXZ Heisenberg model  the directional asymmetry is only exposed with respect to amplitude  (but not  the functional form of the velocity-dependent decay exponent and butterfly velocity).
The obtained velocities are comparable to the group velocities found for the single magnon dispersion. The parameter which characterizes the expansion of the wavefront during the ballistic spreading, ${\alpha=3/2}$, is free-particle like. Thus, the spatial spreading of OTOC is predominantly free-magnon like.
Whether this picture   holds for the effective {\em easy-axis} XXZ model  remains to be investigated.
The obtained results are useful for applications based on the transfer of quantum information through chiral channels and multiferroic spintronics.

{\em Acknowledgments:}
This work was supported by the DFG through SFB TRR227 and the National Science Centre (NCN, Poland) under Grant No. 2019/35/B/ST3/03625 (N.S.). We thank L. Chotorlishvili for numerous discussions.

\appendix

\section{Equivalent expressions for Spin Vector Chirality operators: $\hat{\kappa}^z_j$ and $\e^{\im \hat{\kappa}^z_j}$}
\label{sec:spin_chirality}
In this section, we consider the unitary operators, $\e^{\im \hat{\kappa}^z_j}$, made out of local vector spin chirality
\begin{align}
  \label{eq:kappa}
  \hat{\kappa}^z_j &= \left(\hat{\vec{S}}_{j} \times \hat{\vec{S}}_{j+1}  \right)\ZZ
                    = \spin{x}{j} \spin{y}{j+1} - \spin{y}{j} \spin{x}{j+1}\,
  \nonumber \\
                   &= \frac{1}{4}\sigcroszz{j}{j+1}
                    =\frac{1}{4}\left(\sig{x}{j} \sig{y}{j+1} - \sig{y}{j} \sig{x}{j+1} \right) \,.
\end{align}
The parts on the right-hand side of the Eq.~\eqref{eq:kappa} do commute,
${\left[\sig{x}{j} \sig{y}{k},\sig{y}{j} \sig{x}{k} \right]=0}$.
Therefore,
\begin{align}
 \label{eq:generilized_kappa}
  \e^{\im b\left(\sig{x}{j} \sig{y}{k} - \sig{y}{j} \sig{x}{k} \right)}
  &=
  \e^{\im b\,\sig{x}{j} \sig{y}{k}}\,\e^{-\im b\,\sig{y}{j} \sig{x}{k}}
  \nonumber \\
  &=
   \cos^2\!b\,\mathbb{I}
  +\sin^2\!b\,\,\sig{z}{j}\sig{z}{k}
  \nonumber
  \\
  &
  \quad
  +\frac{\im}{2}\sin(2b) \left(\sig{x}{j}\sig{y}{k} - \sig{y}{j} \sig{x}{k}  \right)
  \,.
\end{align}
Here, we made use of operator identities
\begin{align}
    \e^{\im b \sig{\alpha}{j} \otimes \sig{\beta}{k}}
    =
    \cos b\,\mathbb{I}\otimes\mathbb{I} + \im\sin b\,\sig{\alpha}{j} \otimes \sig{\beta}{k}
\end{align}
and ${\sig{x}{j} \sig{y}{j} =\im \sig{z}{j}}$.
From Eqs.~\eqref{eq:kappa}~and~\eqref{eq:generilized_kappa} follows that
\begin{align}
  \e^{\im \hat{\kappa}^z_j}
  &=
   \cos^2\!\tfrac{1}{4}\,\mathbb{I}
  +\sin^2\!\tfrac{1}{4}\,\sig{z}{j}\sig{z}{j+1}
  +\,\im\,2 \sin\tfrac{1}{2}\kappa^z_j\,.
\end{align}
Hence,
\begin{align}
 \begin{split}
 \left|\!\left[
   \e^{\im \hat{\kappa}^z_j(t)}
   ,
   \e^{\im \hat{\kappa}^z_k}
 \right]\!\right|^2
 &
 =
 \bigg{|}
   \sin^2\!\tfrac{1}{4}
   \left[
   \sig{z}{j}(t)\sig{z}{j+1}(t)
   ,
   \sig{z}{k}\sig{z}{k+1}
   \right]
   \\
   &\quad\ 
   + \im\,2 \sin\!\tfrac{1}{2}
   \Big(\!
     \left[
       \sig{z}{j}(t)\sig{z}{j+1}(t)
       ,
       \kappa^z_k
     \right]
     \\
     &\qquad\qquad\qquad
     +
     \left[
       \kappa^z_j(t)
       ,
       \sig{z}{k}\sig{z}{k+1}
     \right]
   \!\Big)
   \\
   &\quad
   -
   16 \cos^2\!\tfrac{1}{4}
   \left[
     \kappa^z_j(t)
     ,
     \kappa^z_k
   \right]
 \!
 \bigg{|}^2
 \sin^4\!\tfrac{1}{4}.
 \end{split}
\end{align}

\section{Short-time limit, leading and subleading contributions}\label{sec:sigma_2nd_and_3rd_order}

We start with the leading contributions in OTOC, $Q^{(0)}_{mn}(t)$~\eqref{eq:sigma_leading}, for ${d=|n-m|\leqslant2}$ which are quadratic in $t$. 
For $|d| = \pm 1$, $m = n \pm 1$,
\begin{align}
\begin{split}
  {\left[
    [
      \hat{H}
      ,
      \sig{\alpha}{n \pm 1} 
    ]
    ,
    \sig{\alpha}{n}
  \right]}
  &=
  - J_1
  \left(
    \sigmiss{n}{n \pm 1}{\alpha}
  \right)
  \\
  &
  \mp D\,\codel{z}{\alpha}
  \sigcros{n}{n \pm 1}{z}
  \,.
  \label{eq:sigma_first_d_1}
\end{split}
\end{align}
The first term in Eq.~\eqref{eq:sigma_first_d_1} is real-valued whereas the second is pure imaginary. Therefore,
\begin{align}
  Q^{(0)}_{n \pm 1,n}(t)
  &=
  t^2_{\vphantom{1}}
  \bigg(
    J_1^2
    \left(
      \sigmiss{n}{n \pm 1}{\alpha}
    \right)^2
    \nonumber \\
    &\qquad\ \ 
    +
    \codel{z}{\alpha}
    D^2_{\vphantom{1}}
    \left|
      \sigcros{n}{n \pm 1}{z}
    \right|^2
  \bigg)
  \nonumber \\
  &=
  2\,t^2_{\vphantom{1}}\left(J_1^2+\codel{z}{\alpha}D^2_{\vphantom{1}}\right)
  \left(
    \mathbb{I}
    -
    \sig{\alpha}{n} \sig{\alpha}{n \pm 1}
  \right)
  .
\end{align}
For a translationally invariant system
\begin{equation}
  \Delta C^{a/c,(0)}_{n,n+1}(t)
  =
  \Delta C^{b,(0)}_{n,1}(t)
  =
  \Delta C\PHDG_{d=1}(t)
  = 0\,.
\end{equation}
For $|d| = \pm 2$, $m = n \pm 2$,
\begin{align}
  {\left[
    [
      \hat{H}
      ,
      \sig{\alpha}{n \pm 2} 
    ]
    ,
    \sig{\alpha}{n}
  \right]}
  &=
  - J_2
  \left(
    \sigmiss{n}{n \pm 2}{\alpha}
  \right)
  .
  \label{eq:sigma_first_d_2}
\end{align}
As expected, this term vanishes in the case of ${J_2=0}$.
The right-hand side of Eq.~\eqref{eq:sigma_first_d_2} is real valued. Hence,
\begin{align}
  Q^{(0)}_{n \pm 2,n}(t)
  &=
  t^2_{\vphantom{1}}
  J_2^2
  \left(
    \sigmiss{n}{n \pm 2}{\alpha}
  \right)^2
  \nonumber \\
  &=
  2\,t^2_{\vphantom{1}} J_2^2
  \left(
    \mathbb{I}
    -
    \sig{\alpha}{n} \sig{\alpha}{n \pm 2}
  \right)^2
  \,.
\end{align}
For systems with translation invariance 
\begin{equation}
  \Delta C^{a/c,(0)}_{n,n+2}(t)
  =
  \Delta C^{b,(0)}_{n,2}(t)
  =
  \Delta C\PHDG_{d=2}(t)
  = 0\,.
\end{equation}

Finally, for larger values of ${d>2}$, higher-order terms (quartic in $t$, ${\nu\geqslant 2}$) must be considered because, as expected, ${\left[[\hat{H},\sig{\alpha}{n\pm d}],\sig{\alpha}{n}\right]=0}$ in this case.

The subleading correction of $Q^{(0)}_{m,n}(t.)$~\eqref{eq:sigma_leading}, $Q^{(1)}_{m,n}(t)$~\eqref{eq:sigma_subleading},
for ${d=\pm 1}$ (${m=n\pm 1}$) and ${d=\pm 2}$ (${m=n\pm 2}$) are cubic in $t$ and have the following forms
\begin{widetext}
\begin{align}
  Q^{(1)}_{n\pm 1,n}(t)
  &=
  \im
  \frac{t^3}{2}
  \left\{
    [
      [\hat{H},\sig{\alpha}{n \pm 1}]
      ,
      \sig{\alpha}{n}
    ]
    ,
    \left[
      [
        \hat{H}
        ,
        [
          \hat{H}
          ,
          \sig{\alpha}{n \pm 1} 
        ]
      ]
      ,
      \sig{\alpha}{n}
    \right]
  \right\}
  \nonumber \\
\begin{split}
  &=
  \frac{t^3}{2}
  \bigg(
  (J_1^2+\codel{z}{\alpha}\,D^2_{\vphantom{1}})
  \Big(
    J_1^{\vphantom{2}}
    \,
    \sig{\alpha}{n}\!
    \sigcros{n \pm 1}{n \pm 2}{\alpha}
    +J_2^{\vphantom{2}}
    \,
    \sig{\alpha}{n}\!
    \sigcroscomp[+]{n \pm 1}{n \mp 1}{n \pm 3}{\alpha}
    \\
    &
    \qquad\qquad\qquad\qquad\qquad
    \mp
    \codel{z}{\alpha}\,D
    \,
    \sig{z}{n}
    (
      \sigmiss{n \pm 1}{n \pm 2}{z}
    )
  \Big)
  \\
  &
  \qquad\quad
  \pm
  (1 - \codel{z}{\alpha})
  J_1^2 D
  \left(
    \sig{\alpha}{n}
    \sig{\alpha}{n \pm 1}
    \left(
      \sig{z}{n \mp 1}
      +
      \sig{z}{n \pm 2}
    \right)
    +
    2\,
    \sig{\alpha}{n}\sig{z}{n \pm 1}\sig{\alpha}{n \pm 2}
    -
    \sig{z}{n \mp 1}
    -
    2\,
    \sig{z}{n \pm 1}
    -
    \sig{z}{n \pm 2}
  \right)
  \\
  &
  \qquad\quad
  \mp J_2 D J_1
  (
    \mathbb{I}
    -
    \sig{\alpha}{n} \sig{\alpha}{n \pm 1}
  )
  \left(
    \left(
    2
    +
    2\codel{z}{\alpha}
    \right)
    \left(
    \sig{z}{n \mp 1}
    +
    \sig{z}{n \pm 2}
    \right)
    +
    \left(
      1 - \codel{z}{\alpha}
    \right)
    \left(
      \sig{z}{n \mp 2}
      +
      \sig{z}{n \pm 3}
    \right)
  \right)
  \\
  &
  \qquad\quad
  +
  4 B_z J_1
  \left(
    J_1\lev{z}{\alpha}{\beta}
    \sig{\alpha}{n}\sig{\beta}{n\pm 1}
    \mp D
    \left(1-\codel{z}{\alpha}\right)
    \left(\mathbb{I}-\sig{\alpha}{n}\sig{\alpha}{n\pm 1} \right)
  \right)
  \bigg)\,,
 \label{eq:Q_1_n_n}
\end{split}
\end{align}
\\
\begin{align}
  Q^{(1)}_{n\pm 2,n}(t)
  &=
  \im
  \frac{t^3}{2}
  \left\{
    [
      [\hat{H},\sig{\alpha}{n \pm 2}]
      ,
      \sig{\alpha}{n}
    ]
    ,
    \left[
      [
        \hat{H}
        ,
        [
          \hat{H}
          ,
          \sig{\alpha}{n \pm 2}
        ]
      ]
      ,
      \sig{\alpha}{n}
    \right]
  \right\}
  \nonumber
  \\
\begin{split}
  &=
  {t^3}J_2
  \bigg(
    J_1 J_2
    \,
    \sig{\alpha}{n}\!
    \sigcroscomp[+]{n \pm 2}{n \pm 1}{n \pm 3}{\alpha}
    +J_2^2
    \,
    \sig{\alpha}{n}\!
    \sigcros{n \pm 2}{n \pm 4}{\alpha}
    \\
    & \qquad\qquad
    \pm D J_2\,
    \Big(
    \codel{z}{\alpha}
    \,
    \sig{z}{n}\!
    \left(
      \sigmisscomp[-]{n \pm 2}{n \pm 1}{n \pm 3}{z}
    \right)
    \\
    & \qquad\qquad\qquad\qquad
    -(1 - \codel{z}{\alpha})\,
    \sig{\alpha}{n}
    (
      \sig{\alpha}{n \pm 1}
      -
      \sig{\alpha}{n \pm 3}
    )
    \sig{z}{n \pm 2}
  \Big)
    \\
    &
    \qquad\qquad
    \pm
    D J_1
    (1 + \codel{z}{\alpha})
    \sig{z}{n \pm 1}
    (
      \mathbb{I}
      -
      \sig{\alpha}{n} \sig{\alpha}{n \pm 2}
    )
  \\
  &
  \qquad\qquad
  +
  2 B_z J_2\,
  \lev{z}{\alpha}{\beta}
  \sig{\alpha}{n}\sig{\beta}{n\pm 2}
  \bigg)\,.
 \label{eq:Q_2_n_n}
\end{split}
\end{align}
\end{widetext}
In Eq.~\eqref{eq:Q_1_n_n}, the term proportional to ${(1 - \codel{z}{\alpha})}{J_1^2 D}$ and
in Eqs.~\eqref{eq:Q_1_n_n}~and~\eqref{eq:Q_2_n_n}, the terms proportional to ${J_2 D J_1}$ (the pre-last terms) are fully antisymmetric with respect to inversion $\hat{\mathcal{P}}$. Therefore, for translationally-invariant states
\begin{align}
  Q^{(1)}_{n-d,n}(t) \neq Q^{(1)}_{n,n-d}(t)\,,
\end{align}
applies and
\begin{align}
  \Delta C^{a/c,(1)}_{n,n+d}(t)
  &=
  \Delta C^{b,(1)}_{n,d}(t)
  =
  \Delta C\PHDG_{d}(t)
  \neq 0\,.
\end{align}
In the case of ${C^{c,(0)}_{n,n+d}(t)}$, the expectation values have to be taken for the corresponding spatially-inverted states.

\section{Exact Expressions for $L=4$ site system }
\subsection{One(three)-Magnon}\label{sec:L4_app_one_three_magnon}

The OTOC for the states $\ket{\mathrm{Ch}}$ Eq.~\eqref{eq:one_magnon_gs_L4_Ch} and $\ket{\mathrm{W}}$ Eq.~\eqref{eq:one_magnon_gs_L4_W} reads,
\begin{align}
 \label{eq:scrambling4spins}
  C^{\mathrm{Ch/W}}_{n,n \pm 1}(t)
  &=
  \brkt{\left|\!\left[
  \sig{z}{n\pm 1}(t),\sig{z}{n}
  \right]\!\right|^2}_{\mathrm{Ch/W}}
  \nonumber \\
  &=
  A^{\mathrm{Ch/W}}_1(t)\pm B^{\mathrm{Ch/W}}_1(t)\,,
\end{align}
with
\begin{widetext}
\begin{align}
\begin{split}
 \label{eq:scrambling4spins_A_Ch}
  A^{\mathrm{Ch}}_1(t)
  &=
  \sin^2\!\left(J_1 t\right)
  \left(
    1
    +
    \cos\left(2D\,t\right)
    +  
    \sin\left(D\,t\right) \sin\left(\left(D+4 J_2\right)t\right)
  \right)
  +
  \frac{1}{2}
  \sin^2\!\left(D\,t\right)
  \left(
    3 + \cos\left(2D\,t\right)
  \right),
\end{split}
\\
\begin{split}
 \label{eq:scrambling4spins_A_W}
  A^{\mathrm{W}}_1(t)
  &=
  \sin^2\!\left(D\,t\right)
  \left(
    1
    +
    \cos\left(2 J_1 t\right)
    +
    \sin\left(J_1 t\right) \sin\left(\left(J_1+4 J_2\right)t\right)
  \right)
  +
  \frac{1}{2}
  \sin^2\!\left(J_1 t\right)
  \left(
    3 + \cos\left(2 J_1 t\right)
  \right),
\end{split}
\end{align}
and
\begin{align}
\begin{split}
 \label{eq:scrambling4spins_B_Ch}
  B^{\mathrm{Ch}}_1(t)
  &=
  \sin\left(J_1 t\right)
  \Big(
     \sin^2\!\left(J_1 t\right) \cos\left(\left(D+2 J_2\right)t\right)
     \\
     &\qquad\qquad\qquad
    +\frac{1}{2}
     \sin\left(D\,t\right)
     \left(\vphantom{{}^Z_Z}
        \sin\left(2 \left(D\,-J_2\right)t\right)
      -2\sin\left(2 \left(D\,+J_2\right)t\right)
      -5\sin\left(2 J_2 t\right)
     \right)
  \Big),
\end{split}
\\
\begin{split}
 \label{eq:scrambling4spins_B_W}
  B^{\mathrm{W}}_1(t)
  &=
  \sin\left(D\,t\right)
  \Big(
     \sin^2\!\left(D\,t\right) \cos\left(\left(J_1+2 J_2\right)t\right)
     \\
     &\qquad\qquad\qquad
    +\frac{1}{2}
     \sin\left(J_1 t\right)
     \left(\vphantom{{}^Z_Z}
        \sin\left(2 \left(J_1-J_2\right)t\right)
      -2\sin\left(2 \left(J_1+J_2\right)t\right)
      -5\sin\left(2 J_2 t\right)
     \right)
  \Big),
\end{split}
\end{align}
symmetric and antisymmetric contributions, respectively.
Expressions for $\ket{\mathrm{Ch}}$ and $\ket{\mathrm{W}}$ are similar, with only $D$ and $J_1$ exchanged (${D \leftrightarrow J_1}$).
In the case of small $t$, expanding OTOC up-to subleading contribution in $t$, gives:
\begin{align}
\begin{split}
 \label{eq:scrambling4spins_Ser_Ch}
  C^{\mathrm{Ch}}_{n,n \pm 1}(t)
  &=
  2\,t^2
  \left(D^2+J_1^2\right)
  \mp
  t^3
  J_1 \left(J_1^2 + D \left(D+8 J_2\right)\right)
  +
  \mathcal{O}(t^4)\,,
\end{split}
\\
\begin{split}
 \label{eq:scrambling4spins_Ser_W}
  C^{\mathrm{W}}_{n,n \pm 1}(t)
  &=
  2\,t^2
  \left(D^2+J_1^2\right)
  \mp
  t^3
  D \left(D^2 + J_1 \left(J_1+8 J_2\right)\right)
  +
  \mathcal{O}(t^4)\,.
\end{split}
\end{align}

In the case of OTOC with the vector spin chirality operators, ${\hat{V}_n=\exp(\im\kappa^z_n)}$ and ${\hat{W}_m=\exp(\im\kappa^z_m)}$,
\begin{equation}
 \label{eq:scrambling4spins_chiral}
  C^{\mathrm{Ch/W}}_{n,n\pm 1}(t)
  =
  \sin^4\!\tfrac{1}{4}\left( A^{\mathrm{Ch/W}}_{c} \pm B^{\mathrm{Ch/W}}_{c} \right)
\end{equation}
with symmetric
\begin{align}
\begin{split}
  A^{\mathrm{Ch}}_{c}
  &=
  8 \cos^2\!\tfrac{1}{4}
  \left(
  1 + \sin^2\!\left(J_1 t\right)
  \right)
  -
  4 \sin^2\!\left(2 J_2 t\right) \cos^2\!\left(2 J_2 t\right)
  +
  4 \sin^2\!\left(J_1 t\right)
    \sin^2\!\left(\left(D+2 J_2\right)t\right)
  ,
\end{split}
\\
\begin{split}
  A^{\mathrm{W}}_{c}
  &=
  4 \sin^2\!\tfrac{1}{2}
  +
  8 \cos^2\!\tfrac{1}{4}
  \left(
  1
  -
  \sin^2\!\left(J_1 t\right)
     \left(
       1 - 2 \cos^2\!\tfrac{1}{4} \cos\left(2 J_1 t\right)
     \right)
  \right)
  -2 \sin\left(J_1 t\right) \sin\left(\left(J_1+4 J_2\right) t\right) \cos\left(2 D\, t\right)
  \\
  &\quad
  +2 \cos 1
    \left(
      \sin^2\left(\left(J_1+2 J_2\right) t\right)
     +\sin^2\!\left(2 J_2 t\right)
    \right)
\end{split}
\end{align}
and antisymmetric 
\begin{align}
  B^{\mathrm{Ch}}_{c}
  &=
  4 \cos^2\!\tfrac{1}{4} \sin\left(2 J_1 t\right) \cos\left(J_1 t \right) \cos\left(\left(D+2 J_2\right)t\right),
  \\
  B_{c}^{\mathrm{W}}
  &=
  8 \cos^2\!\tfrac{1}{4} \sin\left(D\,t\right) \cos\left(J_1 t\right)
    \left(\vphantom{{}^Z_Z}
        \cos\left(J_1 t\right) \cos\left(\left(J_1-2 J_2\right)t\right)
     +2 \cos\tfrac{1}{2} \sin\left(J_1 t\right) \sin\left(\left(J_1+2 J_2\right)t\right)
    \right)\,,
\end{align}
contributions, respectively.
Again, a nonchiral state exhibits the directional asymmetry only if the Hamiltonian has a nonvanishing DM interaction (the chiral term, ${D\neq 0}$).

\subsection{Two-Magnons in a four site chain, half-filled sector}

In the half-filled case, ${\spin{z}{\mathrm{tot}}=0}$ (two-magnon sector in the ${L = 4}$ site system),
two eigenstates of the Hamiltonian \eqref{eq:Hamiltonian} with energies $-\frac{1}{2}(J_1+2J_2+\sqrt{(J_1-4 J_2)^2 + 8 D^2})$ and ${(J_1+J_2)}$,
\begin{align}
\begin{split}
 \label{eq:two_magnon_gs_L4_ch}
 |\mathrm{Ch}\rangle
 &=
 \mu
 \Big(
 { |\!\downarrow\downarrow\uparrow\uparrow\rangle}
 {-|\!\uparrow\downarrow\downarrow\uparrow\rangle}
 {+|\!\uparrow\uparrow\downarrow\downarrow\rangle}
 {-|\!\downarrow\uparrow\uparrow\downarrow\rangle}
 +\im\lambda
  {|\!\downarrow\uparrow\downarrow\uparrow\rangle}
 -\im\lambda
  {|\!\uparrow\downarrow\uparrow\downarrow\rangle}
 \Big)\,,
\end{split}
\end{align}
and
\begin{align}
\begin{split}
 \label{eq:two_magnon_gs_L4_w}
 |\mathrm{W}\rangle
 &=
 \frac{1}{\sqrt{6}}
 \Big(
 { |\!\downarrow\downarrow\uparrow\uparrow\rangle}
 {+|\!\uparrow\downarrow\downarrow\uparrow\rangle}
 {+|\!\uparrow\uparrow\downarrow\downarrow\rangle}
 {+|\!\downarrow\uparrow\uparrow\downarrow\rangle}
 +{|\!\downarrow\uparrow\downarrow\uparrow\rangle}
 +{|\!\uparrow\downarrow\uparrow\downarrow\rangle}
 \Big)\,,
\end{split}
\end{align}
where
\begin{align}
 &\mu
 =
 \frac{D}{\sqrt{\varXi \left(\varXi+J_1-4 J_2\right)}}\,,
  \qquad
 \lambda
  =
  \frac{\varXi+J_1-4J_2}{2D}\,,
 \qquad
 \varXi
 =
 \sqrt{\left(J_1-4 J_2\right)^2 + 8 D^2_{\phantom{2}}}\,,
 \nonumber 
\end{align}
have a finite (${\braket{\mathrm{Ch}}{\hat{\kappa}^z}{\mathrm{Ch}}={D}/{\varXi}}$) and zero chirality (${\braket{\mathrm{W}}{\hat{\kappa}^z}{\mathrm{W}} = 0}$), respectively.
These are also the eigenstates of the translation operator with the crystal momentum $\pi/2$ and $0$, respectively.
OTOC has only symmetric contribution in both cases
\begin{equation}
 \label{eq:scrambling4spins_M2_appendix}
  C^{\mathrm{Ch/W}}_{n,n \pm 1}(t) = A^{\mathrm{Ch/W}}_2(t)
\end{equation}
with:
\begin{align}
\begin{split}
 \label{eq:scrambling4spins_A_Ch_M2_appendix}
   A^{\mathrm{Ch}}_2(t)
   &=
   2
   -\frac{4D^2}{9\varXi^2}
    \Big(\!
     4 \cos \left(4 J_2 t \right)
    +4 \cos \left(\left(3 J_1-4 J_2\right)t\right)
    +  \cos \left(\left(3 J_1+4 J_2\right)t\right)
    \!\Big)
   \\
   &\!\qquad
   -\frac{\left(\varXi-(J_1-4 J_2)\right)}{3\varXi}
   \left(
    2 \cos \left(\tfrac{1}{2} \left(\varXi-  J_1-4 J_2\right)t\right)
   +  \cos \left(\tfrac{1}{2} \left(\varXi+5 J_1-4 J_2\right)t\right)
   \right)
   \\
   &\!\qquad
   -\frac{\left(\varXi+(J_1-4 J_2)\right)^2}{18\varXi^2}
    \Big(\!
     4 \cos \left(\left(\varXi + 4 J_2\right)t\right)
    +4 \cos \left(\left(\varXi - 3 J_1+4 J_2\right)t\right)
    +  \cos \left(\left(\varXi + 3 J_1+4 J_2\right)t\right)
    \!\Big)
    .
\end{split}
\\
\begin{split}
 \label{eq:scrambling4spins_A_W_M2_appendix}
   A^{\mathrm{W}}_2(t)
   &=
   2
   -\frac{8D^2}{9\varXi^2}
    \Big(\!
     2 \cos \left(4 J_2 t \right)
    -  \cos \left(\left(3 J_1+4 J_2\right)t\right)
    \!\Big)
   \\
   &\!\qquad
   -\frac{2\left(\varXi-(J_1-4 J_2)\right)}{3\varXi}
    \cos \left(\tfrac{1}{2} \left(\varXi-J_1-4 J_2\right)t\right)
   -\frac{2\left(\varXi+(J_1-4 J_2)\right)}{3\varXi}
    \cos \left(\tfrac{1}{2} \left(\varXi+J_1+4 J_2\right)t\right)
   \\
   &\!\qquad
   -\frac{\left(\varXi-(J_1-4 J_2)\right)^2}{18 \varXi^2}
    \left(
      \vphantom{\Big|}
      2 \cos \left(\left(\varXi-4 J_2 \right)t\right)
       +\cos \left(\left(\varXi-3 J_1-4 J_2\right)t\right)
    \right)
   \\
   &\!\qquad
   -\frac{\left(\varXi+(J_1-4 J_2)\right)^2}{18 \varXi^2}
    \left(
      \vphantom{\Big|}
      2 \cos \left(\left(\varXi+4 J_2 \right)t\right)
       +\cos \left(\left(\varXi+3 J_1+4 J_2 \right)t\right)
    \right)
    .
\end{split}
\end{align}
The scrambling is symmetric, $\Delta C^{{\mathrm{Ch}}/\mathrm{W}}_g(t) = 0$.

We will not consider $n\pm 2$ case, because for ${L=4}$ spin chain with PBC, the site on the distance $|d|=2$ could be reached from both sides of the chain from the initial position, hence directional asymmetry is trivially zero.
\end{widetext}

\section{The decay exponent for small time limit}
\label{sec:decay_exp_small_t}

In this section, we will estimate the decay exponent $\lambda(v)$ \eqref{eq:OTOC1_LV},
for the ${d=vt}$ rays at fixed (given) velocity $v$, with ${vt \gg 1}$ and ${t \ll 1}$.
We can define $\lambda(v)$ for arbitrary large $v$ for spin-chains (unlike to local quantum circuits and relativistic field theories, where even exponentially weak signaling is impossible beyond a strict ``light cone'').

At the integer ${vt=2k}$ or ${vt=k}$, ${k \in \mathbb{Z}}$, distances,
the leading in $t$ contribution, following Eq.~\eqref{eq:sigma_leading} (${\nu=k}$), is
\begin{equation}
  C_{d = vt = 2k}(t) = \frac{t^{2k}}{(k!)^2} A(2k) = \frac{t^{v t}}{((v t/2)!)^2} A(vt)
\end{equation}
or
\begin{equation}
  C_{d = vt}(t) = \frac{t^{2k}}{(k!)^2} B(k) = \frac{t^{2 v t}}{((v t)!)^2} B(vt)\,,
\end{equation}
for ${J_2 \neq 0}$ and ${J_2 = 0}$, respectively.
Here ${A(vt)} = {\brkttext{|[[\hat{H},\hat{W}_0]\PHDG_{\nu=k},\hat{V}_{d=\pm 2k}]|^2} \sim \mathcal{O}(1)}$, ${B(vt)} = \brkttext{|[[\hat{H},\hat{W}_0]\PHDG_{\nu=k},\hat{V}_{d=\pm k}]|^2} \sim \mathcal{O}(1)$, and $[\hat{H},\hat{W}_0]\PHDG_{\nu=k}$ is a nested commutator, Eq.~\eqref{eq:nested_commutators}. Note that for free fermions on 1D lattice, $B(k)=1$ \cite{Xu2020}.

Estimates will be the same for both cases (substitute ${v\equiv 2v'}$ in the case of ${J_2 \neq 0}$).
The decay exponent
\begin{align}
 \begin{split}
  \lambda(v)
  &
  = \frac{\ln C_{d=vt}(t)}{t} \\
  &
  = 2 v\ln t - \frac{2\ln((vt)!)}{t} + \frac{\ln(B(vt))}{t}\,.
 \end{split}
\end{align}
For ${vt \gg 1}$ and ${t \ll 1}$, employing Stirling's formula for the factorial ($\ln n! = n \ln n - n + \mathcal{O}(\ln n)$), it can be approximated as
\begin{align}
 \label{eq:v_ln_v_appx}
  \lambda(v)
  &
  \approx
  2 v\ln t
  -
  2\frac{vt \ln (vt) - vt + \mathcal{O}(\ln(vt))}{t}
  \nonumber \\          
  &
  =
  -2v \left(\ln v - 1\right) - \mathcal{O}\left(\frac{\ln(vt)}{t}\right)\,.
\end{align}
Here, we also dropped ${{\ln(B(vt))}/{t}}$ term because ${\mathcal{O}\left({\ln(vt)}\right) > \mathcal{O}\left({\ln(B(vt))}\right)}$ ($B(vt) \sim \mathcal{O}(1)$).
Consequently, for the fixed-velocity rays with ${vt \gg 1}$ and ${t \ll 1}$, ${|\lambda(v)| \approx 2v ( \ln v)}$ grows slower in $v$ than ${2 v^{\alpha}}$ where ${\alpha>1}$.

\section{Early time regime for OTOC with $\e^{\im \hat{\kappa}^z_j}$}
\label{sec:appendix_early_time_reg}

In this appendix we consider the early-time regime of the $C^{\kappa\kappa}_d(t)$.
In all studied cases, we observe a power-law growth of the OTOC, as shown, for example, for ${D=0.5}$, in the center panels of Fig.~{\ref{fig:shorttime_ckkd}, and this is consistent with the discussion in Sec.~\ref{sec:short_time_limit}. At leading order, the OTOC behaves as $t^0$ for ${|d|=1}$ and as $t^{2\nu}$ with ${\nu=\max(1,|d|\ \mathbf{div}\ 2)}$ for ${|d|>1}$.
For the chiral initial state or in a system with a nonvanishing DM interaction, we also observe the asymmetric subleading, $t^1$ for ${|d|=1}$ and $t^{2\nu+1}$ for ${|d|>1}$, corrections to it (see the bottom panel in Fig.~\ref{fig:shorttime_ckkd}).
\vspace{2em}

\begin{figure}[!thpb]
 \includegraphics[width=\columnwidth]{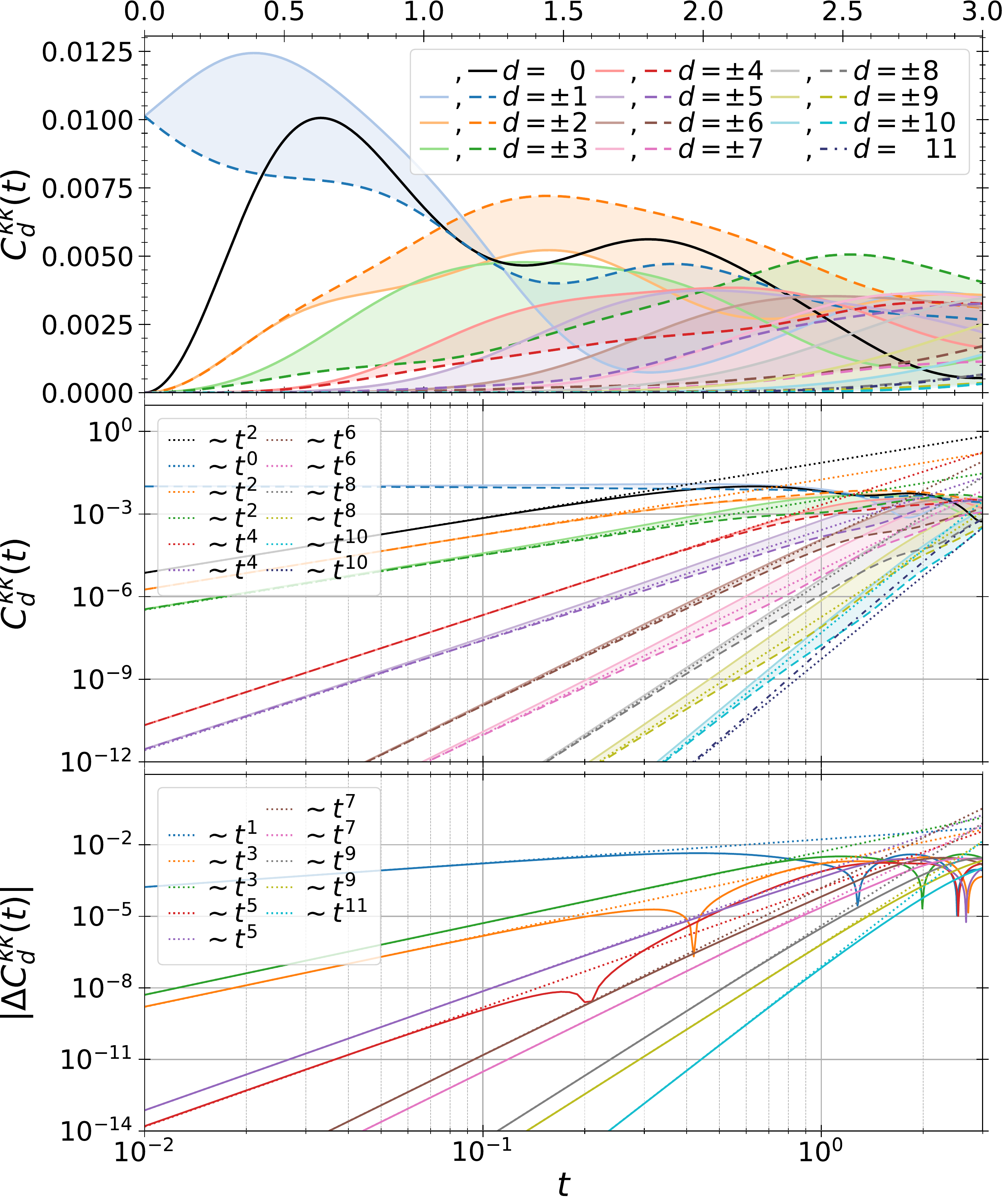}
 \caption{Short-time behavior of OTOC $C^{\kappa\kappa}_d(t)$ from Fig.~\ref{fig:ckkd_ckkd_J2_0}~b) (${D=0.5}$)
          and a corresponding directional asymmetry $|\Delta C^{zz}_d(t)|$.
          Dotted lines in the center and bottom log-log plots correspond to power-law fits for ${t \ll 1}$ .
         }
\label{fig:shorttime_ckkd}
\end{figure}
\section{Saturation Value of OTOC}
\label{sec:apendix_sat_val_OTOC}

The general form of OTOC is given as
\begin{align}
 \label{eq:OTOC_appendix}
 \begin{split}
  C(t)
  &=
  \brkttext{|[\hat{W}(t),\hat{V}]|^2}
  \\
  &=
  \!
  \brkt
  {
    \hat{W}^\dagger\!(t) \hat{W}\!(t)    \hat{V}^\dagger \hat{V} \!
  - \hat{W}^\dagger\!(t) \hat{V}^\dagger \hat{W}\!(t)    \hat{V} \!
  + \mathrm{H.c.}\!
  }\!,
 \end{split}
\end{align}
where the second term:
\begin{equation}
 \label{eq:OTOC1_FT_appendix}
  F(t)
  =
  \brkt
  {
    \hat{W}^\dagger(t) \hat{V}^\dagger   \hat{W}(t)       \hat{V}
  }
\end{equation}
is also known as out-of-time-ordered correlator.

For unitary operators $\hat{W}\!(t)$ and $\hat{V}$ the first term in Eq.~\eqref{eq:OTOC_appendix} reduces to
\begin{equation}
 \label{eq:OTOC_appendix_1st_Unit}
  \brkt
  {
    \hat{W}^\dagger\!(t) \hat{W}\!(t) \hat{V}^\dagger    \hat{V}
  }
  = 1
\end{equation}
otherwise
\begin{widetext}
\begin{align}
 \label{eq:OTOC_appendix_1st_else}
 \begin{split}
  \brkt
  {
    \hat{W}^\dagger\!(t) \hat{W}\!(t) \hat{V}^\dagger \hat{V}
  }
  &=
  \sum_{j,k,p,q,r}
  \e^{-i(E_j-E_p)t}
  W^{*}_{jk} W\PHDG_{kp} V^{*}_{pq} V\PHDG_{qr}
  \rho\PHDG_{jr}
  \\
  &=
  \sum_{j,k,q,r}
  W^{*}_{jk} W\PHDG_{kj} V^{*}_{jq} V\PHDG_{qr}
  \rho\PHDG_{jr}
  +
  \sum_{\substack{j,k,q,r \\ p \neq j}}
  \e^{-i(E_j-E_p)t}
  W^{*}_{jk} W\PHDG_{kp} V^{*}_{pq} V\PHDG_{qr}
  \rho\PHDG_{jr}
  \nonumber
 \end{split}
  \\
  &=
  \sum_{j,r}
  \brakettext{j}{\hat{W}^\dagger\hat{W}}{j}
  \brakettext{j}{\hat{V}^\dagger\hat{V}}{r}
  \rho\PHDG_{jr}
  +
  \sum_{\substack{j, r \\ p \neq j}}
  \e^{-i(E_j-E_p)t}
  \brakettext{j}{\hat{W}^\dagger\hat{W}}{p}
  \brakettext{p}{\hat{V}^\dagger\hat{V}}{r}
  \rho\PHDG_{jr}
  .
\end{align}
Here, ${ \brkttext{\cdot}\equiv\brakettext{\psi}{\cdot}{\psi} }$, ${A_{pq} \equiv \brakettext{p}{\hat{A}}{q}}$,
${\rho_{pq} \equiv \brkttext{\ket{p}\!\bra{q}}}$; $\ket{j}$, $\ket{k}$, $\ket{p}$, $\ket{q}$, and $\ket{r}$,
are the eigenstates of the system Hamiltonian (${H\ket{i} = E_i\ket{i}}$);
the initial state $\ket{\psi}$ resides in the given $S^z_{\mathrm{tot}}$-sector, ${ \brakettext{\psi}{\spin{z}{\mathrm{tot}}}{\psi} = \brkttext{\spin{z}{\mathrm{tot}}} = S^z_{\mathrm{tot}}}$, ${\brktprod{\psi}{\psi}=\Tr \rho=1}$.
For the system Hamiltonian with a non-degenerate energy spectrum, the long-time average of the last term will vanish in the limit of $t\rightarrow \infty$.

The second term in Eq.~\eqref{eq:OTOC_appendix} or equivalently Eq.~\eqref{eq:OTOC1_FT_appendix} can be rewritten as
\begin{align}
 \begin{split}
  F(t)
  &=
  \brkt
  { 
    \hat{W}^\dagger\!(t) \hat{V}^\dagger \hat{W}\!(t) \hat{V}
  }
  \\
  &=
  \sum_{j,k,p,q,r}
  \e^{-i(E_j-E_k+E_p-E_q)t}
  W^{*}_{jk} V^{*}_{kp} W\PHDG_{pq} V\PHDG_{qr}
  \rho\PHDG_{jr}
  \nonumber
 \end{split}
  \\
 \begin{split}
  &=
  \sum_{j,r,p}
  W^{*}_{jj} V^{*}_{jp} W\PHDG_{pp} V\PHDG_{pr}
  \rho\PHDG_{jr}
  +
  \sum_{j,r,p}
  |W\PHDG_{pj}|^2 V^{*}_{pp} V\PHDG_{jr}
  \rho\PHDG_{jr}
  -
  \sum_{j,r}
  |W\PHDG_{jj}|^2 V^{*}_{jj} V\PHDG_{jr}
  \rho\PHDG_{jr}
  \\
  &
  \quad
  +
  \sum_{r}
  \sideset{}{'}\sum_{j,k,p,q}\vphantom{k}
  \e^{-i(E_j-E_k+E_p-E_q)t}
  W^{*}_{jk} V^{*}_{kp} W\PHDG_{pq} V\PHDG_{qr}
  \rho\PHDG_{jr}.
 \label{eq:OTOC_appendix_2nd}
 \end{split}
\end{align}
Here $\sum^{\prime}$ denotes the summation over indices 
where non of the following pairs of indices are equal simultaneously: ${(j,k)}$ and ${(p,q)}$ or
${(j,q)}$ and ${(p,k)}$.
The third sum in Eq.~\eqref{eq:OTOC_appendix_2nd} corresponds to the double-counted term in the second sum.

In the case of Pauli operators, $\sig{\alpha}{}$-s, the first term in Eq.~\eqref{eq:OTOC_appendix} reduces to
\begin{align}
 \begin{split}
 \label{eq:OTOC_aa_appendix_1st}
  \brkt{\sig{\alpha}{n}(t) \sig{\alpha}{n}(t) \sig{\alpha}{m} \sig{\alpha}{m}} = 1
 \end{split}
\end{align}
(Eq.~\eqref{eq:OTOC_appendix_1st_Unit}, $\sig{\alpha}{}$-s are also unitary operators) and for the second term in Eq.~\eqref{eq:OTOC_appendix}, ${F(t)}$, from Eq.~\eqref{eq:OTOC_appendix_2nd} follows
\begin{align}
  F^{\alpha\alpha}_{nm}(t)
  &=
  \brkt
  { 
    \sig{\alpha}{m}(t) \sig{\alpha}{n} \sig{\alpha}{m}(t) \sig{\alpha}{n}(t)
  }
  \nonumber
  \\
 \label{eq:OTOC1_FT_aa_appendix}
 \begin{split}
  &=
  \sum_{j,p,r}
  \brakettext{j}{\sig{\alpha}{m}}{j}
  \brakettext{j}{\sig{\alpha}{n}}{p}
  \brakettext{p}{\sig{\alpha}{m}}{p}
  \brakettext{p}{\sig{\alpha}{n}}{r}
  \rho\PHDG_{jr}
  +
  \sum_{j,p,r}
  \left|\brakettext{p}{\sig{\alpha}{m}}{j}\right|^2
  \brakettext{p}{\sig{\alpha}{n}}{p}
  \brakettext{j}{\sig{\alpha}{n}}{r}
  \rho\PHDG_{jr}
  \\&
  \quad
  -
  \sum_{j,r}
  \left|\brakettext{j}{\sig{\alpha}{m}}{j}\right|^2
  \brakettext{j}{\sig{\alpha}{n}}{j}
  \brakettext{j}{\sig{\alpha}{n}}{r}
  \rho\PHDG_{jr}
  \\&
  \quad
  +
  \sum_{r}
  \sideset{}{'}\sum_{j,k,p,q}
  \e^{-i(E_j-E_k+E_p-E_q)t}
  \brakettext{j}{\sig{\alpha}{m}}{k}
  \brakettext{k}{\sig{\alpha}{n}}{p}
  \brakettext{p}{\sig{\alpha}{m}}{q}
  \brakettext{q}{\sig{\alpha}{n}}{r}
  \rho\PHDG_{jr}
  .
 \end{split}
\end{align}
For the $\spin{z}{\mathrm{tot}}$ conserving Hamiltonian, ${[\spin{z}{\mathrm{tot}},\hat{H}]=0}$,
${\brakettext{i}{\sig{x/y}{n}}{i} = 0}$ 
for any site ${n=1,\dots,L}$ and any eigenstate $\ket{i}$ of the system Hamiltonian,
because ${\sig{x}{n} = \spin{+}{n} + \spin{-}{n}}$ and 
${\sig{y}{n} = \im(\spin{+}{n} - \spin{-}{n})}$ both alter the $S^z_{\mathrm{tot}}$-sector.
Therefore, for $\sig{x/y}{}$-s from Eq.~\eqref{eq:OTOC1_FT_aa_appendix} follows
\begin{equation}
 \label{eq:OTOC1_FT_xx_appendix}
  F^{xx/yy}_{nm}(t)
  =
  \sideset{}{'}\sum_{j,k,p,q,r}
  \e^{-i(E_j-E_k+E_p-E_q)t}
  \brakettext{j}{\sig{x/y}{m}}{k}
  \brakettext{k}{\sig{x/y}{n}}{p}
  \brakettext{p}{\sig{x/y}{m}}{q}
  \brakettext{q}{\sig{x/y}{n}}{r}
  \rho\PHDG_{jr}\,.
\end{equation}
As for $\sig{z}{}$-s, ${\brakettext{i}{\sig{z}{n}}{i}=2 S^z_{\mathrm{tot}}/L}$ 
for the translationally invariant system, for any site ${n=1,\dots,L}$ and any eigenstate $\ket{i}$ of the system Hamiltonian.
Therefore, for the translationally invariant initial state $\ket{\psi}$, which is also an eigenstate of the $z$-component of the total spin, ${ \spin{z}{\mathrm{tot}}\ket{\psi}=S^z_{\mathrm{tot}}\ket{\psi} }$,
\begin{align}
 \begin{split}
  F^{zz}_{nm}(t)
  &=
  \sum_{j,p,r}
  \left( \! \frac{2S^z_{\mathrm{tot}}}{L} \! \right)^2 \!\!
  \brakettext{j}{\sig{z}{n}}{p}
  \brakettext{p}{\sig{z}{n}}{r}
  \rho\PHDG_{jr}
  +
  \sum_{j,p,r}
  \left( \! \frac{2S^z_{\mathrm{tot}}}{L} \! \right)
  \left|\brakettext{p}{\sig{z}{m}}{j}\right|^2
  \brakettext{j}{\sig{z}{n}}{r}
  \rho\PHDG_{jr}
  -
  \sum_{j,r}
  \left( \! \frac{2S^z_{\mathrm{tot}}}{L} \! \right)^3 \!\!
  \brakettext{j}{\sig{z}{n}}{r}
  \rho\PHDG_{jr}
  \\
  &\quad
  +
  \sideset{}{'}\sum_{j,k,p,q,r}
  \e^{-i(E_j-E_k+E_p-E_q)t}
  \brakettext{j}{\sig{z}{m}}{k}
  \brakettext{k}{\sig{z}{n}}{p}
  \brakettext{p}{\sig{z}{m}}{q}
  \brakettext{q}{\sig{z}{n}}{r}
  \rho\PHDG_{jr}
  \\
  &=
  \left( \! \frac{2S^z_{\mathrm{tot}}}{L} \! \right)^2 \!\!
  \brakettext{\psi}{ \sig{z}{n} {\sig{z}{n}} }{\psi}
  +
  \sum_{j,r}
  \left( \! \frac{2S^z_{\mathrm{tot}}}{L}\! \right)
  \brakettext{j}{ \sig{z}{m} \sig{z}{m} }{j}
  \brakettext{j}{\sig{z}{n}}{r}
  \rho\PHDG_{jr}
  -
  \left( \! \frac{2S^z_{\mathrm{tot}}}{L} \! \right)^4 \!\!
  \\
  &\quad
  +
  \sideset{}{'}\sum_{j,k,p,q,r}
  \e^{-i(E_j-E_k+E_p-E_q)t}
  \brakettext{j}{\sig{z}{m}}{k}
  \brakettext{k}{\sig{z}{n}}{p}
  \brakettext{p}{\sig{z}{m}}{q}
  \brakettext{q}{\sig{z}{n}}{r}
  \rho\PHDG_{jr}
  \nonumber
 \end{split}
  \\
 \label{eq:OTOC1_FT_zz_appendix}
  &=
  2 \left( \! \frac{2S^z_{\mathrm{tot}}}{L} \! \right)^2\!\!
  -
  \left( \! \frac{2S^z_{\mathrm{tot}}}{L} \! \right)^4\!\!
  +
  \sum_{r}
  \sideset{}{'}\sum_{j,k,p,q}
  \e^{-i(E_j-E_k+E_p-E_q)t}
  \brakettext{j}{\sig{z}{m}}{k}
  \brakettext{k}{\sig{z}{n}}{p}
  \brakettext{p}{\sig{z}{m}}{q}
  \brakettext{q}{\sig{z}{n}}{r}
  \rho\PHDG_{jr}\,.
\end{align}
Here we used that ${\sig{\alpha}{n}\sig{\alpha}{n}=\mathbb{I}}$,
${ \sum_j \brakettext{j}{\hat{A}}{r}\rho\PHDG_{jr}=\brakettext{\psi}{\hat{A}}{\psi} }$, and ${ \brakettext{\psi}{\sig{z}{n}}{\psi}=2 S^z_{\mathrm{tot}}/L }$ for translationally invariant state $\ket{\psi}$.

For the Hamiltonian with a non-degenerate energy spectrum, if OTOC converges/saturates for long times, from Eqs.~\eqref{eq:OTOC1_FT_xx_appendix} and \eqref{eq:OTOC1_FT_zz_appendix} follows:
\begin{align}
 \label{eq:OTOC1_FT_xx_appendix_limit}
  F^{xx/yy}_{nm}(t\rightarrow \infty)
  &=
  0\,,
  \\
 \label{eq:OTOC1_FT_zz_appendix_limit}
  F^{zz}_{nm}(t\rightarrow \infty)
  &=
  2 \left( \! \frac{2S^z_{\mathrm{tot}}}{L} \! \right)^2 \!\!
  -
  \left( \! \frac{2S^z_{\mathrm{tot}}}{L} \! \right)^4 ,
\end{align}
unless there are some extra symmetries (e.g., symmetric energy spectrum) for which ${(E_j - E_k + E_p - E_q) = 0}$ in the $\sum'$.
Consequently, plugging Eqs.~\eqref{eq:OTOC_aa_appendix_1st}-\eqref{eq:OTOC1_FT_zz_appendix_limit} in Eq.~\eqref{eq:OTOC_appendix} yield:
\begin{align}
 \label{eq:OTOC_xx_appendix_limt}
  C^{xx/yy}_{nm}(t\rightarrow \infty)
  &
  =
  2\,,
  \\
 \label{eq:OTOC_zz_appendix_limit}
  C^{zz}_{nm}(t\rightarrow \infty)
  &
  =
  2\! \left( \! 1-\left( \! \frac{2S^z_{\mathrm{tot}}}{L} \! \right)^2 \right)^2 .
\end{align}
If OTOC does not converge, Eqs.~\eqref{eq:OTOC1_FT_xx_appendix_limit}-\eqref{eq:OTOC_zz_appendix_limit} still give accurate time-averaged values, namely:
\begin{align}
 \label{eq:OTOC1_FT_xx_appendix_limit_TA}
  &
  \lim_{t\rightarrow \infty}
  \frac{1}{t}\int_{0}^{t} F^{xx/yy}_{nm}(t^\prime)\ud t^\prime
  =
  0,
  \\
 \label{eq:OTOC1_FT_zz_appendix_limit_TA}
  &
  \lim_{t\rightarrow \infty}
  \frac{1}{t}\int_0^t \ \ F^{zz}_{nm}(t^\prime)  \ \ \ud t^\prime
  =
  2 \left( \frac{2S^z_{\mathrm{tot}}}{L} \right)^2 
  -
  \left( \frac{2S^z_{\mathrm{tot}}}{L} \right)^4 ,
\end{align}
and
\begin{align}
 \label{eq:OTOC_xx_appendix_limt_TA}
  &
  \lim_{t\rightarrow \infty}
  \frac{1}{t}\int_0^t
  C^{xx/yy}_{nm}(t^\prime)
  \ud t^\prime
  = 2\,,
  \\
 \label{eq:OTOC_zz_appendix_limit_TA}
  &
  \lim_{t\rightarrow \infty}
  \frac{1}{t}\int_0^t
  \ \ C^{zz}_{nm}(t^\prime)\ \ 
  \ud t^\prime
  =
  2 \left(  1-\left( \frac{2S^z_{\mathrm{tot}}}{L} \right)^2 \right)^2 .
\end{align}
\end{widetext}

For the studied system, in the case of a vanishing DM interaction, e.g., in the $2$-excitation sector of $H(J_1=-1,J_2=1,D=0,b)$ of ${L=22}$ site spin chain with PBC, all interior eigenenergies except for a few are doubly degenerate.
In the case of exactly solvable (ferromagnetic) Heisenberg model, $H(J_1=-1,J_2=0,D=0,b)$, the level of energy degeneracy is even higher.
For these cases, there will be the time-independent contributions from the $\sum^{\prime}$ also. In this respect, OTOC is sensitive to the degeneracies in the energy spectrum of the Hamiltonian and can be utilized for the detection of some quantum phase transitions.

\begin{figure*}[!thbp]
 \includegraphics[width=\textwidth]{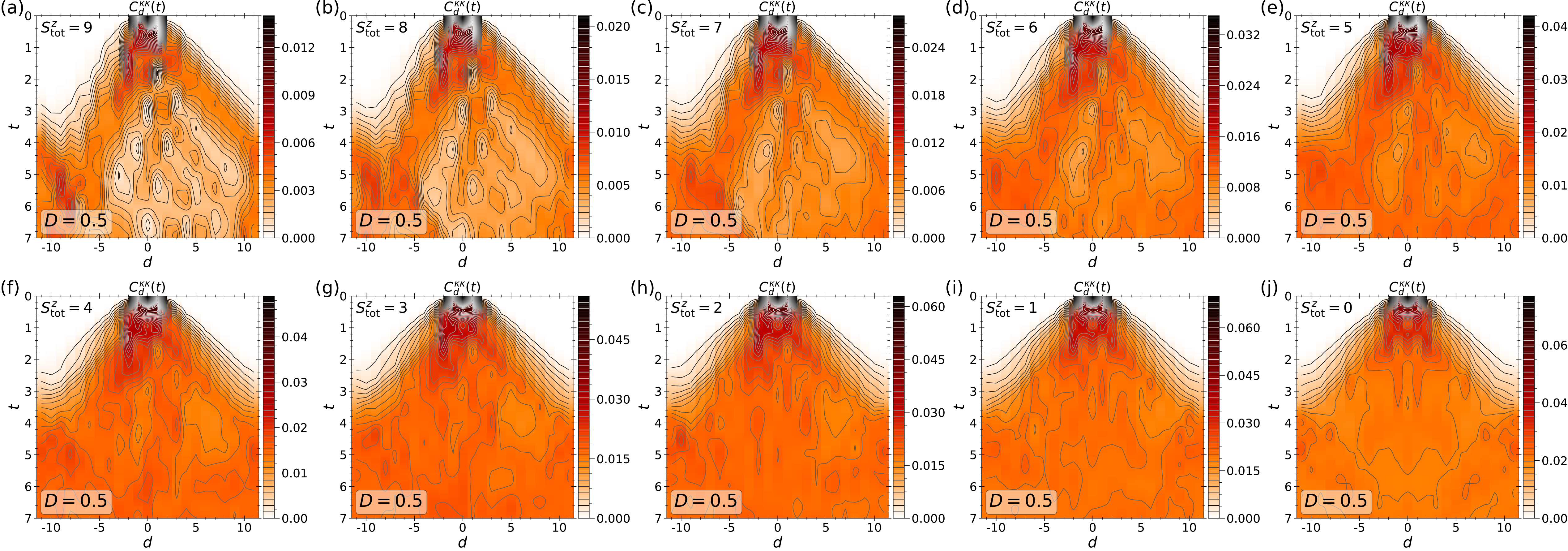}\\
 \includegraphics[width=\textwidth]{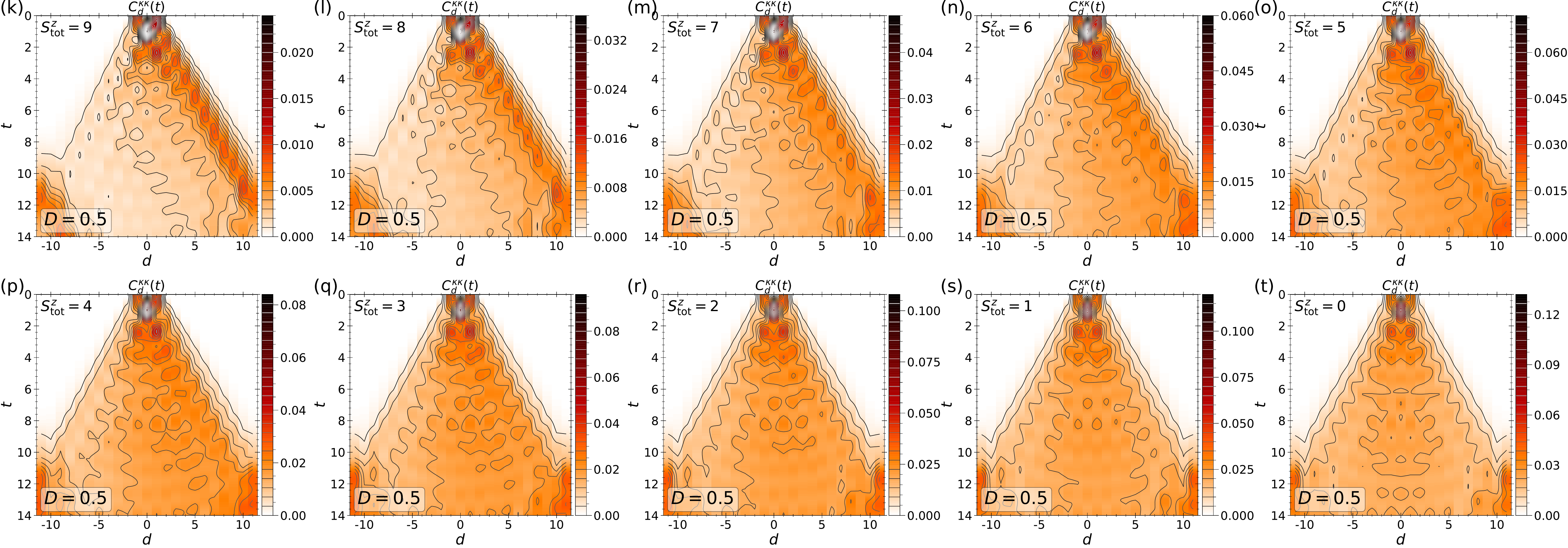}
 \caption{Spatiotemporal evolution of the OTOC $C^{\kappa\kappa}_d(t)$~\eqref{eq:ckkd} for the chiral case,
          in ${2,\ldots,L/2}$-excitation sector (${S^z_{\mathrm{tot}}=9,\ldots,0}$) of ${L=22}$ spin chain with PBC; (a)-(j) ${J_2=-J_1=1}$, ${D=0.5}$, and (k)-(t) ${J_1=-1}$, ${J_2=0}$, ${D=0.5}$ (integrable case);
          The initial chiral-state is prepared as the ground state of the system; Time is measured in units of $|J_1|^{-1}$;
          The upper limit of the color bar, is scaled with the excitation-sector number.
          In each plot, there is the same number of contour lines in the color-bar range.
          A black and white spots around ${d=0}$ ${t=0}$ are due to contour lines.
         }
\label{fig:ckkd_ckkd_pN}
\end{figure*}

\section{OTOC for different magnetization sectors}

In this appendix we show spatiotemporal evolution of OTOC $C^{\kappa\kappa}_d(t)$ for the ${L=22}$ site system and different ${S^z_{\mathrm{tot}}=9,\ldots,0}$ magnetization sectors, for both non-integrable ${J_2=-J_1=1}$ (see Fig.~\ref{fig:ckkd_ckkd_pN}(a)-(j)) and integrable ${J_2=0,\,J_1=-1}$ (see Fig.~\ref{fig:ckkd_ckkd_pN}(k)-(t)) cases.
We chose $C^{\kappa\kappa}_d(t)$ and the asymmetric case ${D=0.5}$ to demonstrate that the directional asymmetry is indeed vanishing as one approaches the ${S^z_{\mathrm{tot}}=0}$ (half-filling) sector, as it was also concluded from the symmetry consideration (see sec.~\ref{sec:anisotropy_measures}).


%

\end{document}